# Causation-guided mechanism identification and interpretable reduced-order modeling of damage-driving grain-boundary stress in creep

Weichen Kong [a], Yanwei Dai [b,1], Yinling Zhang [c], Yinghua Liu [a,1]

[a] Department of Engineering Mechanics, AML, Tsinghua University, Beijing 100084, China

[b] Department of Mechanics, Beijing University of Technology, Beijing 100124, China

[c] Department of Mathematics, University of Wisconsin, Madison, WI 53705, USA

[1] Corresponding authors:

E-mail addresses: ywdai@bjut.edu.cn (Yanwei Dai), yhliu@mail.tsinghua.edu.cn (Yinghua Liu).

## Abstract

Grain-boundary (GB) local stress is central to the initiation and evolution of long-term creep damage in polycrystalline superalloys. Owing to the high-dimensional nonlinear relationships between the GB stress response and multiple crystallographic, microstructural, and micromechanical characteristics, it remains challenging to identify the key characteristics governing GB stress and to elucidate their mechanisms of influence. Dislocation-climb-affected crystal-plasticity finite-element simulations of minimal grain clusters are combined with an integrated causation-guided machine-learning framework, in which mechanics-informed descriptors are analyzed by causation entropy (CE) to identify governing mechanisms and then distilled into a reduced-order regression form for interpretable prediction of GB normal stress. Unlike conventional black-box surrogate models, the proposed framework combines physically meaningful descriptors, CE-based mechanism identification, and an explicit compact regression form, enabling the dominant physical variables and their effects on GB normal stress to be directly interpreted. Among 18 physically motivated characteristics, the GB inclination angle, the slip transmission, the climb-related Schmid-type indicator, and the elastic-modulus mismatch are found to be dominant, revealing the coupled roles of interfacial geometry, crystallographic compatibility, creep stress relaxation, and micromechanical contrast. The identified characteristics hierarchy and functional representation remain effective under multiaxial loading and can be extended to tricrystal systems through physically interpretable nonlocal augmentation when a purely local description becomes insufficient, demonstrating strong physical consistency and robust generalizability across physical conditions. The extracted candidate functions also improve surrogate-model performance across multiple machine-learning model classes, providing supporting evidence for the physical relevance and efficiency of the identified representation. The proposed methods demonstrate strong potential for the development of interpretable machine-learning models and for the study of microscale nonlocal damage.

# 1. Introduction

Creep in superalloys has long posed a persistent challenge in energy and power systems, owing to its multiscale nature, mechanistic complexity, and limited predictability. For long-term high-temperature creep, the characteristic damage mechanism in polycrystalline alloys is intergranular cavitation, including cavity nucleation, coalescence, and growth (Cocks and Ashby, 1982; Sagara et al., 2026; Sham and Needleman, 1983; Shibanuma et al., 2024; Van Der Giessen et al., 1995; Wen et al., 2017).

In models of grain-boundary (GB) cavity growth, the GB normal stress is often regarded as the primary driving force, directly governing the subsequent evolution of intergranular damage. Classical models of GB cavity growth and GB sliding were largely developed within idealized continuum frameworks. In works of Raj and Ashby (1971) and Cocks and Ashby (1982), the GB normal stress was typically treated as an effective or mean driving force for cavity growth, rather than as a spatially resolved local field along the boundary. In the formulation of Needleman and Rice (1980), GB stress was analyzed within a coupled grain-boundary diffusion–creep field, but still under homogeneous and idealized continuum assumptions. These classical studies primarily focused on cavity growth behavior, rather than with explicitly resolving the spatially heterogeneous GB stress state arising in real polycrystals from crystallographic anisotropy, orientation mismatch, and neighboring-grain constraint.

Crystal-plasticity-based modeling provides a natural route for investigating local GB stresses while accounting for microstructural features. From a theoretical perspective, interactions between intragranular dislocations and GBs have also been incorporated into crystal-plasticity frameworks to investigate grain-boundary sliding and migration (Admal et al., 2018; van Beers et al., 2013). From a numerical perspective, Gonzalez et al. (2014) employed crystal plasticity finite element (CPFE) to calculate GB normal stresses in 316 stainless steel during uniaxial loading and unloading, and investigated the respective roles of elastic and plastic mismatch in governing the grain-boundary stress state. The effect of GB sliding mechanisms on GB stresses has also been specifically investigated (Petkov et al., 2021). Chen and Furushima (2024) analyzed the relationship between stress triaxiality and physically based descriptors of intergranular deformation incompatibility. Moreover, CPFE methods have been extended to creep, thereby providing a foundation for the analysis of local stress fields in alloys (Gan et al., 2024; Liang et al., 2024; Zhang et al., 2020). Phan et al. (2017) used CPFE to investigate GB creep damage in the polycrystalline superalloy and statistically analyzed the relationship between GB misorientation and damage. Salvini et al. (2024) coupled CPFE with a phase-field approach to account for GB-misorientation-dependent variations in GB damage initiation and evolution. The authors further incorporated the effects of GB orientation and area into the modeling of GB damage evolution variability, and linked microstructural heterogeneity to creep life predictions (Kong et al., 2025). Although these approaches are built upon certain assumptions, they are physically transparent and offer strong interpretability. However, due to the complexity of micromechanical modeling and the high computational cost, the range of factors that can be systematically analyzed remains limited. When multiple factors are coupled, relying solely on prior intuition or one-at-a-time comparisons often makes it difficult to distinguish the truly governing factors from those that are merely incidental.

In parallel with these computational developments, substantial experimental progress has been made in the identification and characterization of creep cavities. Improved electron-microscopy and image-analysis methods have enabled more reliable detection of grain-boundary cavities (Chen et al., 2011), while two- and three-dimensional multiscale characterization has provided detailed information on cavity morphology, spatial distribution, growth, and coalescence during creep (Burnett et al., 2015). Building on these advances, recent novel studies have employed correlative microscopy to investigate the local microstructural factors associated with creep cavitation. By spatially registering cavities identified from electron-microscopy images with crystallographic and microstructural information obtained from electron backscatter diffraction and complementary characterization techniques, these approaches enable cavitated and uncavitated GBs to be compared in terms of grain orientation, GB misorientation, localized deformation, and the surrounding microstructural environment. For 316H austenitic stainless steel, He et al. (2024) developed a correlative approach to evaluate the relationships between creep-initiated cavities and local parameters including GB misorientation, crystallographic orientation, and deformation heterogeneity. More recently, Galliopoulou et al. (2025) provided an important multimodal experimental extension of this approach to ferritic Grade 91 steel by combining SEM, EBSD, TEM, EDS, and FIB–$XeF_2$ imaging to examine the coupled roles of GB character, localized deformation, MnS inclusion, and interfacial precipitation. In a further notable development, Ashraf et al. (2026) integrated experimentally observed cavity maps with microstructure-resolved crystal-plasticity simulations and classification algorithms, establishing a useful experimental–computational route for identifying mesoscale variables associated with creep-damage initiation. These experimental developments provide an important counterpart to computational investigations of the crystallographic, geometric, and micromechanical factors governing intergranular creep damage.

Statistical data-correlation approaches offer significant advantages in uncovering relationships among high-dimensional variables (Chen et al., 2022; Das and Ponte Castañeda, 2021; Schmelzer et al., 2025). In particular, data-driven methods have been widely used for feature selection, importance ranking, and surrogate modeling, providing powerful tools for mechanistic analysis of micromechanics of alloys (Ashraf et al., 2026; Eghtesad et al., 2023). Typical examples include the identification of microscale stress hotspots (Eghtesad et al., 2023; Mangal and Holm, 2018) and the classification of GB types (Zhang et al., 2022) using tree-based models. One advantage of tree-based models is that they enable feature importance to be ranked using Shapley Additive explanations (SHAP). A key limitation of SHAP is that it provides model-based interpretability rather than physics-based interpretability; a high SHAP importance indicates predictive utility, but does not by itself establish physical causality or identify the governing mechanism. In addition, Hu et al. (2024) employed a temporal graph neural network to capture intergranular interactions and thereby predict the cross-scale mechanical response and orientation evolution of polycrystals under complex loading conditions. Although these methods have provided important insights into the multivariate coupling and nonlinear relationships among microstructural features, they remain primarily based on phenomenological statistics and predictive contributions. The resulting feature rankings depend on the trained surrogate model and do not necessarily correspond directly to governing mechanisms with clear mechanical significance.

Therefore, for the high-dimensional and nonlinear relationships between micromechanical feature variables and microscopic responses such as GB local stress, both mechanics-informed models built upon prior mechanical knowledge and purely data-driven correlation-based models are insufficient for fully revealing the underlying mechanisms and establishing interpretable relationships. Accordingly, the integration of physics-based mechanistic understanding with data-driven modeling represents a natural progression. Recently, feature engineering and interpretable data-driven approaches, particularly those based on machine learning, are gaining increasing attention in multiscale mechanical modeling (Feuerriegel et al., 2024; Hansen et al., 2024; Owens et al., 2025; Tejada-Lapuerta et al., 2025; Yang et al., 2024). Causal discovery and information-theoretic attribution offer a useful framework for distinguishing direct effects, indirect associations, and redundant information in complex systems. Recently, information-theoretic principles (Shannon, 1948) have been introduced into mechanics, providing a useful framework for quantifying information content in mechanical tests and constitutive modeling (Ihuaenyi et al., 2025; Ihuaenyi et al., 2024). In particular, causation entropy analysis, in terms of conditional information, the residual contribution of a candidate variable to a target quantity after conditioning on other variables (AlMomani et al., 2020). It is therefore more suitable than simple correlation analysis or conventional feature-importance ranking for identifying direct informational relevance. Causation entropy was used for attribution analysis of microscale stress evolution and stress concentration (Dunham et al., 2025; Zhang et al., 2023), and for the construction of physics-assisted reduced-order models (Zhang et al., 2025). Therefore, causation entropy analysis provides a principled route for identifying causally relevant features in a physics-guided manner and for constructing interpretable reduced-order models. This framework enables an effective integration of mechanistically grounded dependencies with data-driven association.

Given the central role of local GB stress in creep damage evolution, especially in cavity nucleation, the objective of the present work is to identify the governing microscale mechanisms of GB normal stress and to distill them into an interpretable and transferable reduced-order representation, despite the high-dimensional, nonlinear, and strongly coupled dependence of GB normal stress on local crystallography, interfacial geometry, and micromechanical mismatch. To this end, a mechanics-informed candidate space consisting of 18 physically motivated characteristics is constructed within a crystal-plasticity framework, and causation entropy is employed to isolate the characteristics governing GB normal stress. Using only dislocation-climb-affected CPFE data generated from minimal bicrystal grain clusters under uniaxial tensile creep, the dominant characteristics are identified, and an interpretable reduced-order regression form is derived from the associated candidate functions. It is further shown that what is transferred is not merely a fitted predictor, but the identified characteristic hierarchy and functional representation: these remain effective under multiaxial loading and, more importantly, admit systematic extension to tricrystal systems through physically interpretable nonlocal augmentation, thereby demonstrating that the present framework can accommodate intergranular interaction effects beyond a purely local description. In addition, the effects and uncertainties associated with the dominant characteristics are quantified and, as supporting evidence, improved surrogate-model performance is obtained across multiple machine-learning model classes. Overall, an information-theoretic, mechanics-guided route is established for causal-mechanistic identification and interpretable reduced-order modeling of microscale GB stress in creep.

## 2. Physics-based microscale modelling and characterization

A crystal plasticity model is employed to directly and readily quantify grain-boundary stresses and crystallographic descriptors for the grain clusters. Creep effects are incorporated into the crystal plasticity model. At the microscale, crystallographic, microstructural, and mechanical characteristics are examined. Clusters with bicrystal and tricrystal configurations are used to generate data sets under uniaxial and multiaxial creep conditions. Given the material-specific calibration of the creep crystal-plasticity model, the FCC material, Inconel 617 at 950 °C, is selected as the material modelled here.

## 2.1. Crystal plasticity model considering creep

Crystal plasticity finite element (CPFE) methods are widely used to predict and analyze the plasticity behavior of polycrystals to reveal the microscopic mechanisms of deformation. CPFE has also demonstrated strong capability in modeling creep of polycrystals at elevated temperatures over the long time (Cheng et al., 2024; Gan et al., 2024; Kong et al., 2025; Phan et al., 2017; Zhang et al., 2020). The purpose of the CPFE model here is not to reproduce the full creep-failure trajectory, but to generate physically consistent local GB stress data in the steady creep regime for mechanism identification.

At elevated temperatures during creep, the deformation of metallic polycrystals is governed by the coupled action of dislocation glide and climb. To incorporate the contribution of dislocation climb in high-temperature creep, it is typically necessary to account for both plastic strain accumulation driven by slip and deformation accumulation associated with climb. Similar methodological strategies have been given by Lebensohn et al. (2010), Phan et al. (2017), Bieberdorf et al. (2021) and Cheng et al. (2024). Based on the CPFE framework proposed by Marin and Dawson (1998), the deformation gradient $\mathbf{F}$ is sequentially decomposed into the plastic deformation gradient $\mathbf{F}^{\mathrm{p}}$, elastic rotation $\mathbf{R}^{\mathrm{e}}$, and the left symmetric elastic stretch tensor $\mathbf{V}^{\mathrm{e}}$:

$$\mathbf{F} = \mathbf{V}^{\mathrm{e}} \cdot \mathbf{R}^{\mathrm{e}} \cdot \mathbf{F}^{\mathrm{p}} \tag{1}$$

The schematic of the deformation gradient decomposition is shown in Fig. 1a. It sequentially involves the initial configuration $\mathbb{C}_0$, two intermediate configurations ($\bar{\mathbb{C}}$ and $\tilde{\mathbb{C}}$), and the current configuration $\mathbb{C}$. The unit vectors along the slip direction of the $\alpha$-slip system in different configurations are denoted as $\mathbf{s}^{\alpha}$, $\bar{\mathbf{s}}^{\alpha}$ and $\tilde{\mathbf{s}}^{\alpha}$. The unit vectors normal to the $\alpha$-slip plane are denoted as $\mathbf{m}^{\alpha}$, $\bar{\mathbf{m}}^{\alpha}$ and $\tilde{\mathbf{m}}^{\alpha}$, respectively. In the Marin and Dawson (1998) CPFE framework, the slip-related variables are updated in the current configuration ($\tilde{\mathbb{C}}$). Hence, the velocity gradient $\tilde{\mathbf{L}}^{\mathrm{p}}$ in configuration $\tilde{\mathbb{C}}$ can be decomposed into deformation rate $\tilde{\mathbf{D}}^{\mathrm{p}}$ and spin $\tilde{\mathbf{W}}^{\mathrm{p}}$:

$$\tilde{\mathbf{L}}^{\mathrm{p}} = \tilde{\mathbf{D}}^{\mathrm{p}} + \tilde{\mathbf{W}}^{\mathrm{p}} \tag{2}$$

And the deformation rate and spin rate incorporate both the glide and climb parts (Phan et al., 2017):

$$\tilde{\mathbf{D}}^{\mathrm{p}} = \tilde{\mathbf{D}}^{\mathrm{p}}_{\mathrm{g}} + \tilde{\mathbf{D}}^{\mathrm{p}}_{\mathrm{c}} \tag{3}$$

$$\tilde{\mathbf{W}}^{\mathrm{p}} = \tilde{\mathbf{W}}^{\mathrm{p}}_{\mathrm{g}} + \tilde{\mathbf{W}}^{\mathrm{p}}_{\mathrm{c}} \tag{4}$$

in which the subscripts 'g' and 'c' represent glide and climb. Specifically, these terms are expressed as:

$$\tilde{\mathbf{D}}^{\mathrm{p}}_{\mathrm{g}} = \sum_{\alpha=1}^{n} \dot{\gamma}^{\alpha}_{\mathrm{g}} \left( \tilde{\mathbf{s}}^{\alpha} \otimes \tilde{\mathbf{m}}^{\alpha} \right)_{\mathrm{S}} \tag{5}$$

$$\tilde{\mathbf{W}}^{\mathrm{p}}_{\mathrm{g}} = \dot{\mathbf{R}}^{\mathrm{e}} \cdot \left( \mathbf{R}^{\mathrm{e}} \right)^{\mathrm{T}} + \sum_{\alpha=1}^{n} \dot{\gamma}^{\alpha}_{\mathrm{g}} \left( \tilde{\mathbf{s}}^{\alpha} \otimes \tilde{\mathbf{m}}^{\alpha} \right)_{\mathrm{A}} \tag{6}$$

$$\tilde{\mathbf{D}}^{\mathrm{p}}_{\mathrm{c}} = \sum_{\alpha=1}^{n} \dot{\gamma}^{\alpha}_{\mathrm{c}} \left( \tilde{\mathbf{K}}^{\alpha} \right)_{\mathrm{S}} \tag{7}$$

$$\tilde{\mathbf{W}}^{\mathrm{p}}_{\mathrm{c}} = \sum_{\alpha=1}^{n} \dot{\gamma}^{\alpha}_{\mathrm{c}} \left( \tilde{\mathbf{K}}^{\alpha} \right)_{\mathrm{A}} \tag{8}$$

in which $(\bullet)_{\mathrm{S}}$ and $(\bullet)_{\mathrm{A}}$ represent the symmetric and skew-symmetric parts of the tensor. And $\tilde{\mathbf{K}}^{\alpha}$ is the climb tensor proposed by Lebensohn et al. (2010):

$$\tilde{\mathbf{K}}^{\alpha} = \tilde{\mathbf{s}}^{\alpha} \otimes \tilde{\boldsymbol{\chi}}^{\alpha} \tag{9}$$

$$\tilde{\boldsymbol{\chi}}^{\alpha} = \tilde{\mathbf{m}}^{\alpha} \times \tilde{\mathbf{t}}^{\alpha} \tag{10}$$

in which $\tilde{\mathbf{t}}^{\alpha}$ is the tangent to the dislocation line. As a key component of the crystal plasticity model, the dislocation slip shear deformation rate is defined as (Busso et al.,

2000; Guo et al., 2020; Zhang and Oskay, 2016):

$$\dot{\gamma}_{\mathrm{g}}^{\alpha}=\dot{\gamma}_0 \exp\left\{-\frac{F_0}{k\theta}\left\langle 1-\left\langle\frac{\left|\tau^{\alpha}-B^{\alpha}\right|-S^{\alpha}\mu/\mu_0}{\hat{\tau}_0\mu/\mu_0}\right\rangle^{p_g}\right\rangle^{q_g}\right\}\mathrm{sgn}\left(\tau^{\alpha}-B^{\alpha}\right) \tag{11}$$

in which $\tau^{\alpha}$ is the resolved shear stress $\tau^{\alpha}=\boldsymbol{\tau}:\left(\tilde{\mathbf{s}}^{\alpha}\otimes\tilde{\mathbf{m}}^{\alpha}\right)_{\mathrm{S}}$. Here, $\dot{\gamma}_0$, $k$, $\theta$ and $F_0$ denote the reference shear strain rate, the Boltzmann constant, the absolute temperature and the activation energy, respectively. $\hat{\tau}_0$ represents the threshold stress for thermal dislocation motion. $\mu$ and $\mu_0$ correspond to the shear moduli at current temperature and at 0 K, respectively. $S^{\alpha}$ and $B^{\alpha}$ are the slip resistance to dislocation motion and back stress, respectively. The flow rule for dislocation climb is expressed as:

$$\dot{\gamma}_{\mathrm{c}}^{\alpha}=\dot{\gamma}_0 \exp\left(-\frac{F_0}{k\theta}\right)\left(\frac{\left|\tau_{\mathrm{c}}^{\alpha}-B_{\mathrm{c}}^{\alpha}\right|}{\hat{\tau}_{0\mathrm{c}}}\right)^{p_{\mathrm{c}}}\mathrm{sgn}\left(\tau_{\mathrm{c}}^{\alpha}-B_{\mathrm{c}}^{\alpha}\right) \tag{12}$$

where $\hat{\tau}_{0\mathrm{c}}$ and $p_c$ denote the scalar threshold creep stress and the creep exponent, respectively. The shear stress on the climb system is $\tau_{\mathrm{c}}^{\alpha}=\boldsymbol{\tau}:\left(\tilde{\mathbf{K}}^{\alpha}\right)_{\mathrm{dev}}$. And $B_{\mathrm{c}}^{\alpha}$ denotes the back stress corresponding to the climb-induced shearing deformation. The above CPFE framework and constitutive model were adopted by Phan et al. (2017) and Zhang and Oskay (2016) and shown to be effective. More specifically, the evolution laws and material parameters involved in Eqns. (11) and (12) are provided in Appendix A. As shown in Fig. 1b, the crystal plasticity parameters of Inconel 617 were calibrated against uniaxial tensile creep curves collected at 950 °C under sustained uniaxial applied stresses of 18.5, 24 and 28.6 MPa.

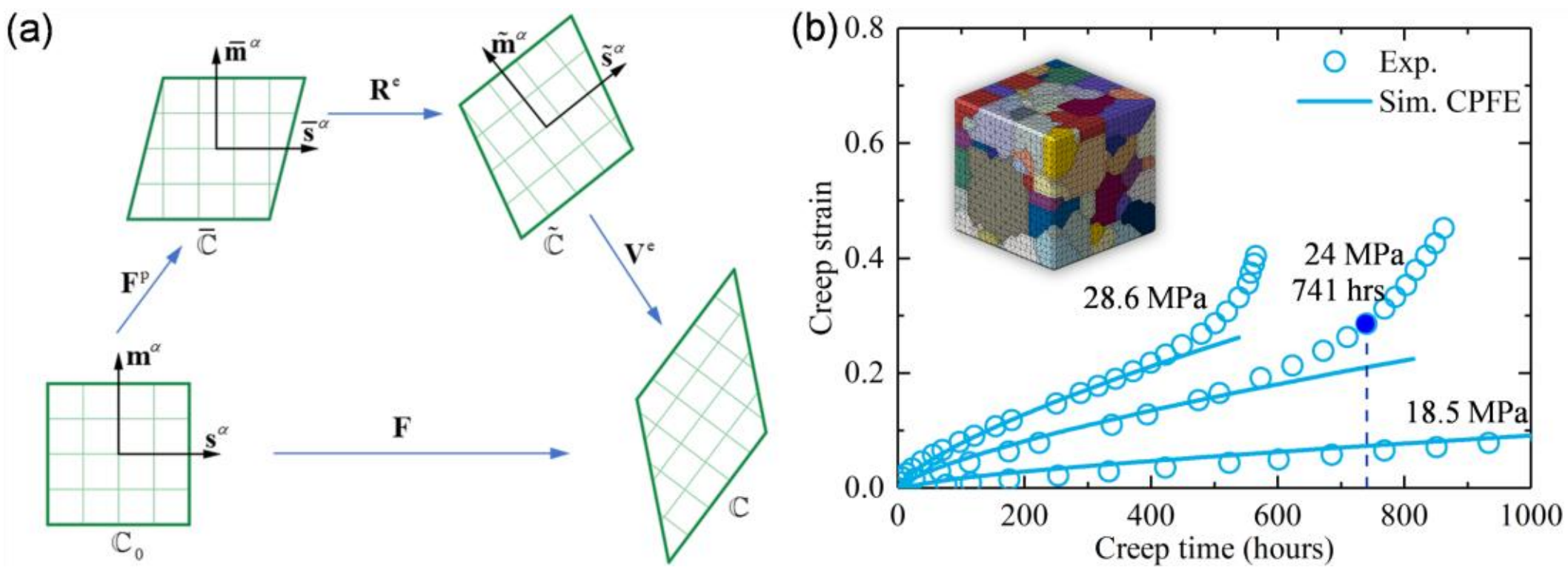

Fig. 1 (a) Kinematics and configurations of single crystals deformation; (b) Comparison between experimental and simulated creep strain curves

As shown in Fig. 1b, calculations are performed on a representative volume element (RVE) containing a sufficient number of grains, with parameters calibrated from uniaxial tensile tests conducted at different stress levels. During high-temperature creep of metals, the primary (transient) stage is typically characterized by a decreasing creep rate. The secondary stage exhibits an approximately constant (minimum) creep strain rate and often accounts for the longest duration, during which stress redistribution becomes quasi-steady. In the tertiary stage, damage accumulates, most notably grain-boundary cavitation, leading to an accelerated creep rate and ultimately material failure. Although the secondary stage of creep is not typically associated with pronounced damage evolution, the sustained local steady-state stresses, particularly stress concentrations at GBs, provide a primary driving force for subsequent damage. Accordingly, this work focuses on the GB stress distribution in secondary creep and its dependence on the microstructural and micromechanical characteristics. Mechanisms dominant in tertiary creep, such as void growth and extensive GB sliding, are not considered here. Therefore, although GB damage evolution has been successfully modelled and validated in previous studies (Kong et al., 2025; Phan et al., 2017), the CPFE model employed here does not explicitly represent GBs. Consequently, as shown in Fig. 1b, the present model is not intended to capture the tertiary creep associated with GB cavitation and sliding. In the following, 'GB stress' refers to the local GB stress in the steady-state creep regime, unless otherwise specified.

## 2.2. Crystallographic, microstructural and micromechanical characteristics

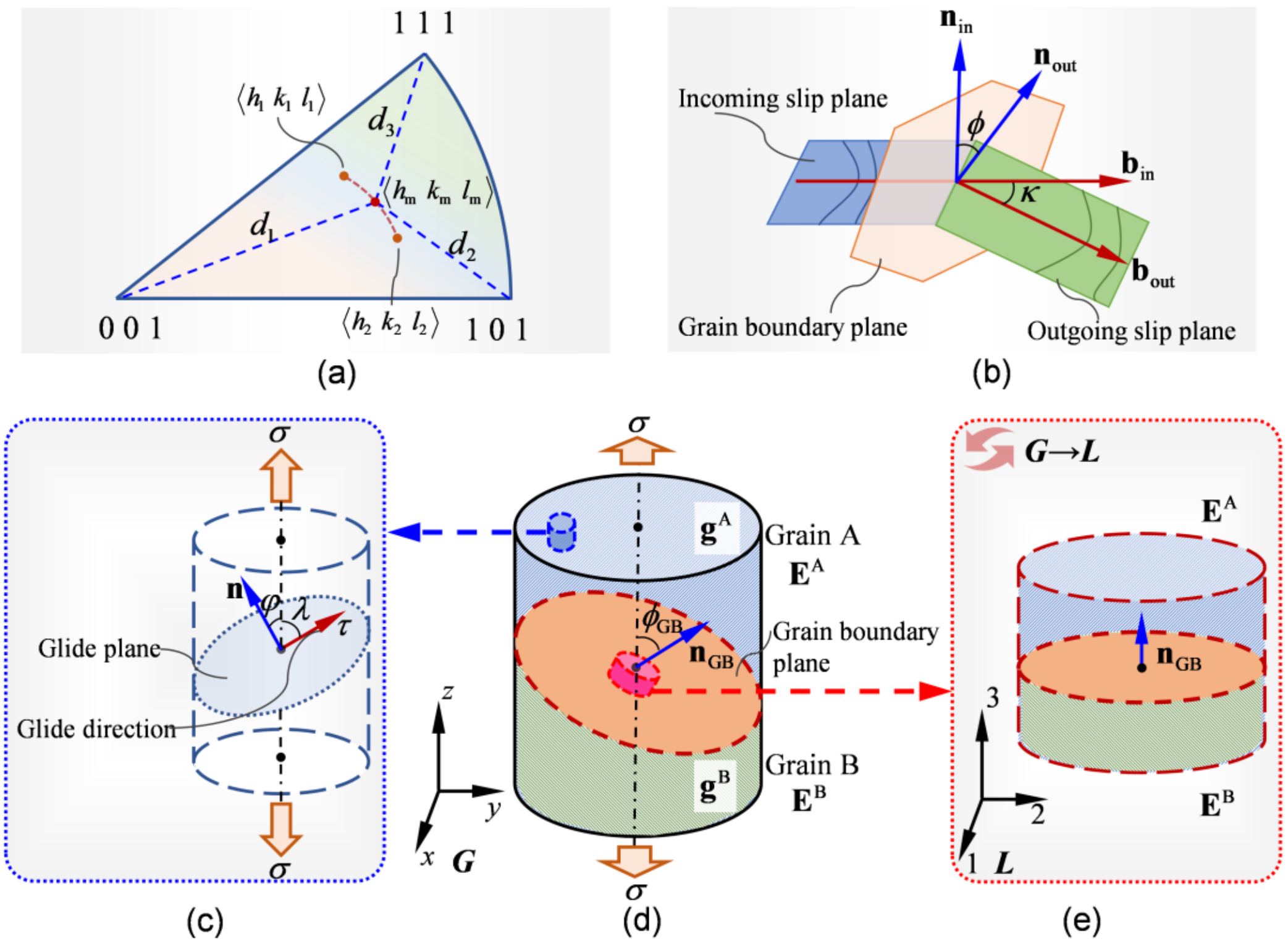

Fig. 2 Schematic illustrations of (a) the average orientation of a bicrystal with [0 0 1] sample direction and Euclidean distances to [0 0 1], [1 0 1] and [1 1 1] in the Inverse Pole Figure (IPF); (b) the slip transmission factor (geometric compatibility factor) $m' = \cos\kappa\cos\phi$; (c) the slip plane normal $\mathbf{n}$ and slip direction $\tau$ in a crystal lattice within a grain; (d) an isolated grain boundary (GB) in global/sample coordinate system and associated physical quantities: orientation matrices ($\mathbf{g}^{\mathrm{A}}$ and $\mathbf{g}^{\mathrm{B}}$), elastic tensors ($\mathbf{E}^{\mathrm{A}}$ and $\mathbf{E}^{\mathrm{B}}$), GB normal vector ($\mathbf{n}_{\mathrm{GB}}$), and GB orientation angle ($\phi_{\mathrm{GB}}$); (e) GB in the local coordinate system.

Beyond the applied macroscopic stress, GB stresses are strongly influenced by microscale features, including crystallographic characteristics, microstructural attributes, and micromechanical properties. As shown in Fig. 2d, for a given GB, the two neighboring grains are labeled as grain A and grain B, respectively. This work examines factors that may affect GB stresses, with descriptors defined primarily at the GB level. These descriptors form a physically motivated candidate space, the objective of the subsequent analysis is to determine which subset carries direct governing relevance to GB normal stress. Specifically, 18 characteristics are considered, labeled $F_1 \sim F_{18}$ and presented in Table 1, which fall into three categories:

Table 1 Key microscale characteristics in fundamental grain clusters

| ID | Characteristics | ID | Characteristics |
|---|---|---|---|
| $F_1$ | $d_1$ | $F_{10}$ | $\left(E_{33}^{\mathrm{G}}\right)_{\mathrm{eff}}^{\mathrm{S}}$ |
| $F_2$ | $d_2$ | $F_{11}$ | $\left(E_{33}^{\mathrm{G}}\right)_{\mathrm{eff}}^{\mathrm{P}}$ |
| $F_3$ | $d_3$ | $F_{12}$ | $\max\left(\Delta E_{11}^{\mathrm{G}}\right)$ |
| $F_4$ | $\theta$ | $F_{13}$ | $\max\left(\Delta G_{12}^{\mathrm{G}}\right)$ |
| $F_5$ | $m'$ | $F_{14}$ | $\Delta E_{33}^{\mathrm{L}}$ |
| $F_6$ | $\bar{m}$ | $F_{15}$ | $\left(E_{33}^{\mathrm{L}}\right)_{\mathrm{eff}}^{\mathrm{S}}$ |
| $F_7$ | $\bar{m}_{\mathrm{n}}$ | $F_{16}$ | $\left(E_{33}^{\mathrm{L}}\right)_{\mathrm{eff}}^{\mathrm{P}}$ |
| $F_8$ | $\phi_{\mathrm{GB}}$ | $F_{17}$ | $\max\left(\Delta E_{11}^{\mathrm{L}}\right)$ |
| $F_9$ | $\Delta E_{33}^{\mathrm{G}}$ | $F_{18}$ | $\max\left(\Delta G_{12}^{\mathrm{L}}\right)$ |

**Category (A) Crystallographic characteristics**

(1) $F_1$, $F_2$, $F_3$: Mean crystal orientation of the bicrystal (grains A and B) with respect to the specimen [001] direction, and the corresponding Euclidean distances (Bieberdorf et al., 2021) in the [001] inverse pole figure (IPF) to the [001], [101], and [111] reference directions, denoted as $d_1$, $d_2$ and $d_3$, respectively. The Euclidean embedding offers an intuitive representation of crystallographic orientations and allows the mean orientation of the two grains adjacent to a GB to be computed and visualized straightforwardly. $\mathbf{R_A}$ and $\mathbf{R_B}$ denote the orientation matrices of grains A and B, respectively, defined as active rotations from the crystal frame to the sample frame. For an FCC crystal with $m\bar{3}m$ point-group symmetry, the physical orientation $\mathbf{R_B}$ is equivalent to the set $\left\{\mathbf{R_B O} \middle| \mathbf{O} \in \mathcal{G}\right\}$, where $\mathcal{G} \subset SO(3)$ is the collection of cubic

symmetry operators. To identify the symmetry-equivalent variant of grain B that is closest to grain A, the evaluation of misorientation is required:

$$\tilde{\theta}(\mathbf{O}) = \arccos\left(\frac{\operatorname{tr}\left((\mathbf{R}_\mathbf{B}\mathbf{O})\mathbf{R}_\mathbf{A}^\mathbf{T}\right)-1}{2}\right) \tag{13}$$

and define the specific operator $\mathbf{O}^*$ corresponding to the minimum misorientation angle:

$$\mathbf{O}^* = \arg\min \tilde{\theta}(\mathbf{O}) \tag{14}$$

Accordingly, the symmetry-aligned orientation of grain B corresponding to the minimum misorientation with respect to $\mathbf{R}_\mathbf{A}$ is defined as:

$$\tilde{\mathbf{R}}_\mathbf{B} = \mathbf{R}_\mathbf{B}\mathbf{O}^* \tag{15}$$

The orientations $\tilde{\mathbf{R}}_\mathbf{B}$ and $\mathbf{R}_\mathbf{B}$ are crystallographically equivalent and therefore represent the same physical orientation. The introduction of $\tilde{\mathbf{R}}_\mathbf{B}$ eliminates the ambiguity arising from crystallographic symmetry, ensuring that the subsequently evaluated mean orientation is uniquely defined and comparable across different GBs. As shown in Fig. 2a, to define a physically meaningful mean orientation for the bicrystal, the midpoint along the geodesic connecting $\mathbf{R}_\mathbf{A}$ and $\tilde{\mathbf{R}}_\mathbf{B}$ is computed. The computation method of the midpoint of geodesic and mean of rotations was proposed by Moakher (2002). Specifically, the relative rotation is given by:

$$\Delta\mathbf{R} = \mathbf{R}_\mathbf{A}^\mathbf{T}\tilde{\mathbf{R}}_\mathbf{B} \tag{16}$$

The shortest geodesic curve connecting $\mathbf{R}_\mathbf{A}$ and $\tilde{\mathbf{R}}_\mathbf{B}$ is expressed as:

$$\mathbf{R}(t) = \mathbf{R}_\mathbf{A}\exp\left(t\log(\Delta\mathbf{R})\right),\ 0 \le \mathrm{t} \le 1 \tag{17}$$

And the mean rotation is obtained as the shortest geodesic midpoint at $t = 1/2$:

$$\bar{\mathbf{R}} = \mathbf{R}_\mathbf{A}\exp\left(\frac{1}{2}\log\left(\mathbf{R}_\mathbf{A}^\mathbf{T}\tilde{\mathbf{R}}_\mathbf{B}\right)\right) \tag{18}$$

Accordingly, $\bar{\mathbf{R}}$ is defined as the rotation matrix representing the mean orientation of the two grains on either side of the GB. The corresponding axis–angle ( $\theta_\mathrm{R}, \hat{r}$ ) representation of $\bar{\mathbf{R}}$ is expressed as:

$$\theta_{\mathrm{R}}=\arccos\left(\frac{\operatorname{tr}\left(\bar{\mathbf{R}}\right)-1}{2}\right) \tag{19}$$

$$r_1^{\mathrm{M}}=\frac{\bar{R}_{23}-\bar{R}_{32}}{2\sin\theta_{\mathrm{R}}},r_2^{\mathrm{M}}=\frac{\bar{R}_{31}-\bar{R}_{13}}{2\sin\theta_{\mathrm{R}}},r_3^{\mathrm{M}}=\frac{\bar{R}_{12}-\bar{R}_{21}}{2\sin\theta_{\mathrm{R}}} \tag{20}$$

in which $\bar{R}_{ij}$ ($i,j=1,2,3$) represents the component of mean rotation matrix. $r_i^{\mathrm{M}}$ ($i=1,2,3$) represents the component of unit vector $\hat{r}$ corresponding to $\bar{\mathbf{R}}$. The Euclidean distances to IPF conners, $d_1$, $d_2$ and $d_3$, are defined as:

$$\begin{cases} d_1=\left\|\hat{r}-\hat{r}^{[001]}\right\| \\ d_2=\left\|\hat{r}-\hat{r}^{[101]}\right\| \\ d_3=\left\|\hat{r}-\hat{r}^{[111]}\right\| \end{cases} \tag{21}$$

in which $\|\bullet\|$ represents the Euclidean ($\ell_2$) norm. The mean orientation of the two grains adjacent to the grain boundary, quantified by $d_1$, $d_2$ and $d_3$, provides an informative descriptor of the deformation capability of the bicrystal system under the prescribed loading condition.

(2) $F_4$: The minimum misorientation $\theta$ of the GB. According to Eqn. (13), $\theta$ is expressed as:

$$\theta=\min\left(\tilde{\theta}\right) \tag{22}$$

The minimum misorientation $\theta$ of the GB characterizes the crystallographic mismatch, which is related to local GB stress concentrations.

(3) $F_5$: The slip transmission factor (geometric compatibility factor) $m'$ of the GB. The factor $m'$ describes the propensity for slip transmission across a GB and is expressed as:

$$m'=\cos\kappa\cos\phi \tag{23}$$

As shown in Fig. 2b, $\phi$ is the angle between the slip plane normals ($\mathbf{n}_{\mathrm{in}}$ and $\mathbf{n}_{\mathrm{out}}$) of the interacting slip systems in the two grains, and $\kappa$ is the angle between their corresponding slip directions ( $\mathbf{b}_{\text{in}}$ and $\mathbf{b}_{\text{out}}$ ). In general, a larger value of $m'$ corresponds to a greater likelihood of slip transmission across the GB. In this work, $m'$ is not used as an additional constitutive parameter. Instead, it is considered as a descriptor of crystallographic compatibility. Since the slip transmission factor $m'$, determined by grain orientations, is explicitly included in CPFE, the resulting stress redistribution near grain boundaries can reflect the first-order effect of crystallographic compatibility.

(4) $F_6$: The mean Schmid factor $\bar{m}$ of the two grains adjacent to the GB. The Schmid factor quantifies the ability of slip systems within a grain to be activated under a specific applied load. As shown in Fig. 2c, for grain A subjected to uniaxial tension, the Schmid factor is expressed as:

$$m_{\text{A}} = \cos\varphi \sin\lambda \tag{24}$$

in which $\varphi$ and $\lambda$ are the angles between the loading direction and the slip-plane normal and slip direction, respectively. The Schmid factor for grain B $m_{\text{B}}$ is defined analogously. Hence, the mean Schmid factor $\bar{m}$ of the two grains adjacent to the GB is defined as the average of the maximum $m_{\text{A}}$ and $m_{\text{B}}$:

$$\bar{m} = \frac{\max\left(m_{\text{A}} + m_{\text{B}}\right)}{2} \tag{25}$$

The mean Schmid factor $\bar{m}$ is used as a reference measure of the overall slip activity in the two grains under the applied loading. Under a prescribed loading direction, the average of the maximum Schmid factors of the two grains serves as a reference metric for the effective softness/hardness tendency of the bicrystal, influencing the mechanical responses.

(5) $F_7$: The mean Schmid-type indicator $\bar{m}_{\text{n}}$ for climb-driven creep of two grains adjacent to the GB. This descriptor is introduced specifically to reflect creep-relevant stress relaxation through climb, which cannot be represented by conventional slip-oriented Schmid measures alone. Dislocation climb plays a crucial role in creep. Unlike

dislocation slip, dislocation climb is not directly controlled by the shear stress on the slip plane, but is closely related to the normal stress component (Jones and Ashby, 2019) on the glide plane and the hydrostatic stress field (Herring, 1950). Hence, for the specific grain A, a Schmid-type geometric indicator for climb-driven creep, defined as the maximum normalized normal stress on the glide planes under uniaxial loading:

$$m_{\mathrm{n}}^{\mathrm{A}} = \cos^2 \varphi \tag{26}$$

The definition is also applicable for grain B. Hence, the mean Schmid-type indicator $\bar{m}_{\mathrm{n}}$ for climb-driven creep is defined as:

$$\bar{m}_{\mathrm{n}} = \frac{\max\left(m_{\mathrm{n}}^{\mathrm{A}} + m_{\mathrm{n}}^{\mathrm{B}}\right)}{2} \tag{27}$$

$\bar{m}_{\mathrm{n}}$ is used as a descriptor of the overall propensity for diffusion-assisted climb and creep compliance.

**Category (B) Microstructural geometry**

$F_8$: The inclination angle $\phi_{\mathrm{GB}}$ of the GB. As shown in Fig. 2d, the GB inclination angle is defined as the angle between the normal of the GB and the loading axis (for multiaxial loading, the direction of the maximum principal stress). The GB inclination angle markedly alters the normal and tangential stress decomposition on the GB plane.

Other microstructural geometric descriptors, such as grain size, grain sphericity, GB area, and GB morphology, may also play important roles. This work focus on the factors that directly govern the local stress state at GBs, and thus these descriptors are not included in the analysis.

**Category (C) Mechanical properties**

The mechanical response of GBs is governed by both plasticity- and elasticity-related factors. In this study, plasticity-associated geometric descriptors related to slip activation, crystallographic compatibility, and deformation-transfer tendency are captured by crystallographic features. Although plastic incompatibility evolves during CPFE simulation, its underlying tendency is influenced by the initial crystallographic arrangement and can therefore be partially characterized through descriptors such as orientation mismatch, Schmid-type indicators, and slip compatibility parameters. Accordingly, elasticity-focused mechanical descriptors are incorporated, primarily based on orientation-dependent elastic moduli. As shown in Fig. 2d, the representation of elastic properties involves three coordinate frames: the crystal (crystallographic) frame, the specimen (global) frame, and a GB-attached local frame. The corresponding stiffness/compliance components are related between these frames through tensor rotation. Although the apparent differences are solely attributable to the transformation of tensor components under a rotation of the reference frame, the relevant relations are expressed in matrix form in what follows to facilitate the subsequent analysis and computational implementation.

To quantify the elastic anisotropy contrast induced by orientation differences in a bicrystal, the single-crystal elastic constants of a cubic material ($C_{11}$, $C_{12}$ and $C_{44}$) are first assembled into the stiffness matrix in the crystal frame, $\mathbf{C}^{\mathrm{c}} \in \mathbb{R}^{6\times6}$:

$$\mathbf{C}^{\mathrm{c}} = \begin{bmatrix} C_{11} & C_{12} & C_{12} & 0 & 0 & 0 \\ C_{12} & C_{11} & C_{12} & 0 & 0 & 0 \\ C_{12} & C_{12} & C_{11} & 0 & 0 & 0 \\ 0 & 0 & 0 & C_{44} & 0 & 0 \\ 0 & 0 & 0 & 0 & C_{44} & 0 \\ 0 & 0 & 0 & 0 & 0 & C_{44} \end{bmatrix} \tag{28}$$

And the corresponding compliance matrix is obtained as $\mathbf{S}^{\mathrm{c}} = \left(\mathbf{C}^{\mathrm{c}}\right)^{-1}$. To account for the average elastic modulus and elastic mismatch of the two grains adjacent to a GB in the global loading frame, the compliance tensor must be rotated into the specimen frame:

$$\mathbf{S}^{\mathrm{G}} = \mathbf{M}\left(\mathbf{R}_1\right)\mathbf{S}^{\mathrm{c}}\mathbf{M}\left(\mathbf{R}_1\right)^{\mathrm{T}} \tag{29}$$

in which $\mathbf{S}^{\mathrm{G}}$ is the compliance matrix in the global frame. $\mathbf{M}$ is a rotation operator for the matrix representation of a fourth-order tensor under the Voigt notation, and is related to the second-order rotation matrix $\mathbf{R}_1$ from crystallographic frame to global frame. The components of $\mathbf{R}_1$ are the direction cosines, whereas the components of $\mathbf{M}$ are given by specific combinations of quadratic terms in these direction cosines. Hence, the Young's moduli along the coordinate axes in the global frame can be expressed as:

$$E_{11}^{G}=\frac{1}{S_{11}^{G}},E_{22}^{G}=\frac{1}{S_{22}^{G}},E_{33}^{G}=\frac{1}{S_{33}^{G}} \tag{30}$$

in which the 3-direction is defined in the global specimen frame and is taken to be aligned with the direction of the maximum principal stress of the applied macroscopic stress state. The 1- and 2-directions lie in the principal plane perpendicular to the 3-direction. To characterize the mismatch and average behavior in mechanical properties across the grain boundary, several potentially dominant features are proposed as $F_9 \sim F_{18}$. These elastic descriptors do not represent direct predictors of local elastic strain. Instead, they characterize the stiffness contrast and constraint conditions between neighboring grains, which influence stress redistribution associated with deformation incompatibility.

(1) $F_9$: The difference in equivalent modulus between the two grains adjacent to the GB, evaluated along the maximum principal stress direction in the specimen frame.

$$\varDelta E_{33}^{\mathrm{G}}=\left(E_{33}^{\mathrm{G}}\right)^{\mathrm{A}}-\left(E_{33}^{\mathrm{G}}\right)^{\mathrm{B}} \tag{31}$$

in which the superscripts $(*)^{\mathrm{A}}$ and $(*)^{\mathrm{B}}$ represent the grains A and B on either side of the GB. $\varDelta E_{33}^{\mathrm{G}}$ is used to quantify the stiffness contrast between the adjoining grains along the loading direction.

(2) $F_{10}$: The effective mean elastic modulus for two grains in series.

$$\left(E_{33}^{\mathrm{G}}\right)_{\mathrm{eff}}^{\mathrm{S}}=\frac{2\left(E_{33}^{\mathrm{G}}\right)^{\mathrm{A}}\left(E_{33}^{\mathrm{G}}\right)^{\mathrm{B}}}{\left(E_{33}^{\mathrm{G}}\right)^{\mathrm{A}}+\left(E_{33}^{\mathrm{G}}\right)^{\mathrm{B}}} \tag{32}$$

It characterizes the effective stiffness of the bicrystal system in which the two grains adjacent to the GB are arranged in series along the loading direction.

(3) $F_{11}$: The effective mean elastic modulus is defined for two grains connected in parallel.

$$\left(E_{33}^{\mathrm{G}}\right)_{\mathrm{eff}}^{\mathrm{P}}=\frac{\left(E_{33}^{\mathrm{G}}\right)^{\mathrm{A}}+\left(E_{33}^{\mathrm{G}}\right)^{\mathrm{B}}}{2} \tag{33}$$

Similarly, it characterizes the effective stiffness of the bicrystal, with the two GB-adjacent grains loaded in parallel along the loading direction.

The above characteristics $F_9 \sim F_{11}$ are used to quantify the elastic modulus mismatch and the effective modulus along the maximum principal stress direction. In addition, the modulus mismatch on planes normal to the maximum principal stress direction may also alter the local stress state near the GB, thereby affecting the GB normal stress. Hence, the in-plane stiffness mismatch is characterized by the maximum difference in elastic (or shear) moduli within the plane normal to the maximum principal stress direction.

Because the critical deformation direction near a GB does not necessarily align with the global in-plane axes, the in-plane elastic contrast is quantified by scanning all directions within the *x*-*y* plane of a chosen frame. A rotation by $\theta \in [0,\pi)$ about the out-of-plane unit vector $\mathbf{e}_3$ is applied, and the rotated compliance is obtained as:

$$\mathbf{S}'(\theta)=\mathbf{M}\left(\mathbf{R}_z(\theta)\right)\mathbf{S}^{\mathrm{G}}\mathbf{M}\left(\mathbf{R}_z(\theta)\right)^{\mathrm{T}} \tag{34}$$

in which $\mathbf{M}$ and $\mathbf{S}^{\mathrm{G}}$ are from Eqn. (29). $\mathbf{R}_z(\theta)$ is the rotation matrix corresponding to $\theta$-rotation. The directional in-plane Young's modulus and in-plane shear modulus are then defined by:

$$E_{11}^{\mathrm{G}}(\theta)=\frac{1}{S_{11}'(\theta)},G_{11}^{\mathrm{G}}(\theta)=\frac{1}{S_{66}'(\theta)} \tag{35}$$

Accordingly, the in-plane "worst-case" elastic mismatches between the two grains A and B are defined as:

(4) $F_{12}$: The maximum in-plane Young's modulus mismatch expressed as,

$$\max\left(\Delta E_{11}^{\mathrm{G}}\right)=\max_{\theta\in[0,\pi)}\left|\left(E_{11}^{\mathrm{G}}(\theta)\right)^{\mathrm{A}}-\left(E_{11}^{\mathrm{G}}(\theta)\right)^{\mathrm{B}}\right| \tag{36}$$

(5) $F_{13}$: The maximum in-plane shear modulus mismatch expressed as,

$$\max\left(\Delta G_{12}^{\mathrm{G}}\right) = \max_{\theta\in[0,\pi)}\left|\left(G_{12}^{\mathrm{G}}(\theta)\right)^{\mathrm{A}} - \left(G_{12}^{\mathrm{G}}(\theta)\right)^{\mathrm{B}}\right| \tag{37}$$

All the mechanical quantities above are defined in the specimen frame. Meanwhile, the mechanical descriptors defined in the GB local coordinate system should also be considered. The local coordinate system of the GB is shown in Fig. 2e. Accordingly, the frame change corresponds to the rotation $\mathbf{R}_2$ from the crystallographic frame to the local GB frame), and the resulting component transformation is still given by Eqn. (29). Therefore, the features corresponding to $F_9 \sim F_{13}$ in the local GB frame are obtained as follows:

(6) $F_{14}$: The difference in equivalent modulus between the two grains adjacent to the GB $\Delta E_{33}^{\mathrm{L}}$, evaluated along the GB normal direction in the local frame.

(7) $F_{15}$: The effective mean elastic modulus $\left(E_{33}^{\mathrm{L}}\right)_{\mathrm{eff}}^{\mathrm{S}}$ for two grains in series evaluated in the GB local frame.

(8) $F_{16}$: The effective mean elastic modulus $\left(E_{33}^{\mathrm{L}}\right)_{\mathrm{eff}}^{\mathrm{P}}$ for two grains in parallel evaluated in the GB local frame.

(9) $F_{17}$ : The maximum in-plane Young's modulus mismatch $\max\left(\Delta E_{11}^{\mathrm{L}}\right)$ evaluated in the GB local frame.

(10) $F_{18}$ : The maximum in-plane shear modulus mismatch $\max\left(\Delta G_{12}^{\mathrm{G}}\right)$ evaluated in the GB local frame.

## 2.3. Generation of datasets from minimal grain clusters

Given that the present work primarily targets the influence of microstructural and micromechanical characteristics on local microscopic stresses, diversity in the externally imposed macroscopic-equivalent stress is not pursued in constructing the datasets. Instead, only representative uniaxial and multiaxial stress states are selected to control the dataset size. As shown in Fig. 3a, a cylindrical micro-specimen is constructed to represent uniaxial tensile creep under an applied macroscopic-equivalent stress $\sigma_a$. To account for the effect of multiaxial stress states, a hydrostatic stress component $\sigma_m$ superimposed on the uniaxial tensile loading $\sigma_a$ is also considered. A cylindrical micro-specimen is adopted primarily to be consistent with the axisymmetric idealization in the GB void-growth-creep-damage model of Cocks and Ashby (1982), thereby facilitating the imposition of compatible boundary conditions and the direct post-processing evaluation of GB damage based on the resulting stress triaxiality and GB-normal stress.

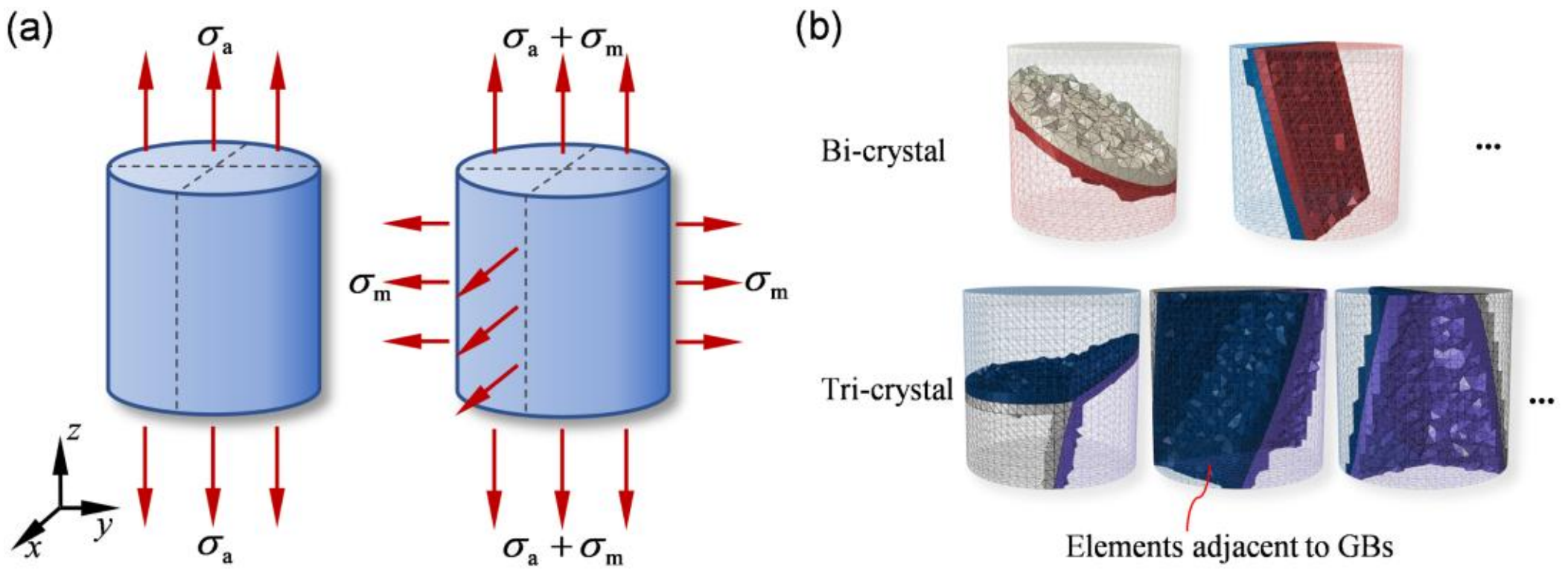

Fig. 3 (a) Uniaxial and multiaxial loading conditions; (b) finite element models of representative bicrystal and tricrystal configurations and elements adjacent to GBs.

Minimal grain clusters, including bicrystal and tricrystal configurations, are employed to evaluate the local response at GBs. A large number of random minimal grain clusters are generated, and several representative configurations are shown in Fig. 3b. Both the grain orientations and the GB positions are assigned randomly. Minimal clusters are used here as mechanism-extraction vehicles rather than simplified surrogates of the full polycrystal. The purpose is to (i) localize the problem and isolate GB-dominated mechanisms by decoupling the near-GB creep stress response from the complexity of polycrystalline networks, enabling direct assessment of the roles of misorientation, GB normal, etc.; (ii) reduce the degrees of freedom to facilitate causal feature extraction and mechanistic interpretation, with bicrystals capturing a single GB between two orientations and tricrystals incorporating junction and constraint effects; and (iii) enable efficient, well-controlled, high-throughput dataset generation by keeping the parameter space and computational cost tractable. Thus, minimal grain clusters are used here not merely for computational convenience, but as a controlled setting for causal-mechanistic distillation before introducing full polycrystalline complexity.

For long-term creep of alloys, both the stress state and the creep rate progressively approach a quasi-steady regime during secondary creep, until tertiary creep is reached, where damage accumulation leads to pronounced strain localization and accelerated creep. Therefore, the quasi-steady GB normal stress near the secondary-to-tertiary transition is examined in this work. Tertiary-creep onset is preceded by a GB stress state that is critical, because it governs the nucleation and growth of typical GB creep cavities. The objective of the present work is to elucidate the mechanisms controlling the dominant GB-normal stress. In practice, a representative stress level for long-term creep is selected (Fig. 1b). Specifically, a macroscopic-equivalent applied stress of $\sigma_{\mathrm{a}} = 24\ \mathrm{MPa}$ is considered, and the GB normal stress is extracted at a creep time of 741 hours. The multiaxial case with von Mises equivalent stress $\sigma_{\mathrm{eq}} = 24\ \mathrm{MPa}$ and stress triaxiality $\eta = 3$, i.e., $\sigma_{\mathrm{a}} = 24\ \mathrm{MPa}$ and $\sigma_{\mathrm{m}} = 64\ \mathrm{MPa}$, is also calculated. For a given GB, the GB-normal stress is taken as the volume average of the values in the finite elements adjacent to the GB, i.e., $\bar{\sigma}_{33}^{\mathrm{L}}$.

To ensure statistical robustness and to assess whether the resulting distributions deviate from Gaussianity, the repeated Kolmogorov Smirnov (K-S) testing procedure proposed by Dunham et al. (2025) and Zhang et al. (2025), is adopted to quantify significance via the corresponding pass-rate metric. The K-S testing results are shown in Fig. 4. Accordingly, the datasets become statistically significant at $N \approx 100$ for the bicrystal case (uniaxial and multiaxial loading) and at $N \approx 300$ for the uniaxial tricrystal case, where $N$ denotes the number of independent samples used to form the empirical distribution. In this work, the dataset sizes used for subsequent statistics and data-driven prediction are $N = 200$ for the bicrystal case and $N = 450$ for the tricrystal case.

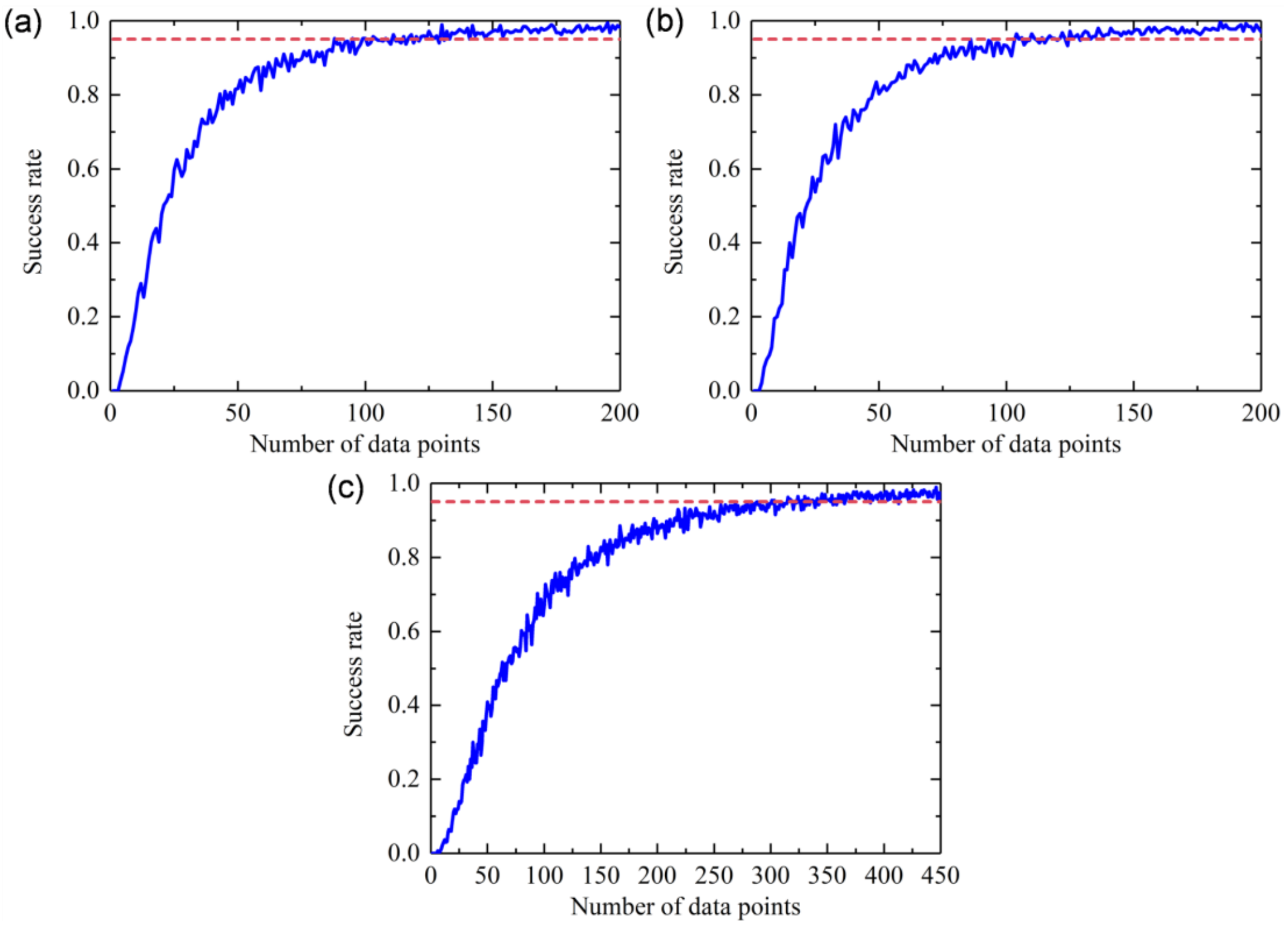

Fig. 4 Kolmogorov–Smirnov test on the GB normal stress in (a) bicrystal configuration under uniaxial tension; (b) bicrystal configuration under multiaxial tension; (c) tricrystal configuration under uniaxial tension.

# 3. Causation-guided machine learning framework for mechanism identification and interpretable prediction of GB local stresses

The framework integrates mechanics-informed characteristics, causation entropy for mechanism identification, and reduced-order regression for interpretable representation of GB stresses. Dominant features are extracted using a causation entropy method, and predictive models are subsequently established via regression based on important candidate functions. Finally, the generalizability of the extracted features from bicrystal uniaxial tension to multiaxial and polycrystalline cases is evaluated, to ensure mechanistic transferability and robustness of the predictive model. Representative machine-learning models are also employed to capture the high-dimensional nonlinear relationships between the characteristics and GB stress.

### 3.1. Causation entropy attribution of microscale physics characteristics and interpretable regression

Identifying the dominant microscale characteristics governing the GB normal stress constitutes the primary objective of this work. Predictive models constructed based these key features are expected to be mechanistically interpretable and to exhibit robust predictive performance. A causation entropy method, combined with interpretable regression, is employed in this work to identify important microscale characteristics and their key candidate functions. In this work, the aim of the presented CE-based framework is not to construct a response-assisted surrogate model using already-evolved mechanical fields as inputs. Instead, the framework aims to identify transferable microstructural descriptors that are available prior to deformation evolution and establish reduced-order models for predicting mechanical responses under creep conditions. Accordingly, quantities such as local stress, strain, and strain rate obtained from CPFE simulations are treated as response variables rather than independent descriptors, because they emerge from the coupled evolution of elastic accommodation and plastic deformation during loading. The purpose of the present framework is to identify the initial microstructural and crystallographic information that governs these responses before deformation evolution.

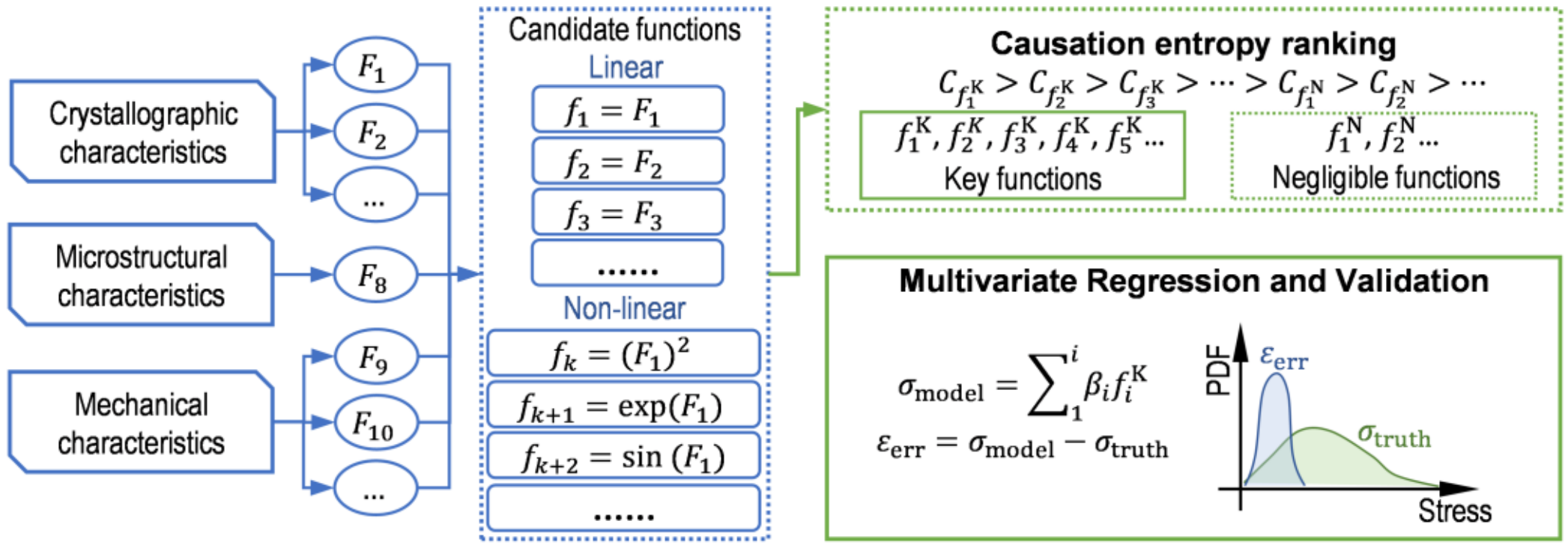

Fig. 5 General procedure for selecting candidate functions from microscale feature based on causation entropy method

The general workflow of the causation entropy method is illustrated in Fig. 5. The crystallographic characteristics $F_1 \sim F_7$, the microstructural characteristic $F_8$, and the local micromechanical characteristics $F_9 \sim F_{18}$, as proposed and defined in Section 2.2, are considered. A corresponding library of candidate basis functions is constructed from these primary characteristics. Specifically, for a given characteristic $F_i$ $(i = 1, 2, ..., 18)$, the designed candidate functions include:

$$\begin{gathered} F_i, F_i^2, F_i^3, F_i^4, \\ \tanh(F_i), \exp(F_i), \exp(-F_i), \\ \log(1+|F_i|), \frac{1}{x}, \frac{F_i}{1+F_i^2}, \\ \sin(F_i) \end{gathered} \tag{38}$$

The compact basis set Eqn. (38) is adopted to enrich the candidate-function library while avoiding an undue expansion of the function space. Baseline functions (low-order polynomials) are included to represent common smooth nonlinear trends. Scale- or boundary-effect functions are incorporated to capture saturating and attenuation-type behaviours with bounded responses and improved numerical stability. Periodic functions (trigonometric terms) are introduced to account for the intrinsic periodicity and symmetry of angular variables. In addition, for the two scalar angular descriptors, namely the GB misorientation angle $\theta$ and the inclination angle $\phi_{\mathrm{GB}}$, trigonometric terms are further included:

$$\cos(F_i), \sin^2(F_i), \cos^2(F_i) \tag{39}$$

These terms are introduced to represent the periodicity and even symmetry of the angular variables. Explicit cross terms between different characteristics are not included in the basis-function set, as ad hoc interactions generally lack clear physical interpretability and may lead to a combinatorial growth of the candidate library. Instead, the coupling between crystallographic and microstructural/micromechanical characteristics is expected to be mediated, in a physically meaningful manner, through the anisotropic local mechanical characteristics, which reflect their combined influence.

As shown in Fig. 5, causation entropy analysis is further performed for the candidate basis functions $f_k$ $(k = 1, 2, 3...)$. In the present work, a total of 204 candidate functions are considered. Key functions are selected according to the causation-entropy ranking, while negligible functions are identified and discarded. Grounded in information-theoretic principles, the causation entropy method effectively identifies physically important characteristics, thereby enabling statistical models with clear physical interpretability (Dunham et al., 2025; Zhang et al., 2025). Specifically, the (differential) Shannon entropy (Shannon, 1948), computed from the probability density function (PDF) $p(x)$, quantifies the uncertainty of the model output.

$$H(\mathbf{X}) = -\int_{\mathbf{X}} p(\mathbf{x}) \ln p(\mathbf{x}) \mathrm{d}\mathbf{x} \tag{40}$$

where $\mathbf{X}$ is the domain of the random variable $\mathbf{x}$. The conditional entropy and the joint entropy for multivariate variables $\mathbf{X}$ and $\mathbf{Y}$ can be expressed, respectively, as:

$$H(\mathbf{Y}|\mathbf{X}) = -\int_{\mathbf{X}}\int_{\mathbf{Y}} p(\mathbf{x},\mathbf{y}) \ln p(\mathbf{y}|\mathbf{x}) \mathrm{d}\mathbf{y}\mathrm{d}\mathbf{x} \tag{41}$$

$$H(\mathbf{X},\mathbf{Y}) = -\int_{\mathbf{X}}\int_{\mathbf{Y}} p(\mathbf{x},\mathbf{y}) \ln p(\mathbf{x},\mathbf{y}) \mathrm{d}\mathbf{y}\mathrm{d}\mathbf{x} \tag{42}$$

with the following relation:

$$H(\mathbf{Y}|\mathbf{X}) = H(\mathbf{X},\mathbf{Y}) - H(\mathbf{X}) \tag{43}$$

For the candidate basis-function list $\mathbf{f}$ corresponding to Eqns.(38)~(39), it can be expressed as:

$$\mathbf{f} = [f_1, f_2, \ldots, f_m, \ldots, f_N] \tag{44}$$

Eqns.(40)~(43) describe the general mathematical framework of information entropy. When evaluating the unique information contribution of a specific candidate basis function $f_m$ in the function library $\mathbf{f}$ for explaining the target variable $\mathbf{Y}$, causation entropy is employed and defined as follows (Dunham et al., 2025; Zhang et al., 2025):

$$C_{f_m \to \mathbf{Y}|[\mathbf{f} \setminus f_m]} = H(\mathbf{Y}|[\mathbf{f} \setminus f_m]) - H(\mathbf{Y}|[\mathbf{f}]) \tag{45}$$

in which $\mathbf{f} \setminus f_m$ represents the set of candidate bases functions excluding $f_m$. It quantifies the reduction in conditional entropy by comparing the conditional uncertainty of $\mathbf{Y}$ given the variables $\mathbf{X}$ without $f_m$ and that given the variables $\mathbf{X}$ with $f_m$ included. For the mechanical responses near GB considered in the present study, $\mathbf{Y}$ represents the specific response quantity of interest, which may include the GB normal stress, equivalent creep strain, equivalent creep strain rate, or other relevant mechanical responses.

For the average GB normal stress $\bar{\sigma}_{33}^{\mathrm{L}}$, i.e., $\mathbf{Y}=\bar{\sigma}_{33}^{\mathrm{L}}$, the corresponding causation entropy is computed as:

$$C_{f_m\to\bar{\sigma}_{33}^{\mathrm{L}}|[\mathbf{f}\setminus f_m]}=H\left(\bar{\sigma}_{33}^{\mathrm{L}}\,|[\mathbf{f}\setminus f_m]\right)-H\left(\bar{\sigma}_{33}^{\mathrm{L}}\,|[\mathbf{f}]\right) \tag{46}$$

The causation entropy quantifies the unique information contributed by the $f_m$ in explaining the GB normal stress $\bar{\sigma}_{33}^{\mathrm{L}}$. The normalized relative value $\tilde{C}$ based on the maximum causation entropy can be expressed as:

$$\tilde{C}_{f_m\to\bar{\sigma}_{33}^{\mathrm{L}}|[\mathbf{f}\setminus f_m]}=\frac{C_{f_m\to\bar{\sigma}_{33}^{\mathrm{L}}|[\mathbf{f}\setminus f_m]}}{\max\limits_{m}\left(C_{f_m\to\bar{\sigma}_{33}^{\mathrm{L}}|[\mathbf{f}\setminus f_m]}\right)} \tag{47}$$

Here, $\max\limits_{m}(\bullet)$ denotes the maximum causation entropy over all candidate functions $f_m$. This normalization rescales the values for relative comparison across candidates, primarily for visualization purposes. After ranking the candidate functions by their causation entropy, a statistical model for $\bar{\sigma}_{33}^{\mathrm{L}}$ can be constructed by performing linear regression on key functions $f_i^{\mathrm{K}}$ of these physics-based features:

$$\bar{\sigma}_{33}^{\mathrm{L}}\left(F_1,F_2,\ldots,F_{18}\right)=\sum_{i=1}^{N_{\mathrm{tr}}}\beta_i f_i^{\mathrm{K}} \tag{48}$$

in which $N_{\mathrm{tr}}$ denotes the number of retained significant candidate functions. $\beta_i$ denotes the coefficients of the model. The model's predictive capability depends on the complexity of the candidate functions; by selecting an appropriate $N_{\mathrm{tr}}$, it can achieve accurate predictions of complex nonlinear relationships while maintaining a controlled computational cost, and explicitly reveal the interpretable relationship between input and output physical variables.

As shown in Fig. 5, causation entropy is employed as an information-theoretic feature identification tool rather than a network-based prediction model. Specifically, the calculated $C_{f_m\to\mathbf{Y}|[\mathbf{f}\setminus f_m]}$ quantifies the unique information contribution of each candidate basis function $f_m$ to the target response $\mathbf{Y}$. Therefore, the CE analysis establishes a directed descriptor-response information relationship, where the strength of each relationship is determined by the corresponding causation entropy value. Unlike black-box machine-learning models, the regression step in the present framework aims to obtain an explicit mathematical representation between the selected physical descriptors and the target response.

## 3.2. Validation of mechanistic representation and transferability assessment

The present section describes the validation procedure and assesses the transferability of the proposed framework. Here, transferability refers to the persistence of governing characteristics and functional representation across loading/configuration changes, rather than the direct reuse of previously fitted coefficients. The validation is used to assess whether the identified representation captures a more stable governing relation and whether its transferability is consistent with a mechanistic basis.

As shown in Fig. 6, the data obtained from uniaxial tensile tests on bicrystals are partitioned into training and test sets. Specifically, 200 random bicrystal configurations are generated and analyzed using CPFE simulations. A total of 80% of the data are used as the training set, while the remaining 20% are used as the test set. Based on the training set, causation entropy (CE) analysis is first performed to identify the key microscale characteristics governing GB normal stress. Subsequently, based on the candidate functional forms constructed from the extracted microscale characteristics, an interpretable multivariate regression is performed according to Eqn. (48). It should be noted that the regression model incorporates the physical information carried by the key microscale characteristics, thereby ensuring its interpretability and generalizability. Based on the key microscale characteristics and the relevant candidate functional forms, machine-learning models with $\bar{\sigma}_{33}^{\mathrm{L}}$ as the output can also be constructed. The model is then validated on the held-out test set. All feature-ranking, candidate-function selection, truncation-level determination, and coefficient identification procedures are performed using the training set only. The held-out test set was used exclusively for the final evaluation of predictive performance and was not involved in any stage of feature screening, model-form selection, or parameter calibration. This protocol was adopted to ensure that the reported test-set performance reflects genuine out-of-sample predictive capability of the identified representation.

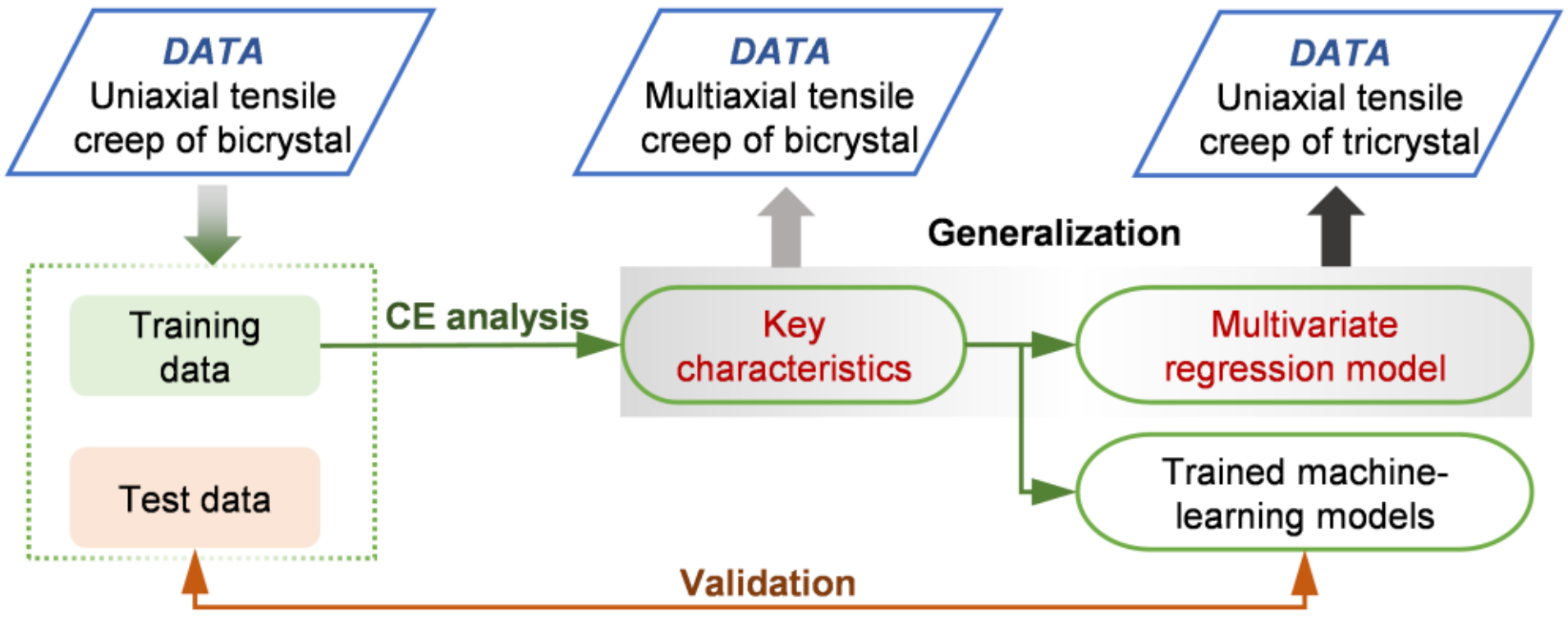

Fig. 6 Training, testing, validation, and generalization: Key characteristics are extracted from uniaxial tensile data to train and validate a predictive model, while the generalizability of both the characteristics and the model is assessed under multiaxial loading and tricrystal configurations.

Although the key microscale characteristics identified by causation entropy analysis and the regression model constructed from the corresponding important candidate functional forms are obtained only from simple uniaxial creep data of minimal grain clusters, i.e., bicrystal configurations, they remain physically meaningful. Therefore, the generalizability of the physical relations extracted from simple loading conditions of minimal grain clusters is further examined. As shown in Fig. 6, the key microscale characteristics and candidate functional forms identified from uniaxial tensile creep data of bicrystals are further examined for their generalizability to multiaxial stress states in bicrystals and to tricrystal configurations. If the identified characteristics and the resulting model remain valid, this implies that the physical relations and information they embody are of a general and transferable nature.

## 3.3. Auxiliary surrogate validation using representative machine-learning models

Although Eqn. (48) is capable of predicting $\bar{\sigma}_{33}^{\mathrm{L}}$, its ability to capture nonlinear behavior entirely depends on the nonlinearity of the chosen candidate functions. Therefore, in this work, representative machine learning methods are also employed to construct predictive models. These machine learning methods can effectively capture the nonlinear relationship between input and output quantities. Machine learning is used only as an auxiliary validation tool to test whether the characteristics and candidate functions selected by the causation-entropy analysis improve predictive accuracy and robustness. Its objective is not to compare the CE-based interpretable regression directly with black-box algorithms, but to determine whether the CE-selected candidate functions provide more informative inputs than the original physical characteristics across different nonlinear surrogate-model classes.

As shown in Fig. 7, the original characteristics are used as inputs to train and validate the machine learning models, employing K-fold cross-validation in combination with Bayesian hyperparameter optimization. To verify the validity of the key physical features and candidate functions identified via causation entropy ranking, the key candidate functions obtained from the causation entropy analysis are used as inputs, and the models are trained and validated within the same machine-learning framework. During training, the mean squared error (MSE) is employed as the loss function, while the coefficient of determination $R^2$ metric on the test set is used to assess the predictive accuracy. By comparing the predictive performance of models trained with different input strategies, the effect of causation entropy analysis can be evaluated.

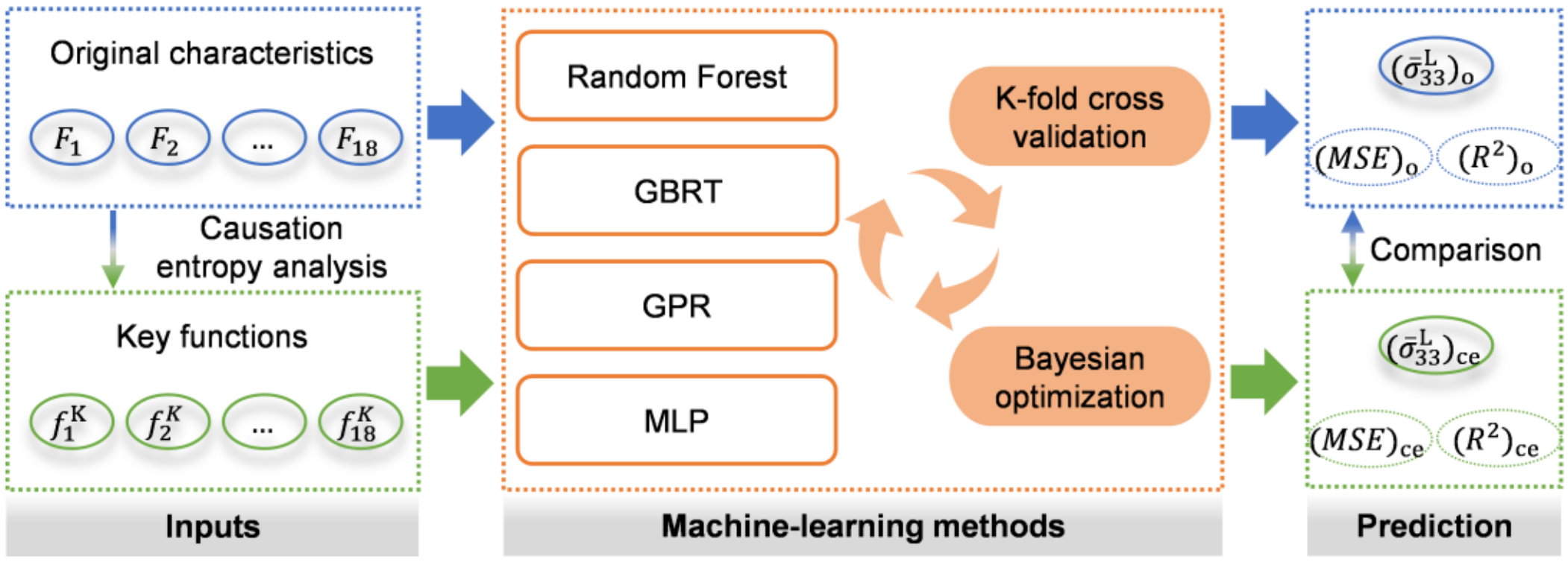

Fig. 7 Prediction of GB normal stress based on machine learning methods: using the initial characteristics and the key candidate functions extracted via causation entropy as inputs, respectively, with models' hyperparameters selected through K-fold cross-validation and Bayesian optimization.

Random Forest (RF), Gradient Boosted Regression Trees (GBRT), Gaussian Process Regression (GPR), and Multi-Layer Perceptron (MLP) are selected to cover a diverse set of machine-learning paradigms, including ensemble methods, probabilistic models, and neural networks, enabling a comprehensive evaluation of predictive performance. Their purpose is not to emphasize algorithmic superiority, but to examine whether the extracted characteristics provide consistent predictive gains across different model classes. The hyperparameters involved in models training are provided in Appendix B.

Both the multivariate regression model given in Section 3.1 and the machine-learning model possess predictive capability for unseen data, but they offer different advantages. The multivariate regression model is characterized by physical interpretability, together with more transparent robustness and generalizability. In contrast, the machine-learning model places greater emphasis on capturing nonlinear input-output relations and provides higher representational flexibility.

# 4. Results

Based on the aforementioned physics-informed micromechanical characteristics and causation-guided machine learning framework, this section presents the key physical characteristics, the interpretable regression model and its transferability across physical conditions, as well as the improvement in machine-learning performance enabled by the use of candidate feature functions.

## 4.1. Key characteristics governing GB normal stress

For the bicrystal system under uniaxial tensile creep, a dataset is constructed from CPFE simulations, with 18 key microstructural and micromechanical characteristics as inputs and the average GB normal stress $\bar{\sigma}_{33}^{\mathrm{L}}$ as the output. The 18 key physical characteristics are also derived from the underlying computational model. As a preliminary examination of the data, Pearson correlation analysis is conducted in Appendix C, which provides only insights into linear correlations. To elucidate, in a physically interpretable manner, how multiple characteristics inform the GB normal stress, a causation-entropy analysis is conducted based on the theoretical framework. One important characteristic of causation entropy is that it offers insights, from an information-theoretic perspective, into the dominant role of key features even before an association model is established.

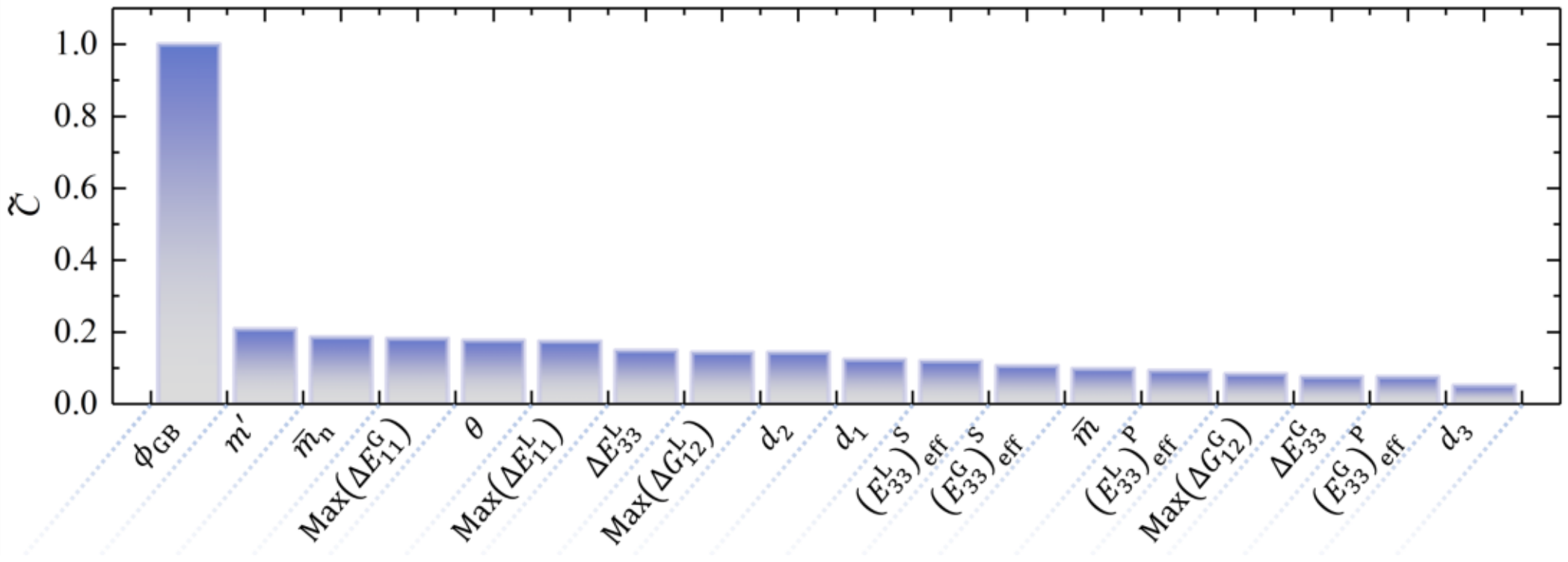

Fig. 8 Normalized causation-entropy ranking of physics-based microscale characteristics, based on all candidate functions involving each characteristic.

The causation entropy is evaluated at the level of each physics-based microscale characteristic. For a given characteristic $F_i$, all candidate functions in the library that involve $F_i$, including its linear and nonlinear terms are collected into a group $G_i = \{ f_k : f_k \text{ involves } F_i \}$ .The group-level causation entropy is then computed as $\tilde{C}_{G_i \to \bar{\sigma}_{33}^{\mathrm{L}} | [\mathbf{f} \setminus G_i]}$, which quantifies the additional information provided by this group conditioned on the remaining candidate library. Therefore, Fig. 8 shows the normalized causation-entropy ranking of the 18 physics-based microscale characteristics.

Among the considered characteristics, the GB inclination angle $\phi_{\mathrm{GB}}$ plays the dominant role. This is consistent with the classical mechanics of stress transformation, according to which the angle $\phi_{\mathrm{GB}}$ governs the normal stress acting on an inclined plane in a homogeneous continuum. The present results indicate that this basic feature is retained even in the heterogeneous polycrystalline system. The second-ranked characteristic is the slip transmission factor $m'$. As it characterizes the ease of slip transmission across the GB, it is expected to play an important role in governing the GB normal stress. The high CE importance of $m'$ does not imply that the present CPFE formulation explicitly models slip transmission resistance. Rather, it indicates that the crystallographic compatibility information represented by $m'$ provides unique information associated with the simulated GB stress response. This observation is consistent with previous full-field crystal plasticity studies (Eghtesad et al., 2023; Zhang et al., 2022), where $m'$ has been used as a microstructural descriptor to interpret localized deformation behavior.

The third-ranked characteristic is the mean Schmid-type indicator $\bar{m}_{\mathrm{n}}$ for climb-driven creep of the two grains adjacent to the GB. This quantity characterizes the tendency of the neighboring grains to undergo local climb-related viscoplastic deformation, and thus plays an important role in the local stress-relaxation process near the GB. The fact that $\bar{m}_{\mathrm{n}}$ is ranked ahead of $\bar{m}$ indicates that climb-assisted creep differs fundamentally from pure plastic slip in its influence on the GB response. In particular, diffusion and climb appear to play a more pronounced role in stress relaxation. This is consistent with the findings of Cheng et al. (2024). Ranking immediately after these are the minimum misorientation angle $\theta$, which characterizes the crystallographic mismatch at the GB, and the micromechanical mismatch represented by the elastic-modulus difference $\max\left(\Delta E_{11}^{\mathrm{G}}\right)$, $\max\left(\Delta E_{11}^{\mathrm{L}}\right)$, $\Delta E_{33}^{\mathrm{L}}$ and $\max\left(\Delta G_{12}^{\mathrm{L}}\right)$. These findings indicate that crystallographic and micromechanical mismatches are important contributors to the local stress at the GB.

Furthermore, the 18 microscale characteristics are grouped into the following categories: (1) the mean crystallographic characteristics of the grains on either side of the GB ($d_1$, $d_2$, $d_3$, $\bar{m}$, $\bar{m}_{\mathrm{n}}$); (2) the crystallographic mismatch features between the grains adjacent to the GB ($\theta$, $m'$); (3) mechanical mean characteristics ($\left(E_{33}^{\mathrm{G}}\right)_{\mathrm{eff}}^{\mathrm{S}}$, $\left(E_{33}^{\mathrm{G}}\right)_{\mathrm{eff}}^{\mathrm{P}}$, $\left(E_{33}^{\mathrm{L}}\right)_{\mathrm{eff}}^{\mathrm{S}}$, $\left(E_{33}^{\mathrm{L}}\right)_{\mathrm{eff}}^{\mathrm{P}}$); (4) mechanical mismatch features ($\Delta E_{33}^{\mathrm{G}}$, $\max\left(\Delta E_{11}^{\mathrm{G}}\right)$, $\max\left(\Delta G_{12}^{\mathrm{G}}\right)$, $\Delta E_{33}^{\mathrm{L}}$, $\max\left(\Delta E_{11}^{\mathrm{L}}\right)$, $\max\left(\Delta G_{12}^{\mathrm{L}}\right)$); (5) microscale geometric characteristics ($\phi_{\mathrm{GB}}$). All candidate functions are classified according to these five categories, and the normalized causation-entropy values and ranking heatmaps of the top 140 candidate functions are presented in Fig. 9a. An examination of the top 20 terms shows that the leading positions are predominantly occupied by candidate functions associated with crystallographic and micromechanical mismatch characteristics, underscoring the important role of microscale mismatch at the GB in governing the local stress. In particular, candidate functions associated with micromechanical mismatch are densely distributed among the top-ranked terms, indicating that micromechanical mismatch serves as an effective descriptor for capturing the key physical information governing the local GB stress.

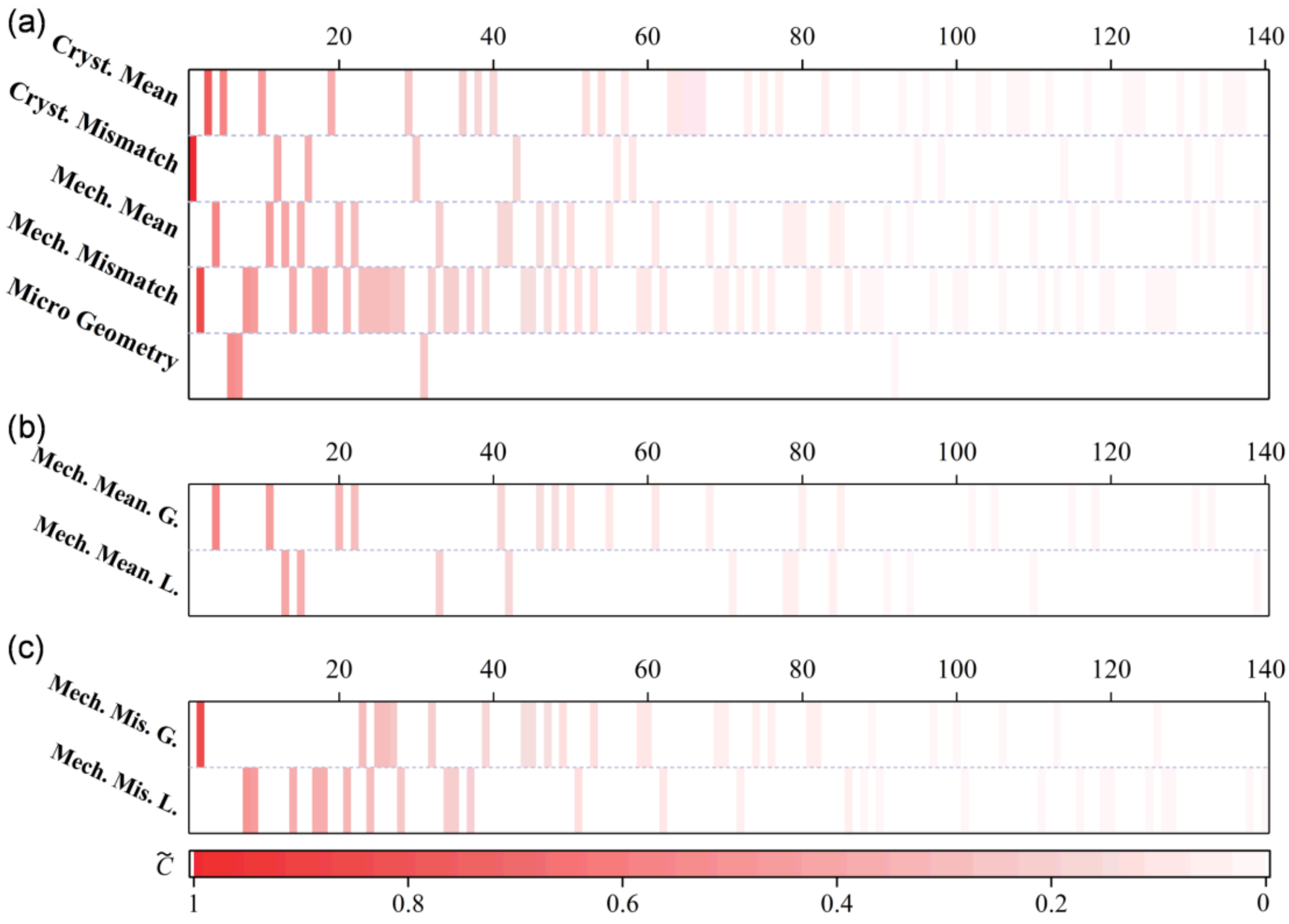

Fig. 9 Heat maps of top 140 candidate feature functions for bicrystal subjected to uniaxial tension creep: (a) grouped into crystallographic mean features, crystallographic mismatch features, mechanical mean features, mechanical mismatch features, and micro-geometric features; (b) grouped into global versus local mechanical mean features; (c) grouped into global versus local mechanical mismatch features.

In addition, the candidate functions associated with micromechanical mean and mismatch characteristics are further distinguished into those defined in the global coordinate system and those defined in the local coordinate system, and the corresponding rankings are shown in Fig. 9b and Fig. 9c. For the micromechanical mean characteristics, the candidate functions associated with the global features are, on the whole, ranked ahead of those associated with the local features in the causation-entropy ordering. The global features are defined in the specimen loading coordinate system, indicating that the influence of averaged mechanical properties on the stress is in fact closely related to the macroscopic loading condition. In contrast, for the micromechanical mismatch characteristics, the candidate functions associated with the local features tend to rank ahead of those associated with the global features, indicating that local-coordinate characteristics are more important for characterizing micromechanical mismatch. This is because micromechanical mismatch is inherently associated with the local stress state and local interfacial geometry at the GB, and is therefore more effectively captured in the local coordinate system.

It should be noted that the extraction of key features and the ranking of candidate functions are conducted entirely on the basis of the bicrystal system under uniaxial tension, and they are expected to remain applicable to more complex multiaxial and tricrystal systems.

## 4.2. Interpretable regression for GB normal stress based on physics-informed candidate functions

### 4.2.1. Bicrystal under a macroscopic uniaxial tension

Based on the datasets of bicrystal, the causation entropy ranking of all candidate functions can be obtained. Next, the top $N_{\mathrm{tr}}$ candidate functions defined by characteristics are selected for linear regression according to Eqn. (48).

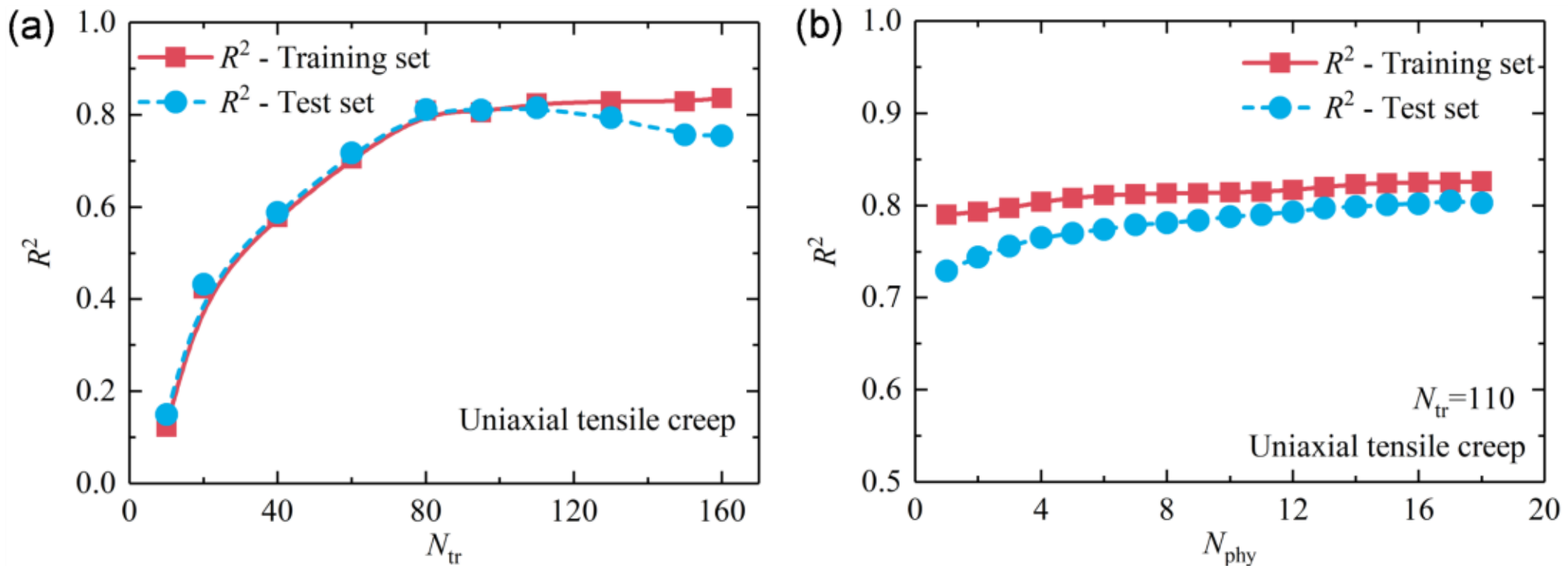

Fig. 10 Effects of (a) the number of retained CE-ranked candidate functions $N_{\mathrm{tr}}$ and (b) the number of physical descriptors $N_{\mathrm{phy}}$ on the $R^2$ of the reduced-order model under uniaxial tensile creep.

To investigate the influence of the number of retained candidate functions in the explicit regression model on the predictive performance, the dependence of the coefficient of determination ($R^2$) for predicting $\bar{\sigma}_{33}^{\mathrm{L}}$ on $N_{\mathrm{tr}}$ is presented in Fig. 10a. Overall, the prediction accuracy gradually improves with increasing number of retained functions, indicating that incorporating more physically ranked candidate functions enables the model to capture additional information governing the GB normal stress response. When $N_{\text{tr}}$ reaches approximately 80, the $R^2$ values for both the training and test sets approach a plateau, suggesting that the dominant information has been effectively captured. When $N_{\text{tr}}$ exceeds 110, the $R^2$ value of the training set continues to increase slightly, whereas that of the test set shows a decreasing trend. This indicates that retaining excessive candidate functions, which corresponds to increasing the number of regression coefficients requiring calibration, may introduce slight overfitting when the model complexity approaches the scale of the training dataset. Nevertheless, the overall predictive performance remains robust. This robustness can be attributed to the CE-based identification of informative candidate functions prior to regression and the physically interpretable nature of the reduced-order representation.

Furthermore, Fig. 8 shows the relative CE values of the physics-based characteristics, among which one characteristic exhibits a clearly dominant contribution. To further investigate the influence of the number of retained physics-based characteristics, denoted by $N_{\text{phy}}$, on the predictive capability of the reduced-order model, candidate-function groups associated with the physics-based characteristics were progressively incorporated according to their group-level CE ranking, from the highest to the lowest. The corresponding prediction accuracy is shown in Fig. 10b. When only the highest-ranked characteristic is considered, the reduced-order model already achieves reasonably good predictive performance, indicating that the dominant physical mechanism has been effectively captured. As additional groups of candidate functions associated with lower-ranked characteristics are included, the coefficient of determination ($R^2$) increases gradually and eventually approaches a plateau after all characteristic-based candidate functions have been incorporated. These results indicate that, although the dominant characteristic accounts for the major variation in GB normal stress, the remaining physics-based characteristics provide complementary mechanistic information. Consequently, incorporating a broader set of physically meaningful characteristics enables the reduced-order model to capture more complete physical effects and further improve its predictive performance.

For the bicrystal system under uniaxial loading, $N_{\mathrm{tr}} = 110$ is found to provide good performance on both the training and test sets defined in Section 3.3. As shown in Fig. 11a, the distribution of the reference $\bar{\sigma}_{33}^{\mathrm{L}}$ values obtained from the CPFE simulations is presented by the blue histogram. Owing to the influence of multiple microscale factors, the distribution exhibits substantial scatter and a pronounced non-Gaussian character. Throughout this work, the CPFE-computed GB stress is regarded as the reference $\bar{\sigma}_{33}^{\mathrm{L}}$, and the corresponding values obtained from the regression are termed the predicted $\bar{\sigma}_{33}^{\mathrm{L}}$. The difference between the reference and predicted $\bar{\sigma}_{33}^{\mathrm{L}}$ is defined as the stress prediction residual. As shown in Fig. 11a, the residual is shown by the red histogram. A marked reduction in scatter is observed, with the variance reduced by 93.7%. In addition, the residual is concentrated around zero and displays an approximately Gaussian distribution. The remaining discrepancy may be attributed to higher-order effects and numerical or modeling errors that are not captured by the linear regression model in Eqn. (48). Fig. 11b shows the performance of the regression model on the training and test sets. Overall, the model exhibits good predictive performance, with a test-set $R^2$ of 0.803. This suggests that the candidate functions constructed from the key characteristics identified by the causation-entropy analysis retain the dominant information governing the GB normal stress. Good predictive performance can be achieved using only nonlinear regression with the prescribed basis functions. This highlights the potential of combining causation-entropy analysis with regression for reduced-order model construction.

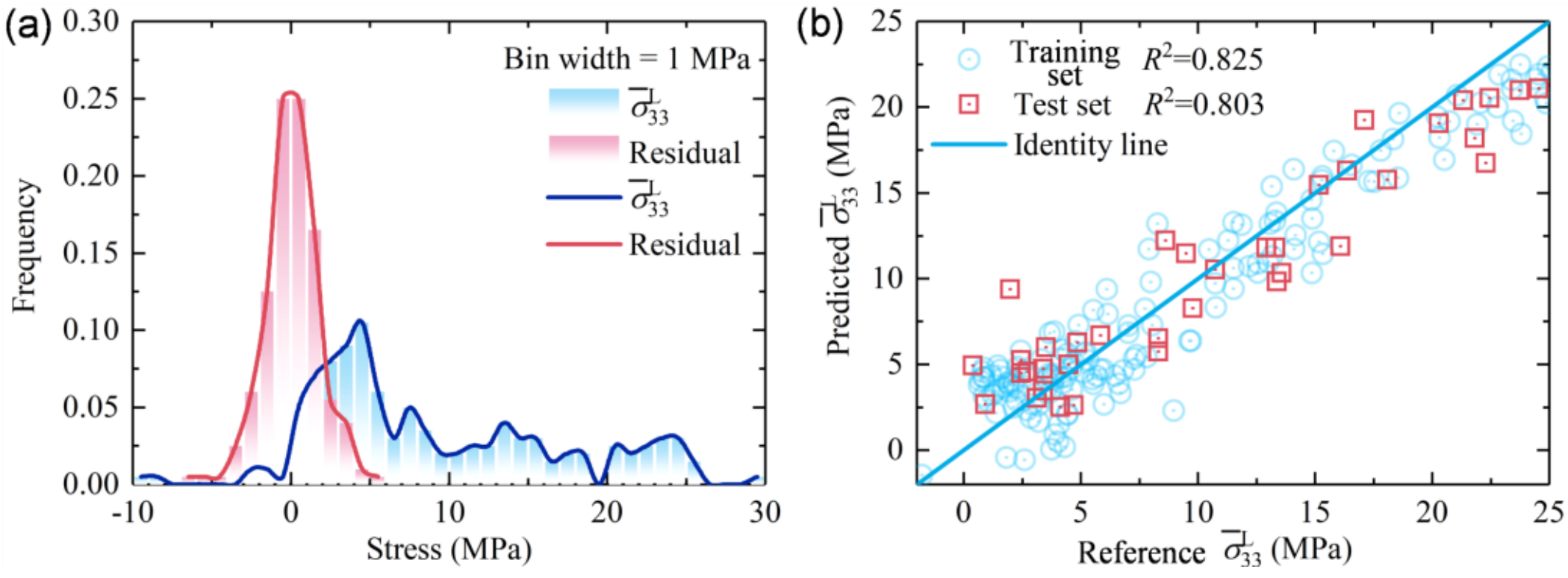

Fig. 11 Bicrystal under uniaxial tensile creep: (a) probability density functions of the original average

GB normal stress and the model residual stress; (b) comparison between the CE-regression model predictions and the corresponding reference values.

### 4.2.2. Bicrystal under a macroscopic multiaxial stress state

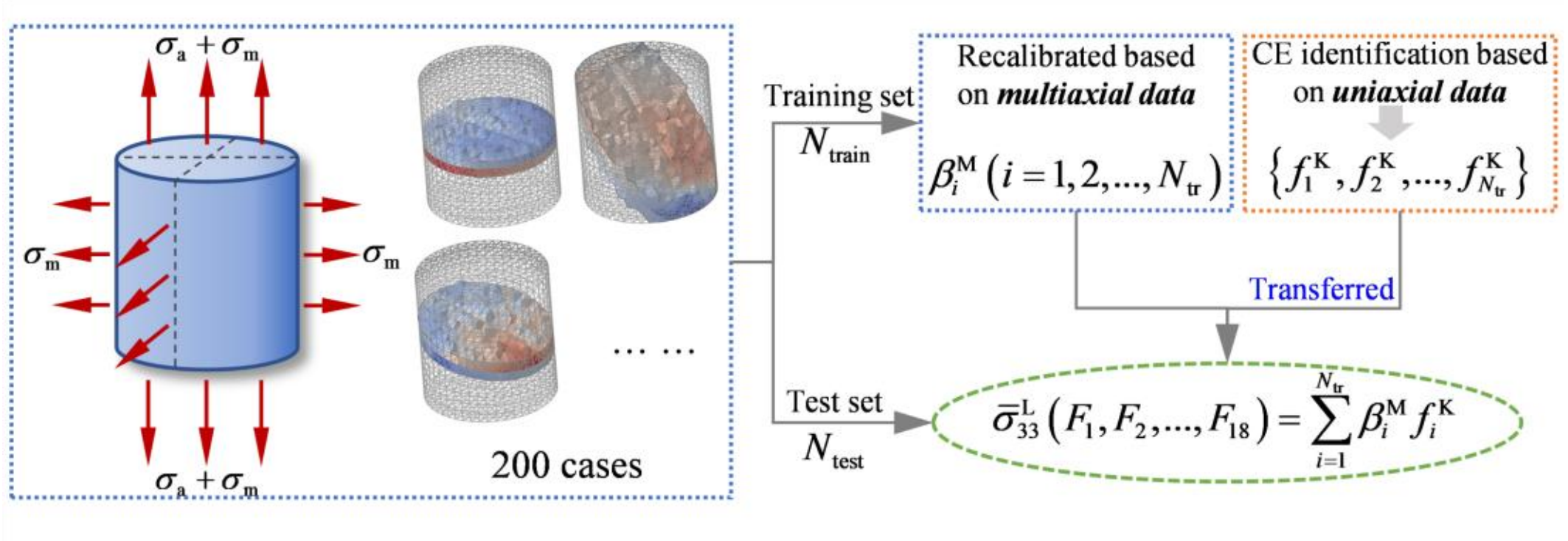

Fig. 12 Schematic of the training and test of cases under multiaxial loading

To evaluate whether the key micromechanical characteristics and the functional form of the regression model identified under uniaxial loading remain applicable under multiaxial stress states, a specific multiaxial loading condition was considered by superimposing a prescribed uniaxial stress component $\sigma_a$ with a hydrostatic stress component $\sigma_m$. The equivalent stress of this multiaxial stress state is 24 MPa, corresponding to a stress triaxiality of 3. The macroscopic stress state is set to a high triaxiality condition ($\eta = 3$), which is markedly different from the low triaxiality state in uniaxial tension. As shown in Fig. 12, a total of 200 multiaxial loading cases were generated. These cases were divided into a training set containing $N_{train}$ cases and a test set containing $N_{test}$ cases. The transferability of the key candidate functions identified from the uniaxially loaded bicrystal configuration and the regression model based on Eqn. (48) is assessed under multiaxial loading. In this multiaxial case, the candidate functions identified under uniaxial tension and the corresponding explicit regression functional form are directly retained and transferred to the multiaxial condition. Only the regression coefficients $\beta_i^M$ are recalibrated using the multiaxial dataset. It should be emphasized that the present analysis concerns the transferability of the extracted micromechanical characteristics and the functional form of the regression model, rather than the direct transfer of a trained predictor or its previously fitted coefficients. Accordingly, under a different macroscopic stress state, the model must be retrained and the regression coefficients must be reidentified. In the training of the new model, the same key candidate functions from bicrystal uniaxial tensile case are adopted as inputs. The functional form of the regression model is kept unchanged, and only the coefficients $\beta_i^{\mathrm{M}}$ are recalibrated.

Similarly, the distribution of the GB normal stress $\bar{\sigma}_{33}^{\mathrm{L}}$ obtained from the CPFE simulations under the multiaxial stress state is shown by the blue histogram in Fig. 13a. The residual is represented by the red histogram and the variance is reduced by 93.9%. The performance of the regression model on the training and test sets is shown in Fig. 13b. An $R^2$ value of 0.804 is obtained on the test set, indicating good predictive accuracy for $\bar{\sigma}_{33}^{\mathrm{L}}$. It indicates that, although they are identified from the simpler uniaxial stress state, the candidate functions associated with the key physical characteristics extracted by the causation-entropy analysis and the regression form remain applicable under multiaxial loading. These results further support the transferability of the extracted candidate functions and the regression form from uniaxial to multiaxial loading, and confirm their effectiveness in describing the GB normal stress $\bar{\sigma}_{33}^{\mathrm{L}}$.

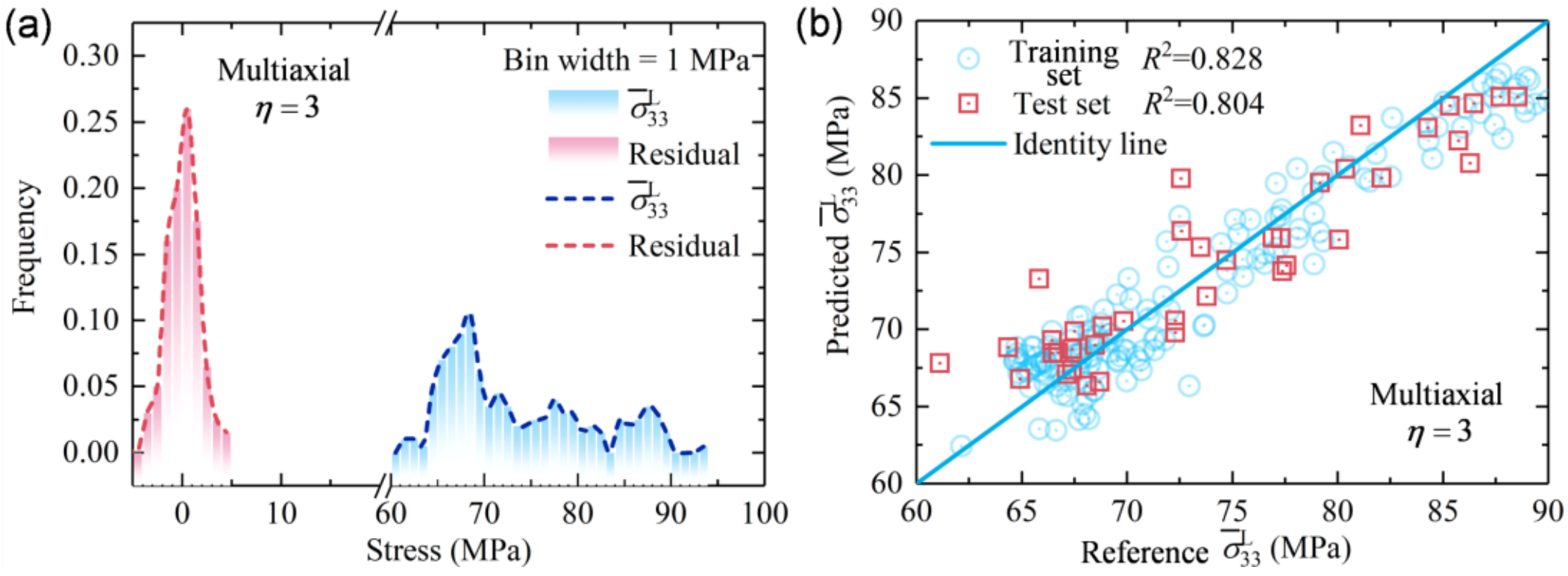

Fig. 13 Bicrystal under multiaxial tensile creep with $\eta = 3$: (a) probability density functions of the original average GB normal stress and the model residual stress, (b) comparison between the CE-regression model predictions and the corresponding ground-truth values.

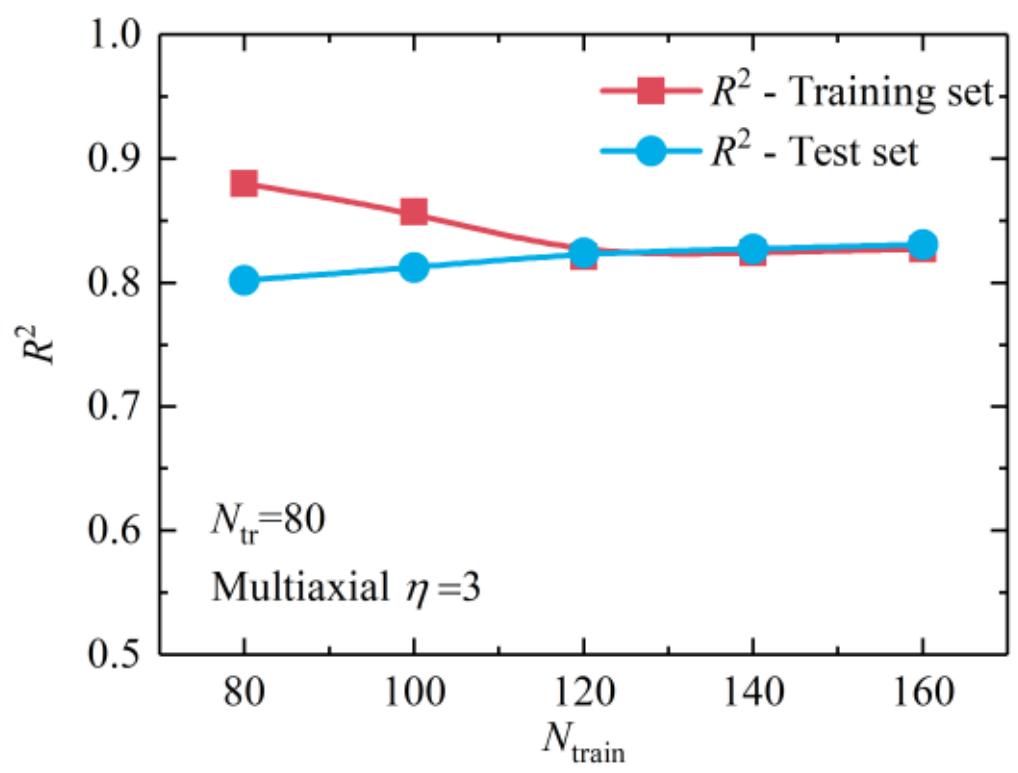

Fig. 14 Effect of different numbers of multiaxial cases used for coefficient recalibration on the coefficient of determination $R^2$

To investigate whether the regression coefficients can be effectively recalibrated using a limited amount of multiaxial data based on the identified governing mechanisms and candidate functional forms, the number of multiaxial cases used for recalibration, $N_{\text{train}}$, was varied. Here, $N_{\text{tr}}$ was fixed as 80, corresponding to the number of retained candidate functions. The regression coefficients were recalibrated using different numbers of multiaxial cases ( $N_{\text{train}} = 80 \sim 160$ ), and the resulting prediction performance was evaluated by the coefficient of determination $R^2$. The $R^2$ values for both the recalibration and independent test datasets under different $N_{\text{train}}$ values are shown in Fig. 14. It can be observed that, based on the identified explicit functional form and retained candidate functions, the prediction of $\bar{\sigma}_{33}^{\text{L}}$ achieves a high $R^2$ even with a limited number of multiaxial cases ( $N_{\text{train}} = 80$ ) for coefficient recalibration. As $N_{\text{train}}$ increases, the predictive performance is further enhanced, while the improvement in $R^2$ gradually becomes less significant.

### 4.2.3. Extension to tricrystal systems with intergranular interaction

This section further assesses the underlying physical mechanism and the physically interpretable representation based on microscale characteristics, together with the regression form, from the bicrystal system to the tricrystal system. Accordingly, for the tricrystal system, the key input functions and the regression form are retained, whereas the regression coefficients are reidentified.

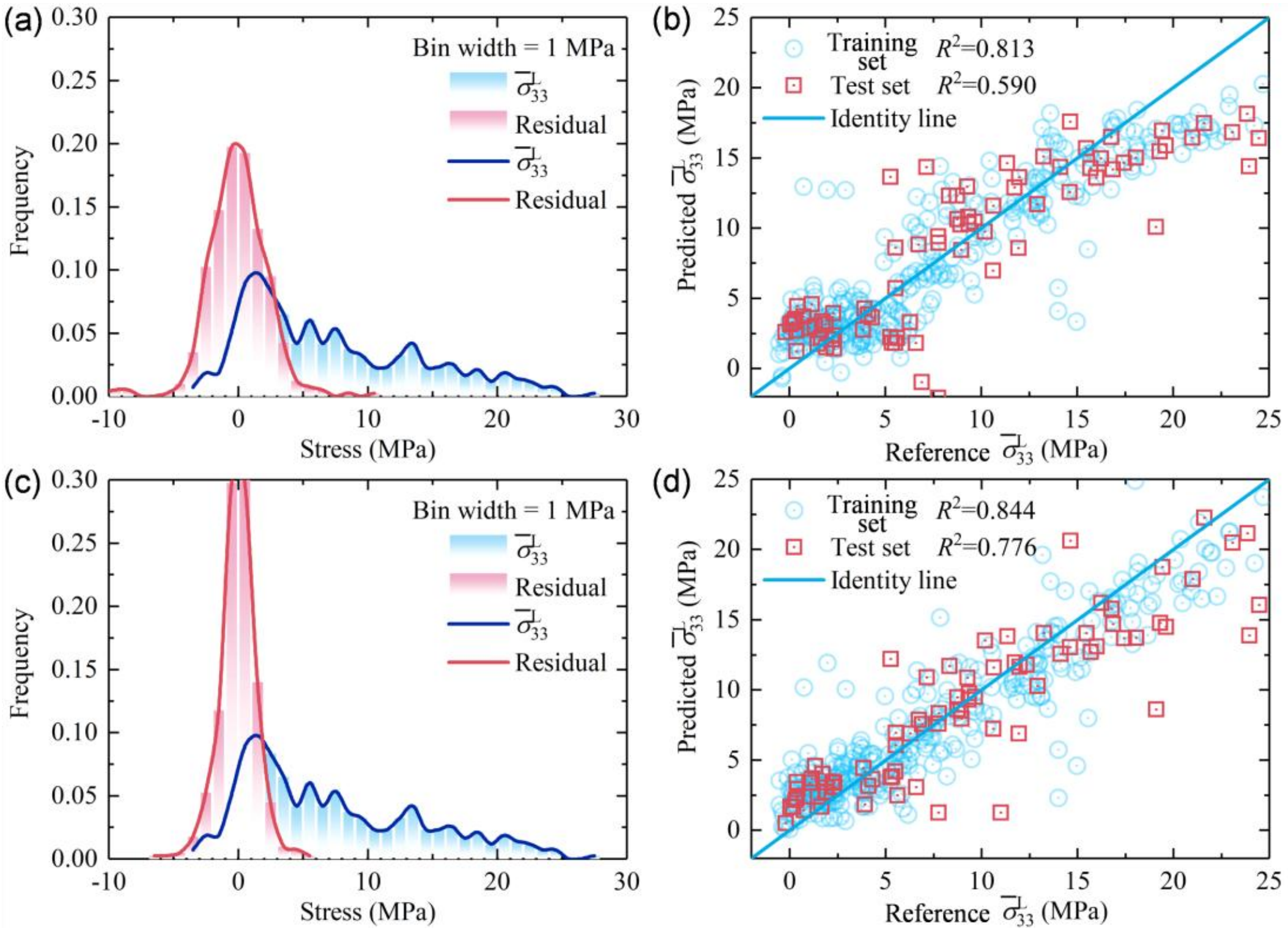

Fig. 15 Tricrystal under uniaxial tensile creep without considering grain-grain interaction: (a) probability density functions of the original average GB normal stress and the model residual stress, (b) comparison between the CE-regression model predictions and the corresponding ground-truth values; Tricrystal under uniaxial tensile creep considering grain-grain interaction: (c) probability density functions of the original average GB normal stress and the model residual stress, (d) comparison between the CE-regression model predictions and the corresponding ground-truth values.

The results obtained using only the regression form constructed from the candidate functions based on local microscale characteristics extracted from the bicrystal system are shown in Fig. 15a and Fig. 15b. In this case, the variance of the residual between the regression prediction and the original CPFE stress is reduced by only 87.5% relative to the variance of the original data, which is noticeably smaller than that achieved in the previous cases. The comparisons between the predicted and reference stresses on the training and test sets more directly reveal the underfitting behavior of the model. In the high-stress regime, the predictive performance of the model deteriorates significantly, and the test-set $R^2$ is only 0.590. This suggests that the candidate functions identified solely from the bicrystal case do not retain sufficient governing information for accurate prediction of the tricrystal response.

Therefore, for the tricrystal system, additional characteristics associated with neighboring grains need to be introduced to capture intergranular interactions that are absent from a purely local description. When new intergranular interactions emerge, the framework reveals the need for an explicit nonlocal extension and accommodates it naturally. In the present work, only the tricrystal configuration shown in Fig. 3b is considered, in which three GBs meet at a triple junction. Therefore, as a simple approximation, the average characteristics of the other two GBs connected to the target GB are introduced as supplementary physical features for the causation-entropy analysis and the subsequent model construction. Hence, 220 truncated candidate functions are retained in the regression based on Eqn. (48), i.e., $N_{\mathrm{tr}} = 220$. With the supplementary characteristics included, the variance of the residual between the regression model and the CPFE results is further reduced by 95.2% relative to the variance of the original data. The predictive capability of the regression model is also further improved, with a test-set $R^2$ of 0.776.

These results suggest that the key characteristics and candidate functions identified from the bicrystal system remain effective in the tricrystal system, indicating a reasonable degree of transferability. Meanwhile, the incorporation of nonlocal information associated with neighboring GBs leads to a marked improvement in predictive performance. It implies that a purely local description is insufficient for the tricrystal system, and that neighboring-GB effects must be incorporated to account for the additional intergranular interactions. Therefore, the enhanced representation provides a necessary extension of the physically informed framework for more complex grain configurations. This further demonstrates that, owing to its clear physical interpretability, the causation-entropy analysis is advantageous for introducing and extending physical characteristics. Additional repeated analyses based on different random data partitions yielded broadly consistent rankings of the dominant characteristics and similar test-set performance, further supporting the robustness of the identified representation.

It should be noted that the present transferability assessment is conducted using minimal grain clusters, with the aim of progressively introducing additional interaction effects while retaining a sufficiently controlled setting for identifying the governing micromechanical characteristics. Accordingly, the current analysis focuses primarily on the overall GB stress response and employs descriptors associated with crystallographic orientation, local GB geometry, and micromechanical mismatch. Although this representation is sufficient to capture the dominant variation in the bicrystal systems and remains effective after introducing nonlocal information in the tricrystal systems, it does not provide a complete description of the morphological and spatial complexity encountered in realistic polycrystalline microstructures.

In particular, the present tricrystal dataset does not contain sufficient variability to systematically quantify the effects of grain morphology and spatial topology, including grain size, grain shape or sphericity, relative grain dimensions, triple-junction geometry, quadruple-junction configurations, and the spatial arrangement of neighboring grains and GBs. These factors may alter the local mechanical constraint, load redistribution, and interaction range around a target GB. Their absence may therefore be one of the reasons why the predictive performance obtained for the tricrystal systems remains lower than that for the bicrystal systems, even after the characteristics of the neighboring GBs are incorporated.

A rigorous evaluation of such effects would require a substantially larger and more representative dataset containing broader variations in grain number, grain size and shape, junction topology, and spatial configuration. Nevertheless, this limitation is primarily associated with the scope and information content of the current minimal-grain-cluster dataset rather than with the causation-guided methodology itself. The proposed framework is not restricted to orientation-related descriptors; morphology-, topology-, and spatial-position-related quantities can be introduced as additional physics-based characteristics and evaluated through the same CE-based identification and reduced-order modeling procedure.

### 4.3. Auxiliary machine-learning assessment of CE-selected candidate functions

As mentioned in Section 3.3, the objective of this section is to quantitatively evaluate whether the candidate functions identified by causation entropy constitute a more informative representation of GB normal stress than the original physical characteristics. This is not a direct comparison between the CE-based interpretable regression and conventional machine-learning algorithms. Instead, for each surrogate-model class, predictions obtained using the original characteristics are compared with those obtained using an equal number of CE-selected candidate functions.

To assess the effectiveness of the candidate functions identified by the causation-entropy analysis, they are used as inputs to construct a surrogate model for predicting the GB normal stress. RF, GBRT, GPR, and MLP are employed as representative machine-learning models spanning different learning paradigms. In the present work, the 18 physical characteristics are taken as the baseline input set. For comparison, models are constructed by directly using these original inputs to predict the GB normal stress. In addition, the top 18 candidate functions are used as inputs as well, in order to ensure a fair comparison. The detailed performance of these machine learning models is shown in Appendix D.

A quantitative comparison of the predictive performance is shown in Fig. 16, where the Root-MSE (RMSE) and $R^2$ values of the four machine-learning methods are summarized. Overall, all four machine-learning models achieve lower RMSE and higher $R^2$ when the extracted key functions are used as inputs, indicating that these inputs provide a more effective representation of the governing relation for the GB normal stress. This confirms that the feature-extraction step prior to model training is effective, and that the identified key candidate functions are physically meaningful and informative for describing the GB normal stress. In other words, the above results suggest that the top 18 candidate functions provide a more efficient representation of the governing information than the original input variables. Their advantage likely arises from the fact that the prescribed functional forms already encode part of the relevant nonlinear dependence among the underlying characteristics. In addition, although the machine-learning models achieve better overall performance in terms of $R^2$, a comparison between Fig. 11b and Fig. D 1b shows that the RF predictions have relatively poor resolution in the low-stress regime. In contrast, the regression model based on the candidate functions performs better in this range. This suggests that the causation-entropy-assisted regression approach may offer greater robustness, especially when local resolution is needed in physically important stress ranges.

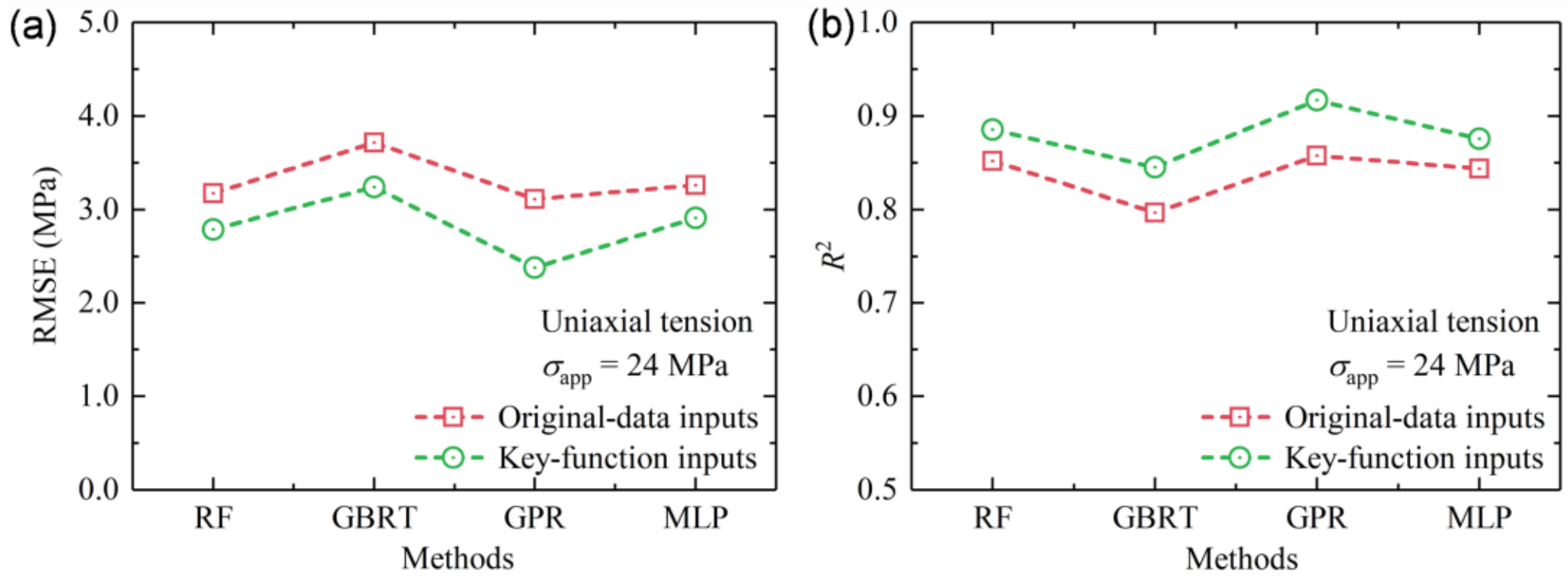

Fig. 16 Comparison of the predictive performance of different models using the original characteristics and the key functions as inputs: (a) root mean square error (RMSE); (b) coefficient of determination ($R^2$).

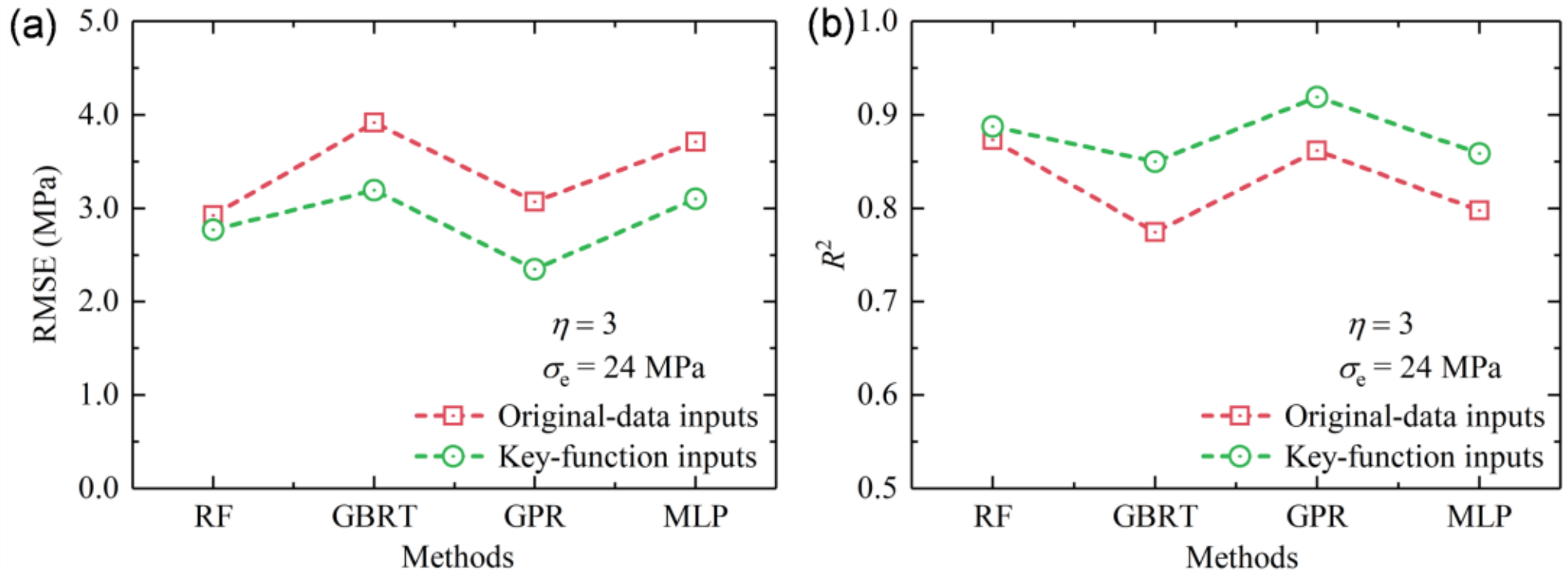

Fig. 17 Comparison of the predictive performance of different models using the original characteristics and the key functions as inputs: (c) root mean square error (RMSE), (d) coefficient of determination ($R^2$).

For the bicrystals under multiaxial loading case, the same machine learning models are trained using either the original inputs or the top 18 candidate functions, and their predictive performance on the test set is compared. As shown in Fig. 17a and Fig.

17b, the models using the candidate functions associated with the key characteristics as inputs achieve overall lower RMSE and higher $R^2$ .This further confirms that the extracted key-function inputs provide a more informative representation than the original inputs under multiaxial loading, and supports the transferability of the causation-entropy-assisted feature-extraction strategy beyond the uniaxial stress state.

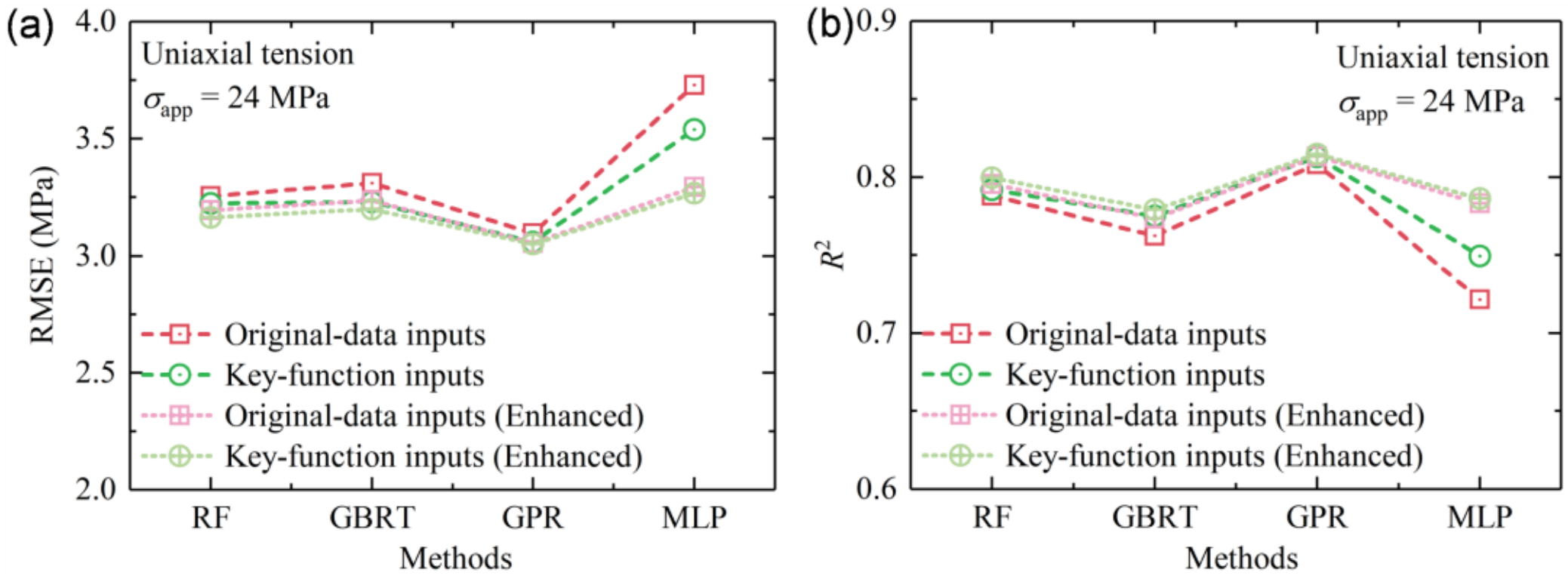

Fig. 18 (a) Comparison of the predictive performance of different models using the original characteristics/key functions as inputs and with/without enhanced features: (a) root mean square error (RMSE); (b) coefficient of determination ($R^2$).

For the tricrystal case, the inclusion of the supplementary characteristics associated with the neighboring GBs increases both the original input dimension and the number of key candidate functions used in the subsequent machine-learning models to 36. The machine-learning models that incorporate the supplementary characteristics of the neighboring GBs are referred to as the enhanced models. Four input sets are considered: the original 18 microscale characteristics, the top 18 candidate functions, the enhanced 36 microscale characteristics, and the top 36 enhanced candidate functions. The training-set performance of the models trained using these four input strategies is shown in Fig. 18a and Fig. 18b. Overall, the models using the enhanced key-function inputs achieve the highest $R^2$ , followed by those using the enhanced original-data inputs, the key-function inputs, and finally the original-data inputs.

Across the four model classes, replacing the original characteristics with the CE-selected candidate functions reduced the test RMSE and increased $R^2$ . The improvement was observed for RF, GBRT, GPR, and MLP, indicating that the result is not specific to one regression architecture. Therefore, the causation-entropy-selected functions consistently improve predictive efficiency across distinct model classes. It indicates that the selected candidate functions provide higher information density in quantifying GB normal stress.

From the perspective of predictive accuracy, the RF, GBRT, GPR, and MLP models generally achieved slightly higher test-set coefficients of determination than the compact CE-based interpretable regression, even when the original physical characteristics were used as inputs. This is expected because these models offer greater flexibility in representing high-dimensional nonlinear relationships. Nevertheless, this result does not imply that the CE analysis is unnecessary. The CE-selected candidate functions provide more informative inputs and consistently improve the performance of all four nonlinear surrogate-model classes. In addition, the CE-based compact regression is designed to satisfy objectives beyond predictive accuracy. Despite its slightly lower predictive performance, it retains an explicit and physically interpretable functional form, shows transferability across the physical conditions examined in this work, and permits systematic incorporation of additional physically meaningful information when a purely local representation becomes insufficient. These properties are particularly desirable for reduced-order mechanical models, for which physical transparency, transferability, and systematic extensibility are important alongside predictive accuracy.

# 5. Discussions

## 5.1. Causal dynamics and information structure shifts in governing mechanisms under variable conditions

In this section, the CPFE data for the multiaxial-loading and tricrystal cases are used to carry out the causation-entropy analysis and the ranking of candidate functions. For the bicrystal system under multiaxial loading, Fig. 19a presents the causation-entropy ranking of the candidate functions constructed from the relevant microscale characteristics. Overall, the ranking of the candidate functions associated with the physical characteristics under multiaxial loading is broadly consistent with that obtained under uniaxial loading. Hence, despite the change in macroscopic loading condition, the dominant physical mechanisms and the governing characteristics underlying the GB normal stress are largely preserved. The multiaxial results further support the transferability of the important microscale characteristics and the regression functional form identified from the uniaxial case. For a more detailed comparison, the top 20 candidate functions ranked by causation entropy under multiaxial loading (see Fig. 19b) are compared with those in the uniaxial case (see Fig. 19c). Under multiaxial loading, the candidate functions associated with crystallographic mean and mismatch characteristics move upward in the ranking. By contrast, those associated with mechanical mean characteristics shift to lower ranks, whereas those associated with micro-geometric characteristics also tend to move upward. With increasing stress triaxiality, particularly under a highly triaxial stress state ($\eta = 3$), the distinction among the three principal loading directions is reduced. Consequently, the GB normal stress becomes more strongly controlled by local microscale characteristics, so that candidate functions associated with local crystallographic and micro-geometric characteristics become more informative, whereas those associated with load-dependent mechanical characteristics become less important.

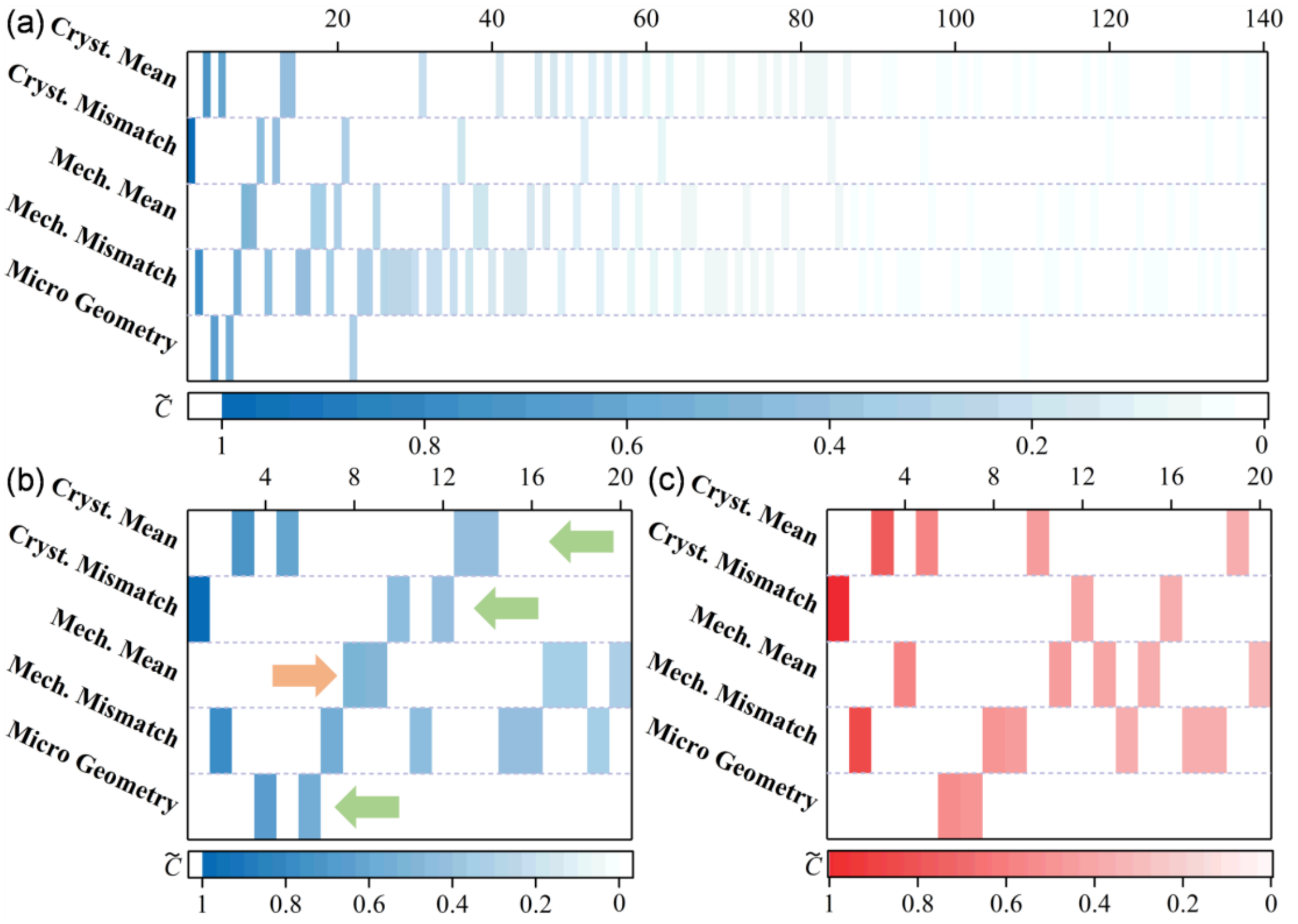

Fig. 19 (a) The heat map of top 140 candidate feature functions for bicrystal subjected to multiaxial tension ( $\eta = 3$ ) creep grouped into crystallographic mean features, crystallographic mismatch features, mechanical mean features, mechanical mismatch features, and micro-geometric features; (b) The heat map of top 20 candidate feature functions for multiaxial tension ( $\eta = 3$ ) creep; (c) The heat map of top 20 candidate feature functions for uniaxial tension creep.

Fig. 20 further presents the ranking among the top 140 candidate functions of the averaged candidate functions associated with the nonlocal characteristics characterizing intergranular interaction for the tricrystal case. Candidate functions associated with nonlocal crystallographic mean and micromechanical mean characteristics are ranked more prominently, indicating that these characteristics carry more information on the GB normal stress. This contrasts with the local GB characteristics, for which the dominant candidate functions are primarily associated with mismatch-related quantities. This is because the nonlocal candidate functions associated with mean crystallographic and micromechanical characteristics incorporate more information from the third grain in addition to the two grains adjacent to the GB, and this additional grain exerts a strong influence on the GB normal stress.

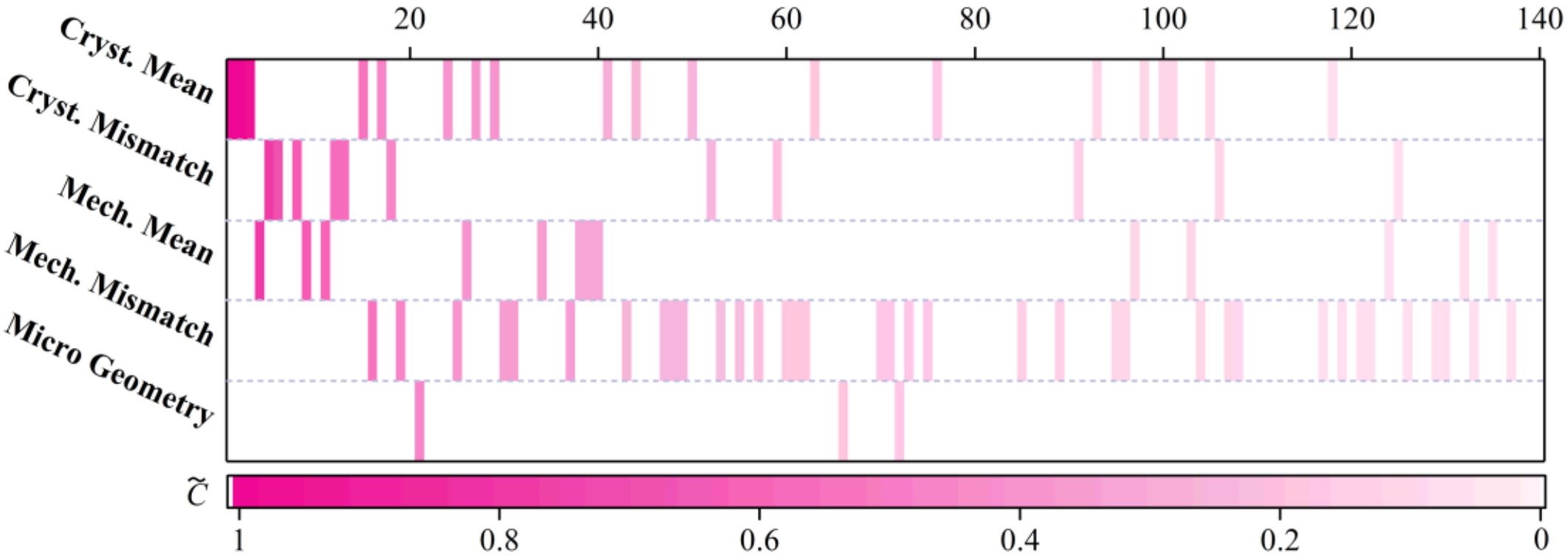

Fig. 20 The heat map of the grain-grain interaction term in top 140 candidate feature functions for tricrystal subjected to uniaxial tension creep grouped into crystallographic mean features, crystallographic mismatch features, mechanical mean features, mechanical mismatch features, and micro-geometric features.

Therefore, the key physical characteristics and the regression functional form extracted by causation entropy from the uniaxial bicrystal case remain reasonably transferable to the multiaxial and tricrystal cases, although the importance ranking of the physical characteristics and candidate functions still shows distinct, mechanism-related shifts. The ranking shifts across loading/configuration changes are mechanism-consistent rather than arbitrary, which supports the physical meaningfulness of the CE-based attribution. These findings provide important insight into the mechanisms governing the microscale GB normal stress and the construction of surrogate models.

The interpretability of the proposed machine-learning methodology originates from the physical construction of the model rather than from post hoc explanation of a black-box predictor. First, the input space is composed of descriptors with explicit physical meanings, including GB geometry, crystallographic compatibility, creep-related deformation tendency, and elastic contrast. Second, causation entropy is used to evaluate the conditional information contribution of each descriptor and candidate function to GB normal stress, allowing physically important variables to be distinguished from redundant correlations. Third, the selected candidate functions are assembled into a compact and explicit regression model, so that both the relative importance of the physical variables and their effects on GB normal stress can be directly inspected. In this sense, the proposed model differs from conventional machine-learning surrogates: it is not only predictive, but also physically traceable.

This interpretability also underpins the generalizability observed in this work. Because the reduced-order representation is built upon physically meaningful descriptors and explicitly selected functional forms, rather than purely data-driven correlations, the identified characteristic hierarchy and functional representation can be transferred across different physical conditions. This explains why the representation extracted from uniaxial bicrystal creep remains effective under multiaxial loading and can be further extended to tricrystal configurations by introducing physically interpretable nonlocal descriptors.

## 5.2. Response dependence of the identified governing mechanisms

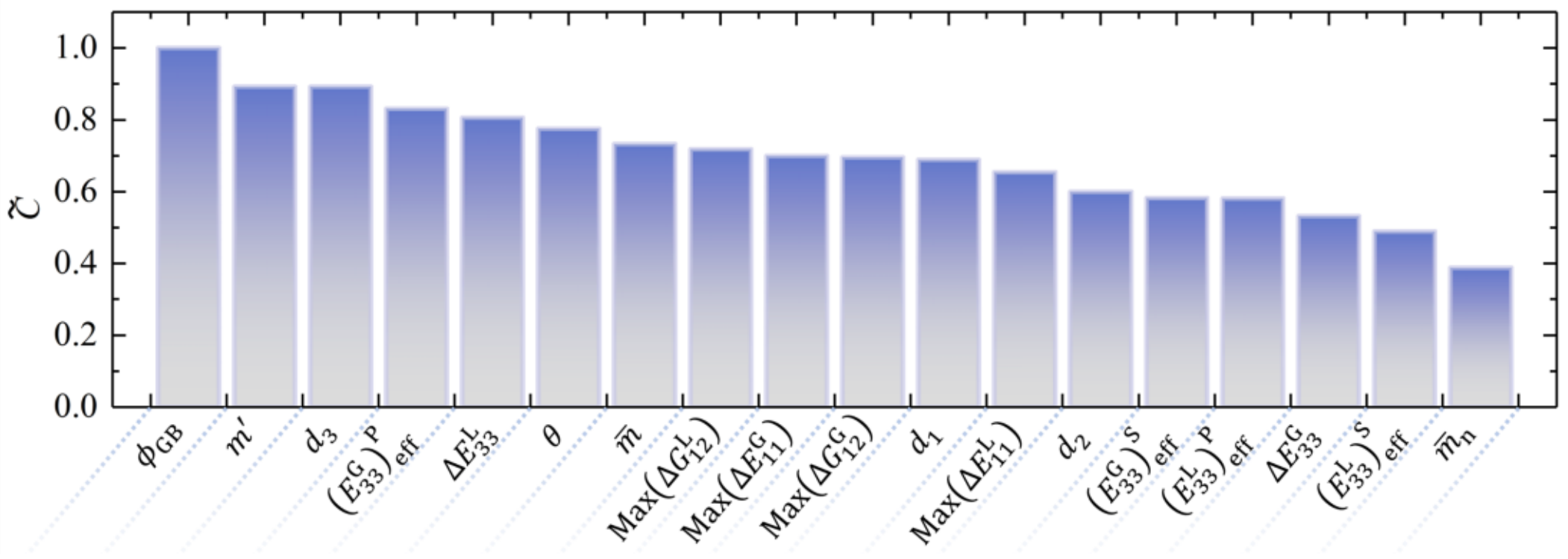

Fig. 21 CE ranking of microscale descriptors for the GB-adjacent equivalent plastic strain.

To examine whether the identified characteristic hierarchy depends on the selected physical response, the causation-entropy analysis was repeated using the equivalent plastic strain adjacent to the GB as the target variable. As shown in Fig. 21, $\phi_{GB}$ remains the dominant descriptor, while $m'$, $d_3$, and several effective-modulus and elastic-mismatch descriptors also carry substantial independent information. However, the relative hierarchy differs from that obtained for GB normal stress. Most notably, the climb-related indicator $\bar{m}_n$ becomes weakly informative for equivalent plastic strain, whereas the mean-orientation and effective-stiffness descriptors gain relative importance.

The persistence of $\phi_{GB}$ and $m'$ indicates that interfacial geometry and cross-boundary crystallographic compatibility affect both local traction and accumulated deformation. Nevertheless, the differences in the remaining hierarchy demonstrate that stress- and deformation-based responses emphasize different aspects of the local micromechanical state. Therefore, changing the target response requires renewed CE attribution and reduced-order identification; the feature ranking obtained for GB normal stress should not be regarded as a universal ranking for creep damage.

In the present study, the GB normal stress was selected as the target response because it represents a direct mechanical indicator of grain-boundary opening tendency during creep cavitation. Therefore, the identified mechanisms primarily describe factors controlling cavity nucleation rather than the complete evolution of creep failure. Creep cavity evolution involves multiple coupled processes, including nucleation, growth, and coalescence. For cavity growth, constrained growth models (Ashby and Dyson, 1984; Cocks and Ashby, 1982; Dyson, 1983) emphasize that the growth rate is not only determined by GB tensile stress but also by the creep deformation accommodation capability of surrounding grains. Therefore, deformation-rate quantities, such as local equivalent creep strain rate, provide an important response for investigating growth-related mechanisms.

To demonstrate the response-dependent capability of the proposed framework, as shown in Fig. 22a, the same CE analysis was additionally applied to the averaged equivalent creep strain rate of GB-adjacent regions in the uniaxial case. The causal hierarchy shows increased contributions from crystallographic compatibility descriptors compared with the GB stress response, which is consistent with the role of strain rate in characterizing local deformation accommodation. For the equivalent creep strain-rate response, the contribution of elastic-modulus-related descriptors is significantly reduced compared with the GB stress response. This indicates that the local creep deformation kinetics are not primarily governed by elastic anisotropy, but rather by crystallographic characteristics associated with plastic deformation and descriptors characterizing intergranular deformation incompatibility. Such a feature hierarchy is physically reasonable because the strain-rate response reflects the instantaneous plastic accommodation capability of neighboring grains, which is controlled by slip activity, crystallographic compatibility, and local deformation constraints. Using the same reduced-order modelling strategy, the predicted equivalent creep strain rate shows good agreement with the original simulation response, with a substantially reduced residual dispersion compared with the raw response variability (see Fig. 22b). As shown in Fig. 22c, the resulting regression model achieves satisfactory predictive performance on randomly selected training and testing datasets, with coefficients of determination of $R^2$=0.846 and $R^2$=0.874, respectively. These results demonstrate that the CE -identified descriptors can effectively represent the dominant factors governing the local creep deformation kinetics and provide an interpretable reduced-order representation of the strain-rate response.

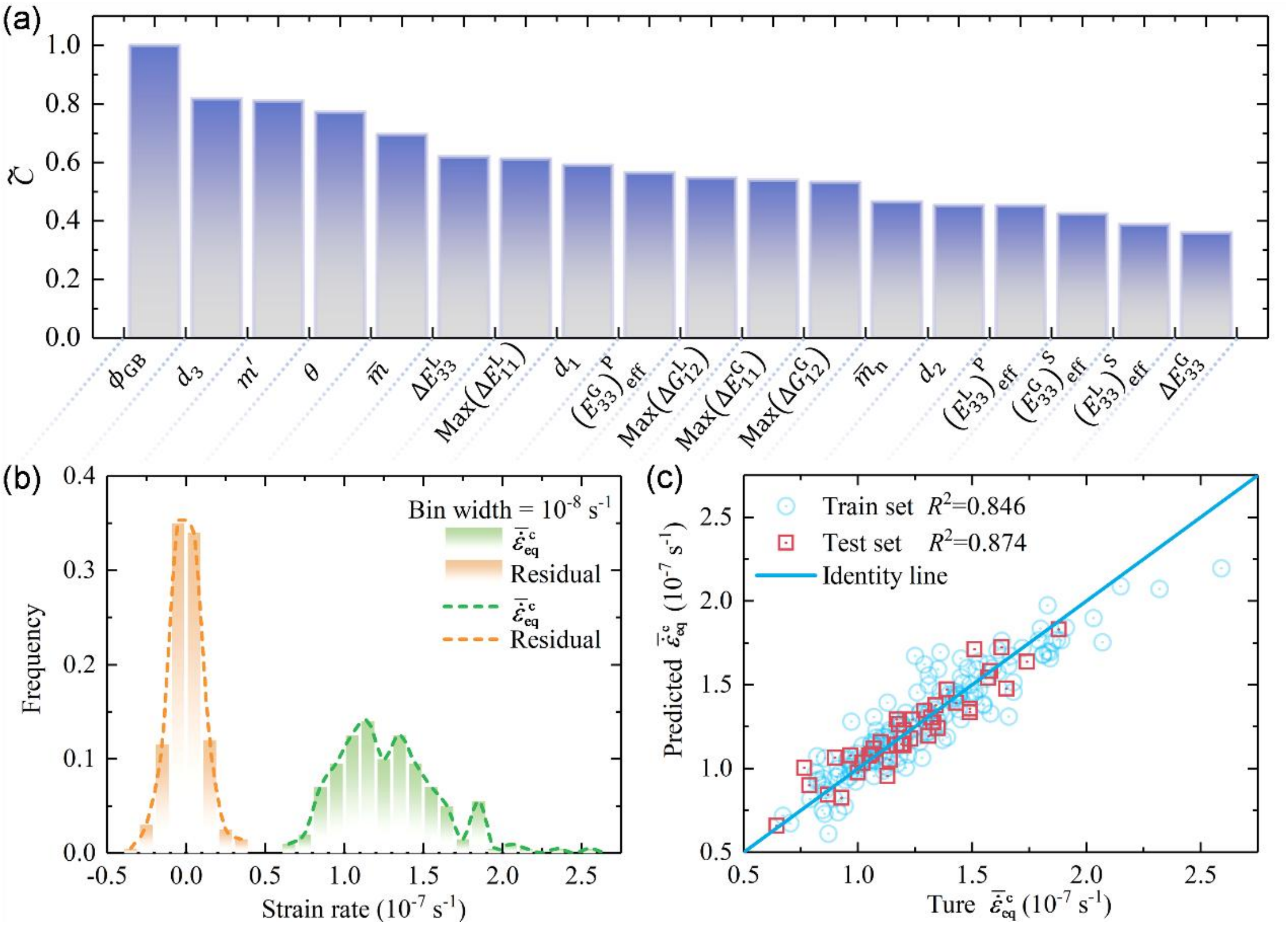

Fig. 22 CE analysis and regression of equivalent creep strain rate (a) CE ranking (b) probability density functions of the original average equivalent creep strain rate and the residual; (b) comparison between the CE-regression model predictions and the corresponding reference values.

Although the present study uses GB normal stress as the target response, the CE formulation is response-dependent and can be applied to other physically meaningful quantities by replacing the target variable. A more comprehensive investigation of creep damage dominated by cavity growth requires an explicit damage-evolution dataset. Hence, the simulation will provide physically resolved cavity growth, which can be further integrated with the proposed CE framework to identify governing mechanisms and establish interpretable reduced-order models.

## 5.3. Statistical characterization of GB normal stress distributions

The regression functional form of Eqn. (48), selected through the causation-entropy ranking of the candidate functions, constitutes a physically informed reduced-order model for the GB normal stress. With the linear regression model in Eqn. (48), the computational cost of predicting $\bar{\sigma}_{33}^{\mathrm{L}}$ from the candidate functions of the key characteristics is extremely low. Therefore, by sampling the values of the candidate functions constructed from a large set of initial characteristics, Monte Carlo calculations can be performed to obtain the variation in the distribution of $\bar{\sigma}_{33}^{\mathrm{L}}$ with the key variables. The GB normal stress results for the uniaxial cases, including the bicrystal and tricrystal systems, are shown in Fig. 23. The blue crosses denote the results obtained from 6000 samples using the causation-entropy (CE)-based regression model. Based on the resulting data cloud, the 95% and 80% confidence intervals (C. I.) are constructed. The scatter points obtained from the CPFE calculations are also shown in Fig. 23. The effects of the four most important physical characteristics ranked at the top in Fig. 8 are examined: GB orientation $\phi_{\mathrm{GB}}$, the slip-transmission factor $m'$, the mean Schmid-type indicator $\bar{m}_{\mathrm{n}}$ for climb-driven creep, and the maximum in-plane Young's modulus mismatch $\max\left(\Delta E_{11}^{\mathrm{G}}\right)$ evaluated in the global sample frame. Overall, the CPFE results fall largely within the spread predicted by the CE-based reduced-order model. The overall decrease in the GB normal stress with increasing misorientation angle is also captured. As $m'$ increases, dislocation transmission across the GB becomes easier, which tends to reduce stress concentration at the GB and thus lowers the GB normal stress. $\bar{m}_{\mathrm{n}}$ characterizes the driving force for climb-driven creep deformation near the GB. A larger $\bar{m}_{\mathrm{n}}$ facilitates creep deformation in the vicinity of the GB, promotes local stress relaxation, and thus reduces the GB normal stress. As a measure of the micromechanical mismatch near the GB, a larger $\max\left(\Delta E_{11}^{\mathrm{G}}\right)$ leads to more pronounced local stress concentration and, consequently, to an increase in the GB normal stress. The CE-based reduced-order model effectively captures the key physical characteristics identified as most influential for the GB normal stress, together with their effects. Fig. 23 also shows that, although the nonlocal characteristics in the tricrystal case exert an additional influence on the GB normal stress, the CE-based model retains strong robustness and transferability in capturing the distribution of the effects of the key characteristics, consistent with the model distilled from the bicrystal case. This highlights an important advantage of the CE-based approach in physical attribution and interpretable model construction, and further supports its ability to extract the key local microscale characteristics.

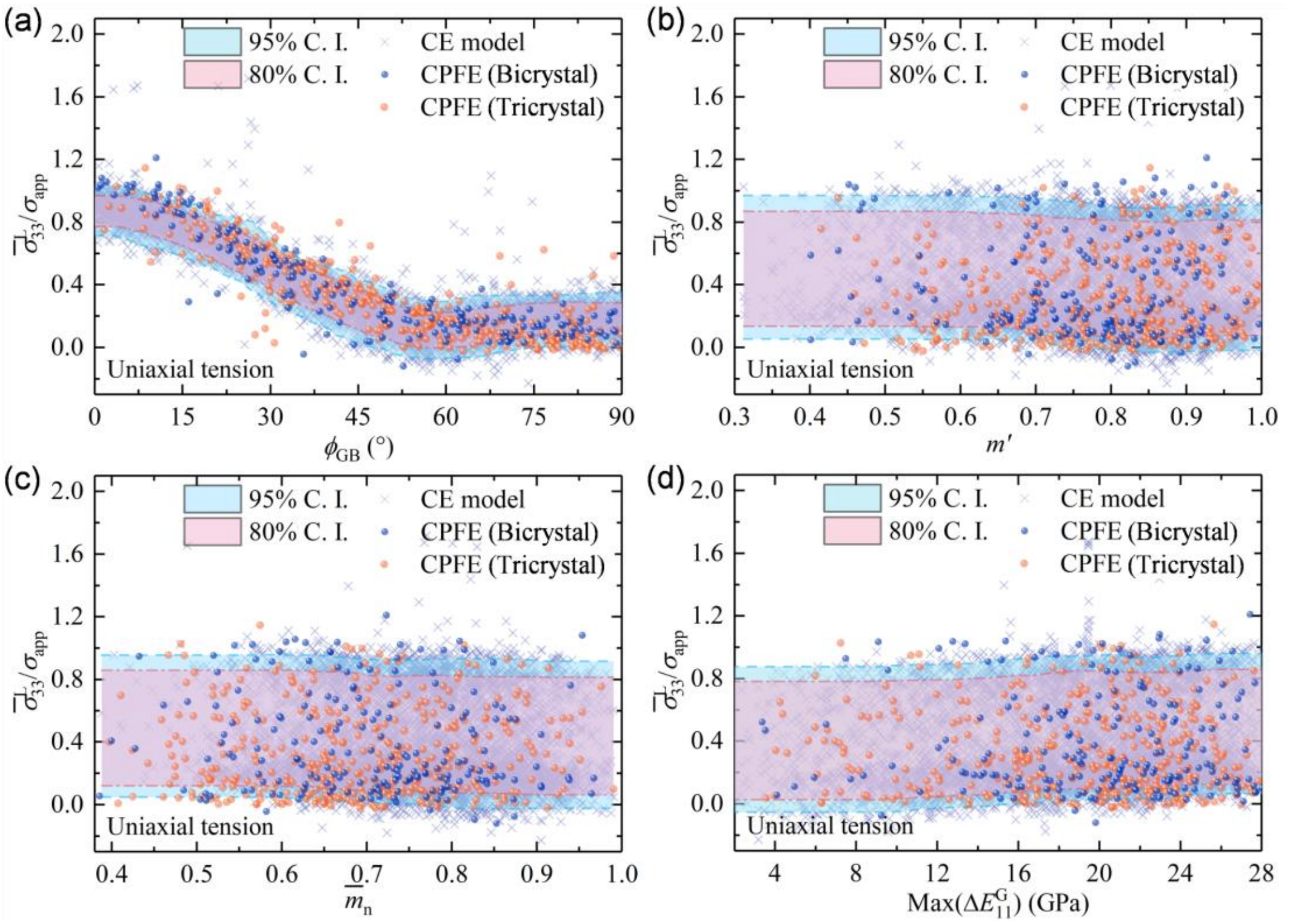

Fig. 23 The variation and fluctuation of grain-boundary normal stress under uniaxial tension with respect to: (a) GB misorientation angle, (b) slip transmission factor of the GB, (c) average Schmid-type factor corresponding to the slip plane normal stress component, (d) global in-plane maximum mismatch elastic modulus.

The effects of the key microscale characteristics on the GB normal stress under multiaxial loading are then examined in the same manner. The results are shown in Fig. 24. Compared with the uniaxial case, the multiaxial stress state alters the quantitative values, but the correlation between $\bar{\sigma}_{33}^{L}$ and the key microscale characteristics remains essentially the same. This indicates that the mechanisms by which the key microscale physical characteristics influence the GB normal stress remain highly consistent between the uniaxial and multiaxial cases. In other words, the influence of the key local microscale characteristics on the local GB normal stress depends only weakly on the macroscopic applied stress state. Such causal mechanics are effectively identified and extracted, and their effects are quantitatively characterized, as also demonstrated in Section 4.2.

Using the same procedure, the effects of additional key physical characteristics on the GB normal stress, as well as the associated features of its distribution, can be obtained with substantial convenience, which is particularly advantageous for stress statistical analysis and uncertainty quantification.

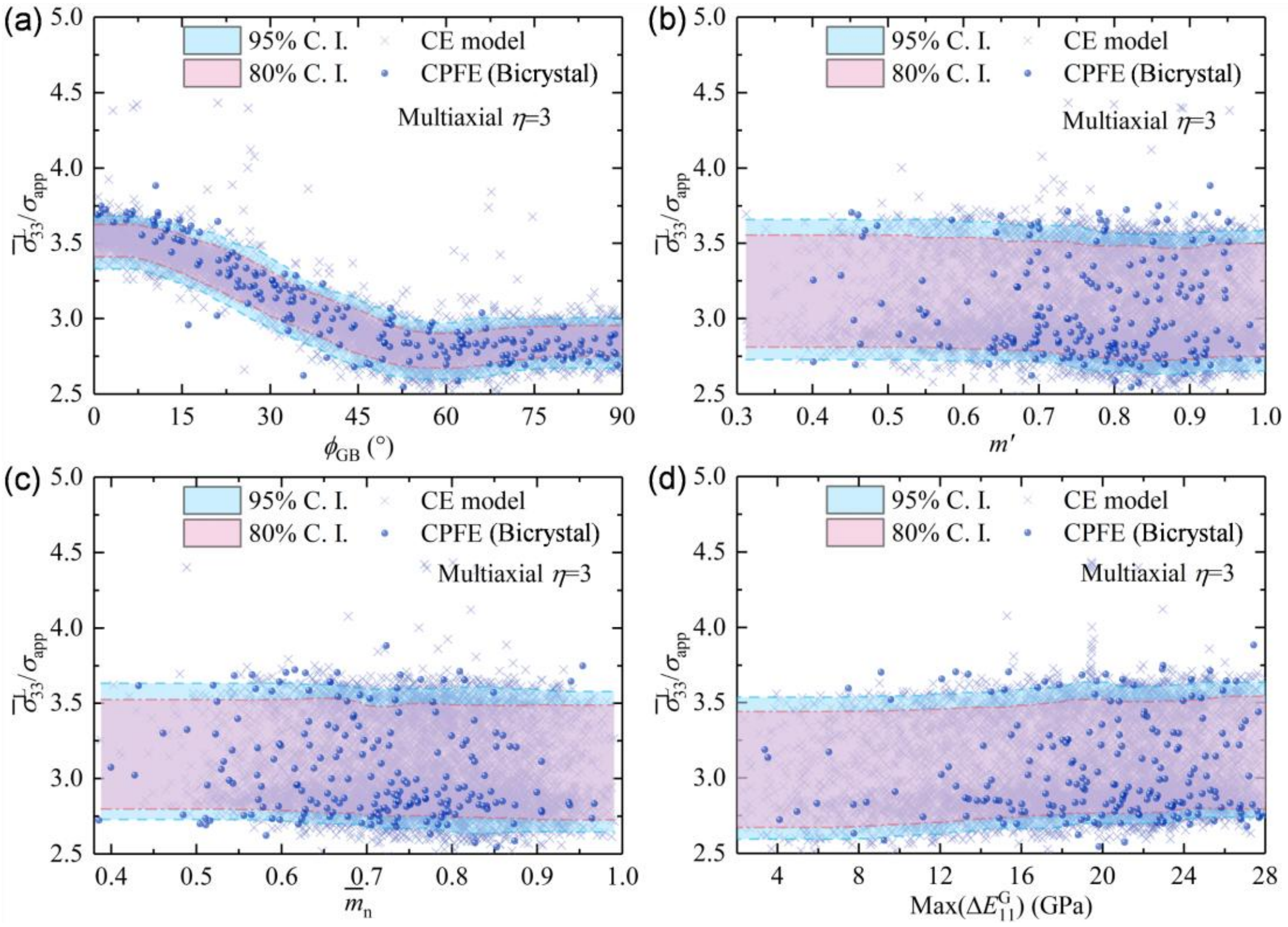

Fig. 24 The variation and fluctuation of grain-boundary normal stress under Multiaxial tension with respect to: (a) GB misorientation angle, (b) slip transmission factor of the GB, (c) average Schmid-type factor corresponding to the slip plane normal stress component, (d) global in-plane maximum mismatch elastic modulus.

## 5.4. Experimental consistency and scope of the identified mechanisms

Recent correlative-microscopy and material-specific creep studies provide experimental support for the mechanisms identified in the present work. Although these experiments generally characterize cavity occurrence or local deformation rather than the quasi-steady GB stress considered here, several consistent physical trends can be identified.

A particularly clear correspondence concerns the orientation of GBs relative to the loading direction. Correlative observations in 316H stainless steel and creep experiments on IN617 show that cavities preferentially occur on boundaries favorably oriented for tensile opening (He et al., 2024; Wang et al., 2022). This is consistent with the dominant role of GB inclination identified in this work, since the inclination directly governs the normal and tangential traction components acting on the boundary plane.

Experimental studies also emphasize the importance of localized deformation and deformation heterogeneity. High-resolution microscopy has associated cavitation with local orientation gradients and strain localization, while combined experimental-CPFE analysis identified plastic deviatoric strain and loading-direction normal stress as important variables associated with damage initiation (Ashraf et al., 2026; Galliopoulou et al., 2025; He et al., 2024). These findings support the present treatment of stress- and deformation-based quantities as complementary target responses, and are consistent with the response-dependent feature hierarchy obtained from the additional equivalent-plastic-strain analysis.

In contrast, conventional single-grain Schmid factors have shown limited or statistically insignificant correlations with cavity formation (Galliopoulou et al., 2025; He et al., 2024). This agrees with the present result that the mean Schmid factor is not dominant, whereas the GB compatibility factor $m'$ carries greater independent information. This correspondence suggests that deformation compatibility across a GB is more informative than the slip tendency of an isolated grain. The importance of $m'$ identified by CE is also consistent with experimental observations. Bieler et al. (2019) demonstrated that the slip transfer behavior across GBs is strongly influenced by crystallographic compatibility between neighboring grains, supporting the physical relevance of $m'$ as a descriptor of grain-boundary deformation compatibility.

The importance of neighborhood constraints is also supported experimentally. Creep cavities in IN617 have been observed along GBs and at triple junctions, particularly where intragranular deformation and intergranular sliding are insufficiently coordinated (Wang et al., 2022). This observation is consistent with the need for nonlocal augmentation when extending the bicrystal representation to tricrystal configurations.

Overall, these experimental observations support the identified roles of GB inclination, localized deformation, cross-boundary compatibility, neighborhood constraints, and climb-mediated creep. Experimental studies also show that precipitate evolution, inclusions, and oxidation may contribute to cavity formation in real alloys (Galliopoulou et al., 2025; Wang et al., 2018; Wang et al., 2022). It should be noted that the present CPFE framework does not explicitly incorporate $m'$ -dependent slip-transfer resistance into the constitutive formulation; introducing such GB-aware mechanisms may provide a higher-order and more non-local description of interfacial deformation and represents an important direction for future development. These additional factors provide natural directions for extending the present descriptor space in future damage-oriented applications.

# 6. Conclusions

A mechanics-guided and information-theoretic machine-learning framework has been established for identifying the governing microscale mechanisms of GB normal stress in creep and for representing them in an interpretable reduced-order form. By combining dislocation-climb-affected CPFE simulations of minimal grain clusters with causation-entropy analysis, the dominant characteristics governing local GB normal stress have been extracted from a high-dimensional, nonlinear, and strongly coupled characteristics space. The main conclusions are summarized as follows.

(1) Methodologically, this work establishes an interpretable causal mechanics-informed framework for identifying the governing structure of GB normal stress from high-dimensional, nonlinear, and strongly coupled microscale creep data. Minimal grain clusters are employed as controlled vehicles for mechanism distillation, and a candidate space of 18 physically motivated characteristics describing local crystallography, interfacial geometry, and micromechanical contrast is constructed within crystal plasticity. Causation entropy is then used to isolate the characteristics and candidate functions with direct governing relevance to GB normal stress, thereby moving beyond linear correlation analysis or purely black-box surrogate fitting. The resulting framework provides a systematic route for causal-mechanistic identification and reduced-order representation of GB stress.

(2) Mechanistically, the dominant characteristics governing GB normal stress in the present FCC superalloy creep setting are identified as the GB inclination angle, the slip-transmission factor, a climb-related Schmid-type indicator, and elastic-modulus mismatch. These results show that GB normal stress is governed by the coupled effects of interfacial geometry, crystallographic compatibility, climb-enabled stress relaxation, and local micromechanical mismatch. Among the candidate functions, mismatch-related quantities are found to be especially informative, indicating that local crystallographic and mechanical incompatibilities play a central role in governing the GB stress state. The characteristics hierarchy identified from bicrystals under uniaxial creep is further found to remain broadly consistent under multiaxial loading, although mechanism-consistent ranking shifts are observed.

(3) At the reduced-order modeling, the characteristics identified by causation entropy are distilled into interpretable candidate functions and a compact regression form for GB normal stress. What is shown to transfer is not merely a fitted predictor, but the identified characteristics hierarchy and functional representation: the same reduced-order form remained effective under multiaxial loading after recalibration of the coefficients. For tricrystal systems, the deterioration of a purely local representation revealed the limit of locality rather than a failure of the framework itself. By incorporating physically interpretable nonlocal characteristics associated with neighboring GBs, the model is systematically extended to account for intergranular interaction effects, leading to a marked recovery in predictive performance. These results indicate that the present framework not only yields an interpretable reduced-order representation of GB stress, but also admits physically transparent extension when additional mechanisms emerge in more complex grain configurations.

(4) In addition, the reduced-order representation enabled efficient statistical characterization and uncertainty-aware analysis of GB normal stress over broad characteristics ranges. The extracted candidate functions also improved surrogate-model performance across multiple machine-learning model classes, indicating that they encode governing information more efficiently than the original characteristics set. This improvement should be viewed as supporting evidence for the physical relevance of the identified characteristics and functions, rather than as the primary contribution of the study.

Overall, the present work establishes a causal-mechanistic and interpretable route for linking microscale crystallography, geometry, and micromechanics to GB normal stress in creep. The framework provides a basis for further study of nonlocal effects, GB damage initiation, and reduced-order constitutive descriptions for creep damage in complex polycrystalline microstructures. Future studies may further extend the present framework by incorporating additional physically meaningful descriptors associated with grain morphology and spatial topology when more comprehensive polycrystalline datasets become available.

## Acknowledgments

The study was supported by the Key Project of the National Natural Science Foundation of China (12332005) and the Postdoctoral Fellowship Program of CPSF under Grant Number GZC20261057. The authors gratefully acknowledge Prof. Nan Chen at the University of Wisconsin–Madison and Prof. Juner Zhu at Northeastern University for helpful discussions and assistance regarding causation entropy and writing.

## Appendix A: The CPFE model and parameters

In Eqn.(11), $S^\alpha$ is the slip resistance to dislocation motion proposed by Zhang and Oskay (2016) for Inconel 617 at high temperature, and its evolution obeys:

$$\dot{S}^\alpha = \left[h_\mathrm{S} - d_\mathrm{D}\left(S^\alpha - \bar{S}^\alpha\right)\right]\left|\dot{\gamma}^\alpha\right| - h_2\left(S^\alpha - S_0^\alpha\right)H\left(\dot{\gamma}_\mathrm{th} - \sum_{\alpha=1}^{n}\left|\dot{\gamma}^\alpha\right|\right) \tag{A.1}$$

in which $h_\mathrm{S}$ and $d_\mathrm{D}$ denote the hardening and dynamic recovery parameters. $\bar{S}^\alpha$ is the steady flow strength parameter of the $\alpha$ -slip system. $h_2$ is the static recovery parameter. $S_0^\alpha$ and $\dot{\gamma}_\mathrm{th}$ represent the initial slip resistance and the threshold rate for static recovery, respectively. $H(\bullet)$ represents the Heaviside function. The back stress $B^\alpha$ is proposed by Lin et al. (2010) and expressed as:

$$\dot{B}^\alpha = \left[h_\mathrm{B} - D^\alpha B^\alpha \operatorname{sgn}\left(\dot{\gamma}^\alpha\right)\right]\dot{\gamma}^\alpha \tag{A.2}$$

where $h_\mathrm{B}$ and $D^\alpha$ are the hardening and dynamic recovery parameters for back stress with $D^\alpha$ expressed as:

$$D^\alpha = \frac{h_\mathrm{B}\mu_0}{S^\alpha}\left(\frac{\mu_0'}{f_\mathrm{co}} - \mu\right)^{-1} \tag{A.3}$$

where $\mu_0'$ is the local slip shear modulus at 0 K. $f_\mathrm{co}$ denote a coupling parameter for the internal slip variables. In Eqn. (12), the evaluation of back stress is given as follows

(Staroselsky and Cassenti, 2011):

$$\dot{B}_{\mathrm{c}}^{\alpha}=h_{\mathrm{c}}\left(\dot{\gamma}_{\mathrm{c}}^{\alpha}B_{\infty}-\left|\dot{\gamma}_{\mathrm{c}}^{\alpha}\right|B_{\mathrm{c}}^{\alpha}\right) \tag{A.4}$$

In which $B_{\infty}$ is a saturation value and $h_{\mathrm{c}}$ is a material parameter. The parameters of CPFE glide and climb models are given in Table A 1.

Table A 1 The parameters of the CPFE model of Inconel 617 at 950°C (Zhang and Oskay (2016))

| Parameter | Value | Parameter | Value |
|---|---|---|---|
| $C_{11}$ (GPa) | 170.64 | $p_g$ | 0.181 |
| $C_{12}$ (GPa) | 108.39 | $q_g$ | 1.633 |
| $C_{44}$ (GPa) | 77.82 | $h_{\mathrm{S}}$ (MPa) | 397.73 |
| $\dot{\gamma}_0$ ($\mathrm{s}^{-1}$) | $2.288\times10^{-3}$ | $d_{\mathrm{D}}$ (MPa) | 5073.62 |
| $F_0$ (J) | $5.148\times10^{-19}$ | $S_0^{\alpha}$ (MPa) | 143.41 |
| $\mu_0$ (GPa) | 265.33 | $\bar{S}^{\alpha}$ (MPa) | 18.03 |
| $\mu$ (GPa) | 77.82 | $h_{\mathrm{B}}$ (MPa) | 104.31 |
| $\mu_0'$ | 31.13 | $h_2$ (MPa) | 0.015 |
| $\hat{\tau}_0$ (MPa) | 268.2 | $\dot{\gamma}_{\mathrm{th}}$ ($\mathrm{s}^{-1}$) | $1.0\times10^{-6}$ |
| $f_{\mathrm{co}}$ | 0.36 | $\psi_{\mathrm{c}}$ (°) | $\pi/4$ |
| $p_{\mathrm{c}}$ | 3 | $\hat{\tau}_{0\mathrm{c}}$ (Pa) | 7750 |
| $h_{\mathrm{c}}$ | 32 | $B_{\infty}$ (MPa) | 4.7 |

# Appendix B: Hyperparameters for machine-learning models

This appendix provides the implementation details of the auxiliary machine-learning comparison reported in Section 4.3. For each model class, the original-characteristic and CE-selected-function inputs were evaluated using identical data partitions, preprocessing procedures, and performance metrics. In this work, five-fold cross-validation is adopted for performance evaluation, and Bayesian optimization is employed to determine the hyperparameters of the random forest, GBRT, GPR, and MLP models for different cases. The resulting hyperparameters are summarized as follows.

Table B 1 Hyperparameters used in the random forest (RF) model under different cases

| Hyperparameter | Optimal Value | | | | | |
|---|---|---|---|---|---|---|
| | Bicrystal Uniaxial Original | Bicrystal Uniaxial Function | Bicrystal Multiaxial Original | Bicrystal Multiaxial Function | Tricrystal Uniaxial Original | Tricrystal Uniaxial Function |
| Number of decision trees in the ensemble | 1200 | 1200 | 1200 | 1200 | 1200 | 800 |
| Minimum number of samples required to perform a node split | 10 | 10 | 10 | 10 | 10 | 2 |
| Minimum number of samples required at a leaf node | 1 | 1 | 1 | 1 | 4 | 2 |
| Fraction of features considered when searching for the best split | 0.7 | 0.7 | 0.7 | 0.7 | 0.7 | 0.7 |
| Maximum depth of each decision tree | 6 | 6 | 6 | 6 | 10 | 6 |
| Indicates whether bootstrap sampling is applied | True | True | True | True | True | True |

Table B 2 Hyperparameters used in the GBRT model under different cases

| Hyperparameter | Optimal Value | | | | | |
|---|---|---|---|---|---|---|
| | Bicrystal Uniaxial Original | Bicrystal Uniaxial Function | Bicrystal Multiaxial Original | Bicrystal Multiaxial Function | Tricrystal Uniaxial Original | Tricrystal Uniaxial Function |
| Proportion of training samples used for fitting each base learner | 0.8 | 0.6 | 0.8 | 0.6 | 0.8 | 0.6 |
| Total number of boosting iterations | 200 | 300 | 800 | 300 | 200 | 200 |
| Minimum number of samples required to perform a split | 5 | 2 | 10 | 2 | 5 | 5 |
| Minimum number of samples required at a leaf node | 1 | 2 | 2 | 2 | 1 | 2 |
| Number of features considered when searching for the best split | None | None | Sqrt | None | None | None |
| Maximum depth of the | 2 | 3 | 3 | 3 | 2 | 5 |
| individual regression trees | | | | | | |
| Learning rate controlling the contribution of each tree | 0.1 | 0.05 | 0.2 | 0.05 | 0.1 | 0.2 |

Table B 3 Hyperparameters used in the GPR model under different cases

| Hyperparameter | Optimal Value | | | | | |
|---|---|---|---|---|---|---|
| | Bicrystal Uniaxial Original | Bicrystal Uniaxial Function | Bicrystal Multiaxial Original | Bicrystal Multiaxial Function | Tricrystal Uniaxial Original | Tricrystal Uniaxial Function |
| Composite kernel capturing nonlinear and linear structure | Matern(ν =2.5) & Linear & Noise | Matern (ν = 1.5) & WhiteKernel | Matern (ν = 2.5) & DotProduct & WhiteKernel | Matern (ν = 2.5) & WhiteKernel | Matern (ν = 1.5) & WhiteKernel | Matern (ν = 2.5) & WhiteKernel |
| Controls overall function amplitude | ~1000 | 2.37 | ~1000 | 2.34 | 2.34 | 1.25 |
| Determines relevance of each feature | Feature-dependent | Feature-dependent | Feature-dependent | Feature-dependent | Feature-dependent | Feature-dependent |
| Observation noise variance | 1 | 0.229 | 1 | 0.233 | 0.185 | 0.185 |
| Number of optimizer restarts | 8 | 10 | 5 | 10 | 10 | 8 |
| Target normalization applied | False | True | False | True | True | True |

Table B 4 Hyperparameters used in the MLP model under different cases

| Hyperparameter | Optimal Value | | | | | |
|---|---|---|---|---|---|---|
| | Bicrystal Uniaxial Original | Bicrystal Uniaxial Function | Bicrystal Multiaxial Original | Bicrystal Multiaxial Function | Tricrystal Uniaxial Original | Tricrystal Uniaxial Function |
| Number of neurons in each hidden layer | (100) | (100) | (100) | (100) | (50, 50) | (50, 50) |
| Activation function of hidden layers | ReLU | ReLU | ReLU | ReLU | ReLU | ReLU |
| L2 regularization strength | $5.28\times10^{-4}$ | $5.28\times10^{-4}$ | $5.28\times10^{-4}$ | $5.28\times10^{-4}$ | $4.38\times10^{-5}$ | $4.38\times10^{-5}$ |
| Number of samples per training batch | 16 | 16 | 16 | 16 | 64 | 64 |
| Learning rate adjustment strategy | adaptive | adaptive | adaptive | adaptive | adaptive | adaptive |
| Initial learning rate | 0.0905 | 0.0905 | 0.0905 | 0.0905 | 0.0674 | 0.0674 |

# Appendix C: Pearson correlation coefficients among characteristics and GB normal stress

For an intuitive understanding of the dataset, the Pearson correlation coefficients (PCC) among these variables are first computed, and the resulting correlation heat map is presented in Fig. C 1. The PCCs are given by:

$$r = \frac{\mathrm{Cov}(X,Y)}{\sigma_X \sigma_Y} \tag{C.1}$$

in which $\mathrm{Cov}(X,Y)$ denotes the covariance, $\sigma_X$ and $\sigma_Y$ denote the standard deviations of $X$ and $Y$, respectively. It should be noted that the PCC characterizes only linear correlation and may fail to capture nonlinear dependence. It therefore provides only a partial view of the overall dependence between variables.

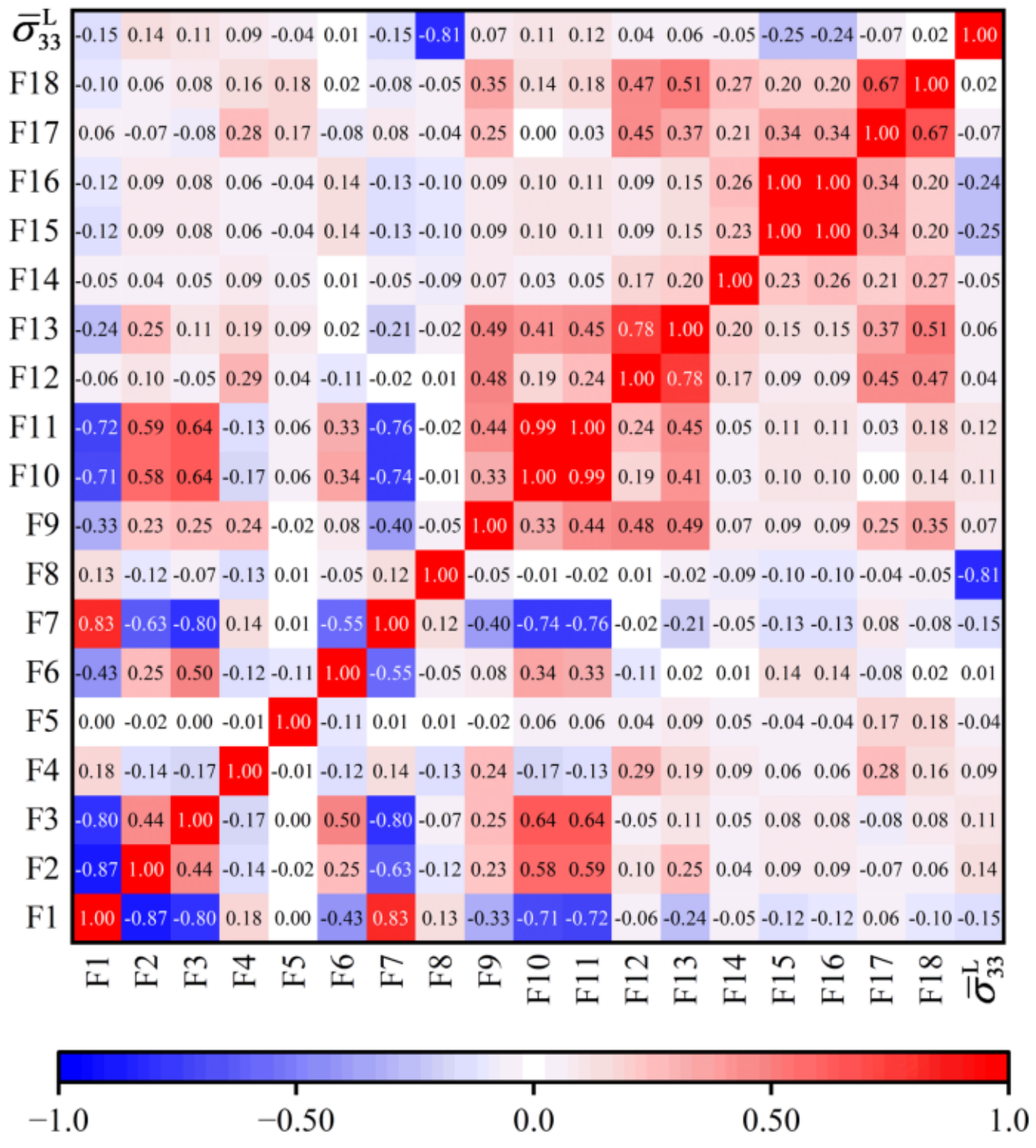

| ID | Feature | ID | Feature |
|---|---|---|---|
| $F_1$ | $d_1$ | $F_{10}$ | $(E^{\mathrm{G}}_{33})^{\mathrm{S}}_{\mathrm{eff}}$ |
| $F_2$ | $d_2$ | $F_{11}$ | $(E^{\mathrm{G}}_{33})^{\mathrm{P}}_{\mathrm{eff}}$ |
| $F_3$ | $d_3$ | $F_{12}$ | $\mathrm{Max}(\Delta E^{\mathrm{G}}_{11})$ |
| $F_4$ | $\theta$ | $F_{13}$ | $\mathrm{Max}(\Delta G^{\mathrm{G}}_{12})$ |
| $F_5$ | $m'$ | $F_{14}$ | $\Delta E^{\mathrm{L}}_{33}$ |
| $F_6$ | $\bar{m}$ | $F_{15}$ | $(E^{\mathrm{L}}_{33})^{\mathrm{S}}_{\mathrm{eff}}$ |
| $F_7$ | $\bar{m}_{\mathrm{n}}$ | $F_{16}$ | $(E^{\mathrm{L}}_{33})^{\mathrm{P}}_{\mathrm{eff}}$ |
| $F_8$ | $\phi_{\mathrm{GB}}$ | $F_{17}$ | $\mathrm{Max}(\Delta E^{\mathrm{L}}_{11})$ |
| $F_9$ | $\Delta E^{\mathrm{G}}_{33}$ | $F_{18}$ | $\mathrm{Max}(\Delta G^{\mathrm{L}}_{12})$ |

Fig. C 1 PCC correlation heatmaps for microscale features and $\bar{\sigma}^{\mathrm{L}}_{33}$ in bicrystal cases

As shown in Fig. C 1, $\bar{\sigma}_{33}^{\mathrm{L}}$ exhibits a strong negative linear correlation with the GB inclination angle $\phi_{\mathrm{GB}}$. In other words, statistically, the GB normal stress increases as it becomes more perpendicular to the direction of the maximum principal stress. This trend is consistent with the general understanding of internal forces in solid mechanics. $\bar{\sigma}_{33}^{\mathrm{L}}$ also exhibits a negative linear correlation with the local effective mean elastic modulus $\left(E_{33}^{\mathrm{L}}\right)_{\mathrm{eff}}^{\mathrm{S}}$ and $\left(E_{33}^{\mathrm{L}}\right)_{\mathrm{eff}}^{\mathrm{P}}$ of the two grains adjacent to the GB, evaluated in the GB local frame for both series and parallel configurations. This behavior can be interpreted in terms of the local effective stiffness of the bicrystal system. A lower effective elastic modulus implies a more compliant mechanical response under the applied load, which promotes larger local deformation and results in higher normal stresses on the GB. Moreover, the PCC analysis also reveals a negative correlation between $\bar{\sigma}_{33}^{\mathrm{L}}$ and the mean Schmid-type indicator $\bar{m}_{\mathrm{n}}$ for climb-driven creep of the two grains adjacent to the GB. This indicates that when the adjacent grains are more favorably oriented for climb-driven creep, stress relaxation becomes more effective, thereby reducing the stress accumulated on the GB. A negative correlation is also observed with the Euclidean distance $d_1$ in the IPF from the [001] reference direction. It should be noted that the influence of $d_1$ on the GB normal stress arises through multiple mechanisms, including its effects on the plastic deformability (Wei et al., 2021) of the adjacent grains and on the effective elastic modulus. Therefore, this effect arises from a combination of competing mechanisms and is not straightforward to interpret. Furthermore, as a measure of linear correlation, the PCC is insufficient to characterize more complex nonlinear mechanisms and can therefore only serve as a preliminary qualitative indication.

# Appendix D: Representative machine-learning models used in prediction of GB normal stress

The top 18 candidate functions are listed in Table D 1. These 18 candidate functions involve only nine of the features listed in Table 1, while still providing broad coverage of crystallographic, microstructural geometric, and local micromechanical characteristics.

Table D 1 Causation entropy ranking associated with GB normal stress under uniaxial tensile creep conditions

| Candidate Functions | $C$ (bits) | $\tilde{C}$ | Candidate Functions | $C$ (bits) | $\tilde{C}$ |
|---|---|---|---|---|---|
| $\log(1+\theta)$ | 0.0767 | 1.000 | $\tanh(d_2)$ | 0.0348 | 0.453 |
| $1/\mathrm{Max}(\Delta E_{11}^{\mathrm{G}})$ | 0.0662 | 0.863 | $\exp\left(\left(E_{33}^{\mathrm{G}}\right)_{\mathrm{eff}}^{\mathrm{S}}\right)$ | 0.0338 | 0.440 |
| $d_2/(1+d_2{}^2)$ | 0.0604 | 0.788 | $\theta$ | 0.0330 | 0.431 |
| $1/\left(E_{33}^{\mathrm{G}}\right)_{\mathrm{eff}}^{\mathrm{S}}$ | 0.0447 | 0.583 | $\log\left(1+\left(E_{33}^{\mathrm{L}}\right)_{\mathrm{eff}}^{\mathrm{P}}\right)$ | 0.0322 | 0.420 |
| $1/\bar{m}_{\mathrm{n}}$ | 0.0435 | 0.567 | $\left(\mathrm{Max}(\Delta E_{11}^{\mathrm{L}})\right)^2$ | 0.0296 | 0.386 |
| $\log(1+\phi_{\mathrm{GB}})$ | 0.0404 | 0.527 | $1/\left(E_{33}^{\mathrm{L}}\right)_{\mathrm{eff}}^{\mathrm{P}}$ | 0.0291 | 0.379 |
| $\phi_{\mathrm{GB}}/(1+\phi_{\mathrm{GB}}{}^2)$ | 0.0385 | 0.502 | $\tanh(\theta)$ | 0.0291 | 0.379 |
| $\exp\left(\mathrm{Max}(\Delta E_{11}^{\mathrm{L}})\right)$ | 0.0366 | 0.477 | $1/\left[1+\left(\mathrm{Max}(\Delta G_{12}^{\mathrm{L}})\right)^2\right]$ | 0.0289 | 0.376 |
| $1/\mathrm{Max}(\Delta G_{12}^{\mathrm{L}})$ | 0.0352 | 0.459 | $\mathrm{Max}(\Delta E_{11}^{\mathrm{L}})$ | 0.0288 | 0.375 |

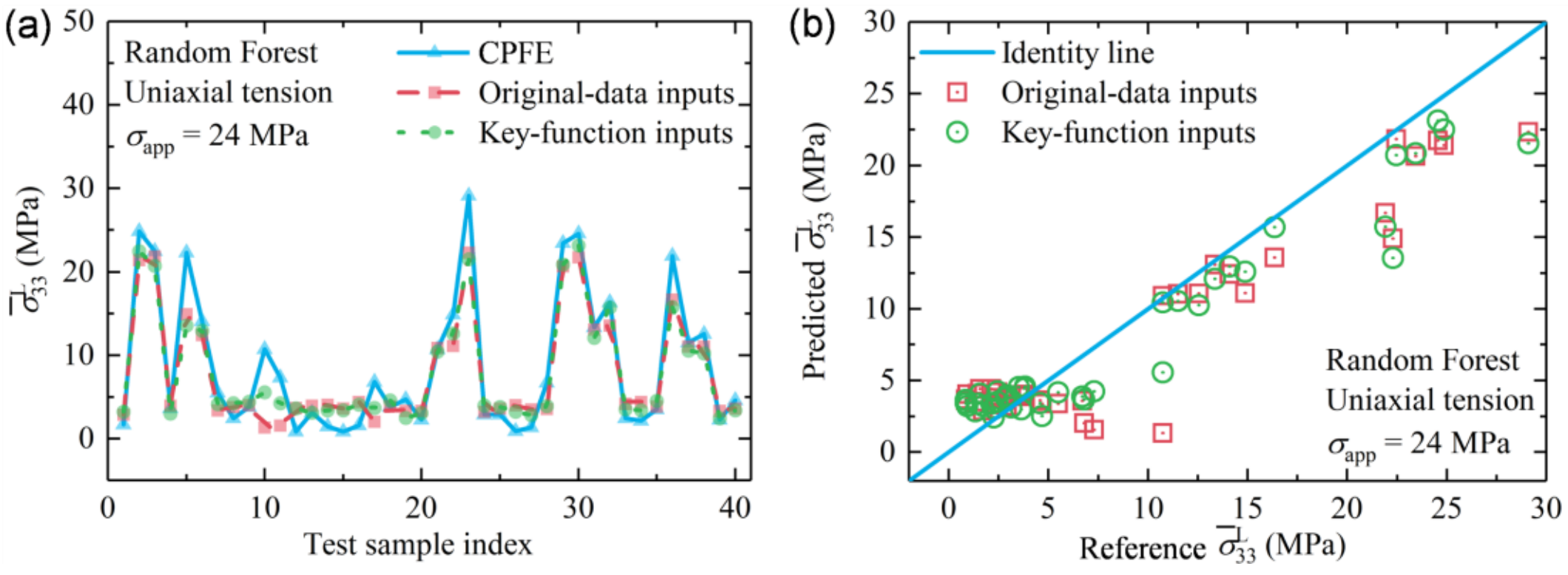

Fig. D 1 Comparison of the performance in predicting the average GB normal stress using random forest models trained with the original characteristics and the key functions as inputs: (a) Performance of the trained models on the test set; (b) comparison between the model predictions and the reference

values.

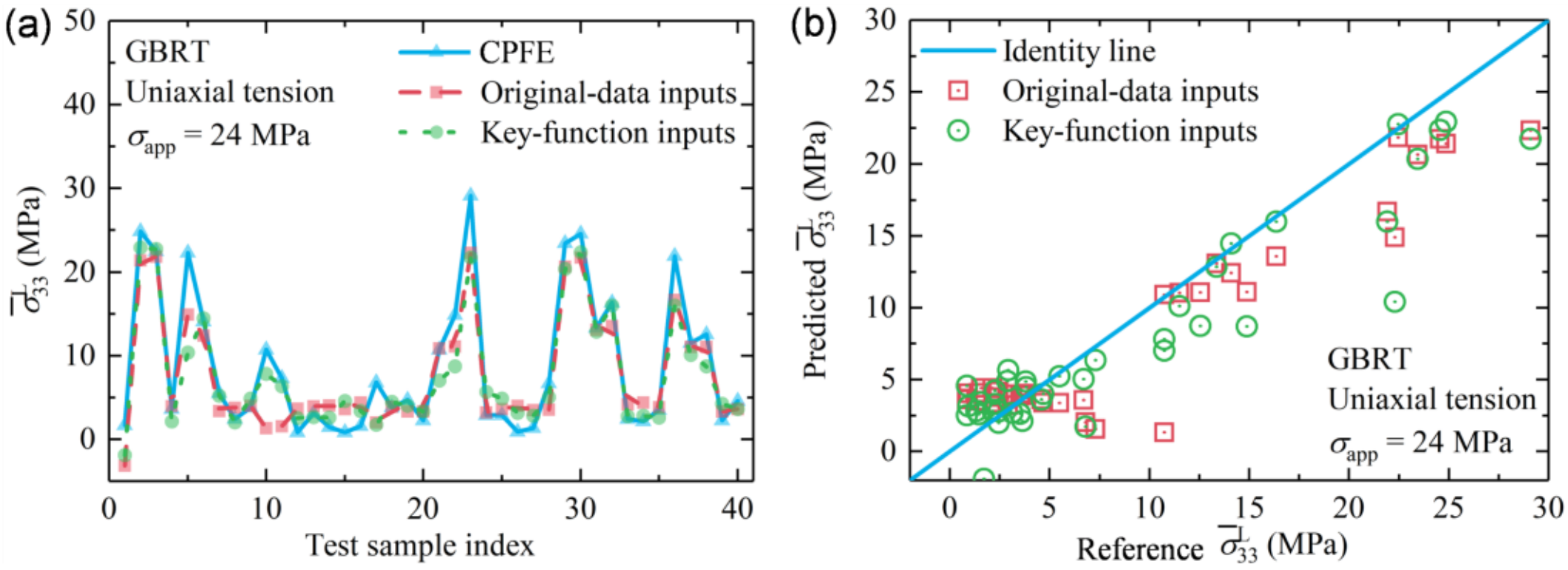

Fig. D 2 Comparison of the performance in predicting the average GB normal stress using gradient boosting regression tree models trained with the original characteristics and the key functions as inputs: (a) Performance of the trained models on the test set; (b) comparison between the model predictions and the reference values.

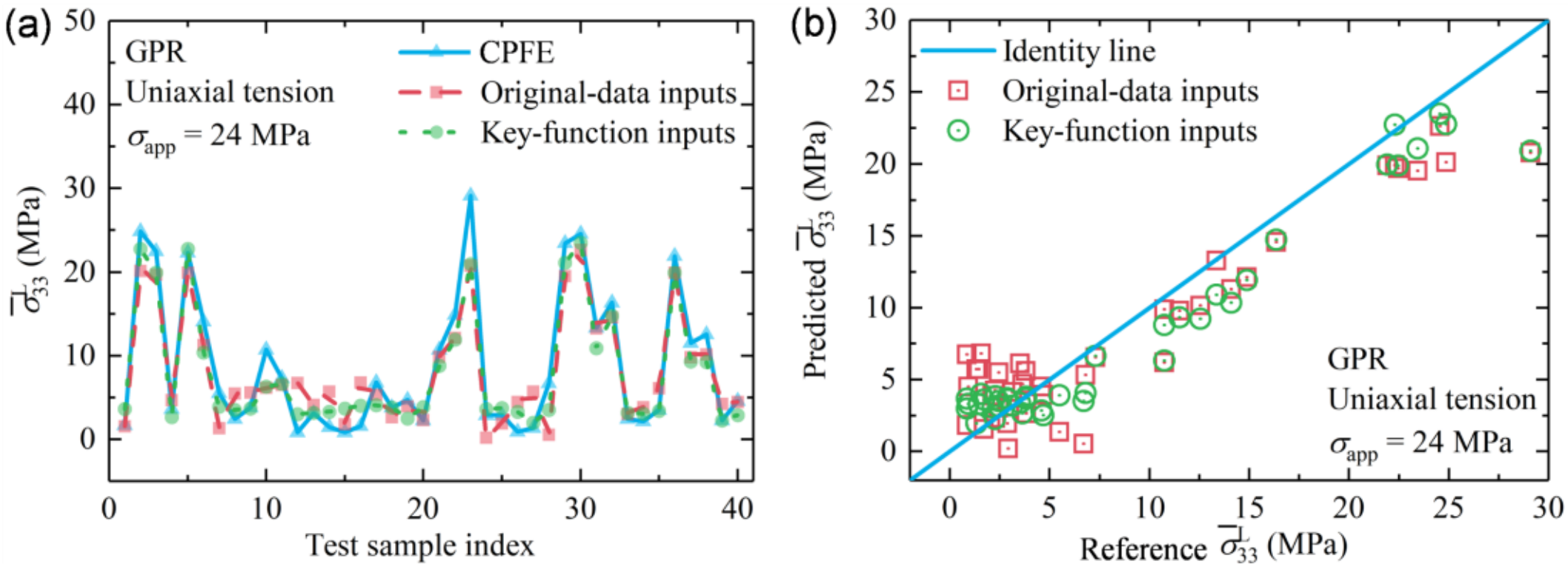

Fig. D 3 Comparison of the performance in predicting the average GB normal stress using gaussian process regression models trained with the original characteristics and the key functions as inputs: (a) Performance of the trained models on the test set; (b) comparison between the model predictions and the reference values.

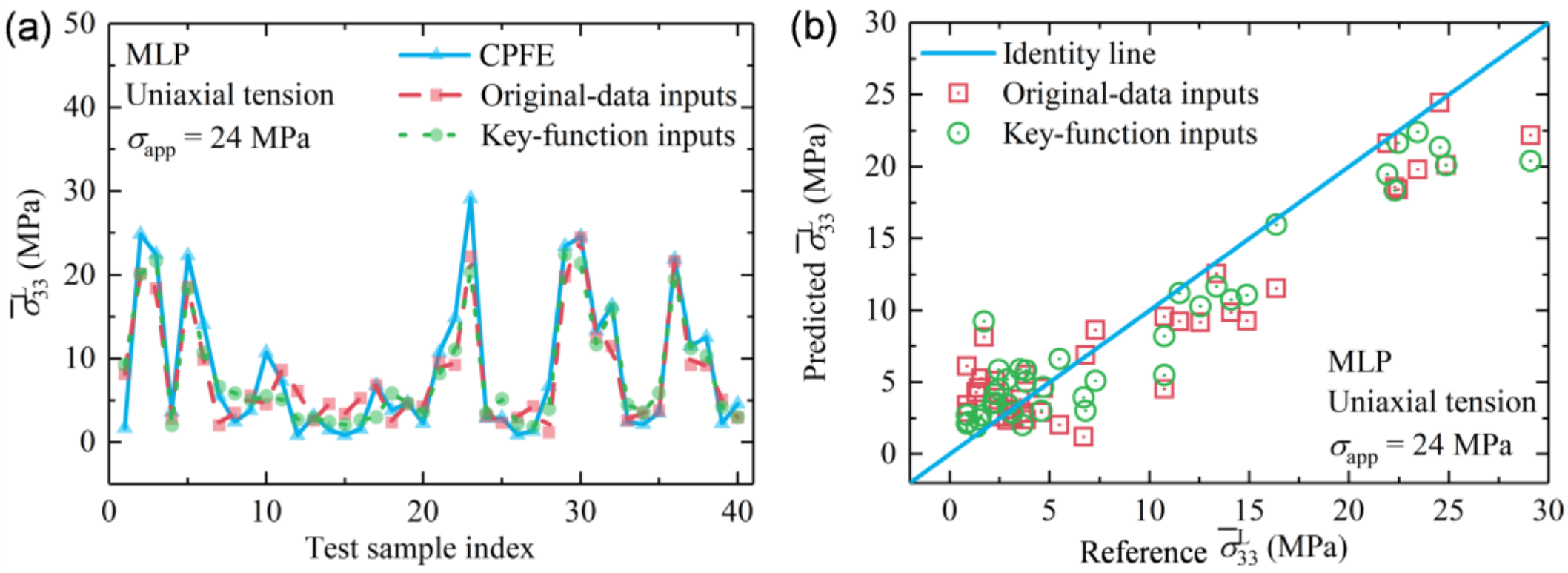

Fig. D 4 Comparison of the performance in predicting the average GB normal stress using multilayer perceptron models trained with the original characteristics and the key functions as inputs: (a) Performance of the trained models on the test set; (b) comparison between the model predictions and the reference values.

The test-set performance of the models trained with RF, GBRT, GPR, and MLP is presented in Figs. Fig. D 1~Fig. D 4, respectively. Panels (a) compares the predicted values obtained using the original inputs and the key candidate functions with the CPFE results for the individual test samples. Panels (b) compares the distributions of the predicted stresses from the two input strategies around the identity line with respect to the CPFE reference stress. As seen from these figures, both the original-data inputs and the key-function inputs yield reasonable predictive accuracy for the GB normal stress $\overline{\sigma}_{33}^{L}$.